\documentclass[notitlepage,showkeys,superscriptaddress]{revtex4-1} 

\usepackage{amssymb}
\usepackage{subfigure}
\usepackage{amsmath}

\usepackage{graphicx}
\usepackage{epstopdf}
\usepackage{color}
\usepackage{ulem}

\newcommand{\scl}{0.2} 
\newcommand{\scla}{0.25} 

\begin{document}

\title{Continuous families of solitary waves in non-symmetric complex potentials: \\
A Melnikov theory approach}

\date{\today}	

\author{Yannis Kominis}
\affiliation{School of Applied Mathematical and Physical Science, National Technical University of Athens, Athens, Greece	}

\author{Jes\'us Cuevas-Maraver}
\affiliation{Grupo de F\'{\i}sica No Lineal, Universidad de Sevilla, Departamento de F\'{i}sica Aplicada I, Escuela Polit\'ecnica Superior.
C/ Virgen de \'{A}frica, 7, 41011-Sevilla, Spain \\
Instituto de Matem\'{a}ticas de la Universidad de Sevilla (IMUS). Edificio Celestino Mutis. Avda. Reina Mercedes s/n, 41012-Sevilla, Spain}

\author{Panayotis G. Kevrekidis}
\affiliation{Department of Mathematics and Statistics, University of Massachusetts, Amherst, MA 01003-4515, USA}

\author{Dimitrios J. Frantzeskakis}
\affiliation{Department of Physics, National and Kapodistrian University of Athens, Panepistimiopolis, Zografos, Athens 15784, Greece}

\author{Anastassios Bountis}
\affiliation{Department of Mathematics, School of Science and Technology, Nazarbayev University, Astana 010000, Republic of Kazakhstan}

\begin{abstract}
  The existence of stationary solitary waves in symmetric and non-symmetric complex potentials is studied by means of Melnikov's perturbation method. The latter provides analytical conditions for the existence of such waves that bifurcate from the homogeneous nonlinear modes of the system and are located at specific positions with respect to the underlying potential. It is shown that the necessary conditions for the existence of continuous families of stationary solitary waves, as they arise from Melnikov theory, provide general constraints for the real and imaginary part of the potential, that are not restricted to symmetry conditions or specific types of potentials. Direct simulations are used to compare numerical results with the analytical predictions, as well as to investigate  the propagation dynamics of the solitary waves. 
\end{abstract}
\keywords{solitary waves; complex potentials; symmetry breaking; PT-symmetry; Melnikov's method; homoclinic bifurcations}
\maketitle

\section{Introduction}
Wave self-localization in spatially inhomogeneous, nonconservative and nonlinear systems emerges as a subject of increasing research interest from both a theoretical and a technological point of view. The formation of solitary waves in nonconservative systems has distinct properties, as compared to the case of conservative systems; the most important difference is that, generally, such waves correspond to isolated stationary solutions in the former case, whereas they form continuous solution families in the latter case \cite{Akhmediev_book}. From a practical point of view,  significant technological applications related to non-Hermitian photonics necessitate the consideration of the interplay between nonlinearity and refractive index as well as gain/loss inhomogeneity, with respect to light localization and propagation \cite{Feng_2017, Longhi_2017, Christodoulides_2018}. In this context, the formation and propagation of solitary waves has been studied under spatially periodic modulation of the linear refractive index with homogeneous \cite{Sakaguchi_2008, Kominis_2012a, Kominis_2012b} or inhomogeneous \cite{Musslimani_2008, Abdullaev_2010, Abdullaev_2011, Miri_2012, Mihalache_2012a, Nixon_2012, He_2013} gain and loss, as well as in aperiodic configurations where defect \cite{Barak_2008, Lam_2009, Kartashov_2010a, Tsang_2010, Zhou_2010, Wang_2011, Hu_2011, Kartashov_2011, Borovkova_2012, Wang_2012, Ye_2013, Abdullaev_2013, Devassy_2017} and surface \cite{Kartashov_2007, Kartashov_2010b, He_2012a} localized modes have been shown to be supported by localized gain distributions. Moreover, cases of spatial modulation of the nonlinear refractive index and gain/loss properties have been considered \cite{He_2012b, Mihalache_2012b, Nath_2015, Kartashov_2016, Zhou_2017}.\

The spatial inhomogeneity of the refractive index and the gain/loss properties of such media correspond to the real and the imaginary part of a complex potential in a nonlinear Schr\"{o}dinger (NLS) equation governing the wave profile formation and propagation. Based on concepts originating from the context of quantum mechanics, the case of Parity-Time (PT) symmetric complex potentials has been initially considered \cite{Konotop_2016,Dmitriev2016}. In such cases, the spatial profiles of the refractive index and the gain/loss are, respectively, even and odd functions. However, in several realistic photonics applications such complex potentials can be either inherently or intentionally asymmetric, necessitating the extension of studies on solitary wave formation and dynamics in non-symmetric complex potentials \cite{Kominis_2015a, Kominis_2015b, Cuevas_2018}. Although wave localization in PT-symmetric complex potentials has been well-studied, and the existence of continuous families of solitary waves has been shown \cite{Konotop_2016}, the existence of such families in non-symmetric complex potentials is still under investigation. Recently, continuous families of solitary waves were found to exist \cite{Tsoy_2014, Konotop_2014, Yang_2014, Yang_2015, Chen_2016} in non-PT-symmetric potentials of a special form (so-called Wadati-type) \cite{Wadati_2008}; it has been shown that such form is a necessary condition for the existence of continuous solitary wave families bifurcating from the linear modes of the system \cite{Nixon_2016}.\

In this work, we revisit the conditions for the formation of solitary waves in the most general, either symmetric or non-symmetric, linear and nonlinear complex potentials, starting from the homogeneous limit. The wave formation  is governed by a two-degree-of-freedom non-autonomous dynamical system, with the solitary wave solutions corresponding to homoclinic points of the system. The existence of such points is investigated by means of Melnikov's perturbation theory \cite{Gruendler_1985, Yamashita_1992, Chow_1992, Bountis_1997, Bountis_1999} that provides analytical results elucidating conditions for the existence of multiple continuous families of solitary waves, located at specific points within the spatially inhomogeneous structures. While such persistence conditions have often been used in the presence of Hamiltonian perturbations~\cite{kapitula,jones}, 
their application is less common, to the best of our knowledge, in setups such as the present one with (especially unbalanced) gain and loss.
The analytical results and the estimations for the specific wave locations are utilized in the numerical solution of the dynamical system towards identifying such analytically predicted waves. Notably, the conditions for the existence of continuous families of solitary waves are much more general than those corresponding to PT-symmetry or to the aforementioned special form of the complex potential. The latter appears as a necessary condition for the {\it existence} of a homoclinic orbit, corresponding to a solitary wave, when starting from the linear modes of the inhomogeneous system \cite{Nixon_2016}. On the contrary, in our approach, the starting point consists of the homogeneous nonlinear system which is integrable and features a homoclinic orbit, and our conditions ensure its {\it persistence} under inhomogenous perturbations, so that the resulting continuous families of solitary waves bifurcate from the nonlinear modes of the homogeneous system.\

The paper is organized as follows. In Section II, we present the model of solitary wave formation and propagation and the utilization of Melnikov's perturbation method for the existence of homoclinic points corresponding to stationary solitary solutions of the system. In section III we utilize the analytical results of Melnikov's theory in order to provide conditions for solitary wave existence, as well as to pinpoint their location within the inhomogeneous structure for linear and nonlinear, symmetric or asymmetric complex potentials; the spatial profiles and the propagation dynamics of the stationary solutions are presented. Finally, the conclusions of this work and some possible future directions
are summarized in Section IV.\

\section{NLS with complex potential and Melnikov's method}
Wave propagation in a nonlinear optical medium with spatially inhomogeneous refractive index and gain/loss modulation is described by the inhomogeneous NLS equation:
\begin{equation}
i\psi_z+ \psi_{xx}+2|\psi|^2\psi+\epsilon \left[U_1(x)+U_2(x)|\psi|^2\right]\psi=0, 
\label{NLS}
\end{equation}
where $\psi$ is the wave field envelope, $z$ the normalized propagation distance, and $x$ the scaled transverse coordinate. $U_i=V_i(x)+iW_i(x)$ $(i=1,2)$ are the linear and nonlinear complex potentials with their real and imaginary parts corresponding to the inhomogeneity of the linear ($i=1$) and the nonlinear ($i=2$) refractive index and gain/loss, respectively, and $\epsilon$ is a dimensionless parameter related to the strength of the modulation. The solitary wave solutions that can be supported in such configurations can be found as the stationary nonlinear modes of the system having the form:
\begin{equation}
 \psi(x,z)=\left[u(x)+iv(x) \right]e^{i\beta z}, 
 \label{stationary}
\end{equation}
with  $\beta$ being the real propagation constant and $u(x), v(x)$ being real functions describing the complex transverse profile of the stationary mode. Substitution of the stationary solutions (\ref{stationary}) in the NLS Eq.~(\ref{NLS}) leads to the following system of coupled ODEs:
\begin{eqnarray}
 u_{xx}-\beta u +2(u^2+v^2)u+\epsilon \left\{ V_1(x)u-W_1(x)v+[V_2(x)u-W_2(x)v](u^2+v^2) \right \}&=&0, \nonumber \\
 v_{xx}-\beta v +2(u^2+v^2)v+\epsilon \left \{ V_1(x)v+W_1(x)u+[V_2(x)v+W_2(x)u](u^2+v^2) \right \}&=&0. \label{p-Manakov}
\end{eqnarray}
In order to study solitary waves bifurcating from the respective nonlinear modes of the homogenous system, we assume that $\epsilon$ is sufficiently small, so that the terms of the inhomogeneous complex potentials $U_i(x)$ can be considered as perturbations of the unpertubed system:
\begin{eqnarray}
 u_{xx}-\beta u +2(u^2+v^2)u&=&0, \nonumber \\
 v_{xx}-\beta v +2(u^2+v^2)v&=&0. \label{Manakov}
\end{eqnarray}
These coupled ODEs are the equations of motion of a two-degree-of-freedom integrable Hamiltonian system, with Hamiltonian:
\begin{equation}
 H(u,u_x,v,v_x)=\frac{1}{2}\left(u_x^2+v_x^2\right)-\frac{1}{2}\beta\left(u^2+v^2\right)+\frac{1}{2}\left(u^2+v^2\right)^2. \label{H}
\end{equation}
The Hamiltonian is an invariant of the system along with the quantity
\begin{equation}
 F(u,u_x,v,v_x)=uv_x-vu_x. \label{F}
\end{equation}
The dynamical system (\ref{Manakov}) corresponds to a two-degree of freedom nonlinear oscillator, with $x$ playing the role of time, for which $H$ is the total energy and $F$ is the angular momentum. Using the transformation $u=r\cos\theta, v=r\sin\theta$, the quantities $H$ and $F$ become:
\begin{equation}
 H=\frac{1}{2}\dot{r}^2-\frac{1}{2}\beta r^2+\frac{1}{2}r^4+\frac{F^2}{2r^2},
\end{equation}
and
\begin{equation}
 F=r^2\dot{\theta}.
\end{equation}
For $F\neq0$ the Hamiltonian goes to infinity when $r, \dot{r}$ go to zero, therefore solitary wave profiles correspond to the case where $F=0$ and have the following form
\begin{eqnarray}
 u_0(x)&=&p_0(x-x_0)\cos\theta_0, \nonumber \\
 v_0(x)&=&p_0(x-x_0)\sin\theta_0, \label{homoclinic}
\end{eqnarray}
where 
\begin{equation}
 p_0(x;\beta)=\sqrt{\beta}\mbox{sech}\left(\sqrt{\beta}x\right),
\end{equation}
and $\theta_0$, $x_0$ being arbitrary constants. The arbitrariness of $\theta_0$ is related to the invariance of the perturbed (and the unperturbed) NLS equation (\ref{NLS}) under the transformation $\psi \rightarrow \psi\exp(i\theta_0)$ and $\theta_0$ can be set equal to zero, without loss of generality, whereas the arbitrariness of $x_0$ reflects the translational invariance of the unperturbed NLS equation. These solutions are members of a two-parameter $(\beta,x_0)$ family of orbits homoclinic to the saddle located at the origin in the four-dimensional phase space of the system. \

The presence of the inhomogeneous perturbation in Eq.~(\ref{p-Manakov}) removes the translational invariance of the system; the stable and unstable manifolds of the origin are no longer joined smoothly to form homocilinc orbits, but may intersect transversely. The orbits corresponding to such transverse intersections are the stationary solitary wave profiles of the perturbed system, with the discrete set of $x_0$ parametrizing the respective solutions and providing the transverse positions where solitary waves are located with respect to the inhomogeneous complex potential \cite{Kominis_2008}. For sufficiently small perturbations, Melnikov's method provides analytical information for the existence of such homoclinic points \cite{Gruendler_1985, Yamashita_1992, Chow_1992, Bountis_1997, Bountis_1999}, in terms of the simple zeros of the Melnikov vector:
\begin{equation}
 \vec{M}=[M_1(x_0;\theta_0,\beta),M_2(x_0;\theta_0,\beta)],
\end{equation}
with
\begin{eqnarray}
 M_1(x_0;\theta_0,\beta)&=&\int_{-\infty}^{+\infty}dH(\gamma_0(x;\theta_0,\beta))g(x-x_0,\gamma_0(x;\theta_0,\beta))dx,\\
 M_2(x_0;\theta_0,\beta)&=&\int_{-\infty}^{+\infty}dF(\gamma_0(x;\theta_0,\beta))g(x-x_0,\gamma_0(x;\theta_0,\beta))dx.
\end{eqnarray}
Here, $dH=\partial H / \partial \vec{X}$, $dF=\partial F / \partial \vec{X}$, $\vec{X}=(u,v,\partial u/\partial x,\partial v/\partial x)$, and
\begin{equation}
 g=g_1+g_2(u^2+v^2),
 \end{equation}
with $g_i=-[0, 0, V_i(x)u-W_i(x)v, V_i(x)v+W_i(x)u]$ being the perturbative part evaluated at the unperturbed homoclinic orbit (\ref{homoclinic}),  $\gamma_0(x;\theta_0,\beta)=(u_0,v_0,\partial u_0/\partial x,\partial v_0/\partial x)$, according to \cite{Yamashita_1992, Chow_1992}. Therefore, the  components of the Melnikov vector are given as
\begin{eqnarray}
M_1(x_0;\theta_0,\beta)&=&\int_{-\infty}^{+\infty} \left[\frac{\partial V_1(x-x_0)}{\partial x}+\frac{\partial V_2(x-x_0)}{\partial x}p_0^2(x;\beta)\right]p_0^2(x;\beta)dx, \\
M_2(x_0;\theta_0,\beta)&=&-\int_{-\infty}^{+\infty}\left[W_1(x-x_0)+W_2(x-x_0)p_0^2(x;\beta)\right]p_0^2(x;\beta) dx.
\end{eqnarray}
This form of the Melnikov vector is general and valid for all types of inhomogeneity of the system, including spatially localized or extended, periodic or quasiperiodic, symmetric or asymmetric transverse profiles of the complex potentials. The set of $x_0$, defining the homoclinic points, are determined as the intersection of the two sets of $x_0^{(i)}$ defined by $M_i(x_0^{(i)})=0$, $(i=1,2)$. The conditions for existence of common solutions of the two equations for $x_0$ determine the forms of the complex potentials that support the existence of solitary waves, in terms of relations between their real and imaginary parts, in the most general way. \

For the case of only linear complex potential $(U_2(x)\equiv0)$, the two equations
\begin{eqnarray}
M_1(x_0;\theta_0,\beta)&=&\int_{-\infty}^{+\infty} \frac{\partial V_1(x-x_0)}{\partial x}p_0^2(x;\beta)dx =0 \\
M_2(x_0;\theta_0,\beta)&=&-\int_{-\infty}^{+\infty}W_1(x-x_0)p_0^2(x;\beta) dx=0 \label{M_12}
\end{eqnarray}
correspond to balance conditions for the ``force'' $(-\partial V_1/ \partial x)$ exerted to the wave by the real part (refractive index) of the potential and the gained/lost wave ``mass'' ($\int|\psi|^2dx$) due to the imaginary part (gain/loss) of the potential.
The first condition implies that the position of the solitary wave with respect to the profile of the real part of the potential must be such that the overlap integral of the wave profile with the ``force'' is zero, suggesting that $x_0$ is related to a sign change of $(-\partial V_1/ \partial x)$, corresponding to a local extremum of $V_1(x)$. The second condition implies that the overlap integral of the wave profile with the linear gain and loss is zero, suggesting that $x_0$ is related to a sign change, that is a zero, of $W_1$. The necessary condition for the existence of a homoclinic point, and therefore a solitary wave, is that the spatial forms of the real and imaginary part are such that an $x_0$ fulfilling both requirements can be found. It is worth mentioning that PT-symmetric linear complex potentials, for which $V_1(-x)=V_1(x)$ and $W_1(-x)=-W_1(x)$,
given the symmetry of $p_0$, fulfill the condition for the existence of a common solution of Eqs. (\ref{M_12}) at $x_0=0$, rendering PT-symmetry a sufficient (but not necessary) condition for the existence of solitary waves bifurcating from the nonlinear modes of the homogeneous system. More generally, for any type of inhomogeneity, the two sets of solutions for $x_0$ coincide under the condition:
\begin{equation}
 W_1(x)=C\frac{\partial V_1(x)}{\partial x}, \label{condition}
\end{equation}
with $C$ being an arbitrary constant, which has also been found as a condition for the existence of an invariant of soliton dynamics, under an effective particle approach \cite{Kominis_2015a, Kominis_2015b}. The above arguments suggest that {\it it is the relation between the real and the imaginary part that determines the existence of solitary waves in complex potentials and not their spatial symmetry properties}, as we will also show in several examples. When nonlinear complex potentials $U_2(x)$ are also considered the respective conditions involve the interplay between the real and imaginary parts of both $U_i(x)$ and can be fulfilled in even more general asymmetric confgurations. In all cases we will show that the zeros of the Melnikov vector will provide suitable initial guesses for the relative position of the solitary waves with respect to the underlying complex potentials, analogously to the case of NLS with real potential \cite{Kominis_2008}.

\section{Existence and propagation properties of solitary waves}
In the following, we consider characteristic cases, where the potentials can be periodic or quasiperiodic and spatially symmetric or asymmetric, given by the following form:
\begin{eqnarray}
 V_i&=&\sum_l v_{i,l}\cos\left(K_{i,l}x+\phi_{i,l}\right),\\
 W_i&=&\sum_l w_{i,l}\sin\left(L_{i,l}x+\xi_{i,l}\right),
\end{eqnarray}
with $i=1,2$ corresponding to the linear and nonlinear complex potential. For such complex potentials, the integrals involved in the Melnikov vector can be calculated analytically as follows:
\begin{eqnarray}
 M_1(x_0;\beta,\theta_0)&=& \sum_l v_{1,l} K_{1,l} F_1(K_{1,l},\beta)\sin\left(K_{1,l}x_0-\phi_{1,l}\right) +\sum_l v_{2,l}K_{2,l} F_2(K_{2,l},\beta)\sin\left(K_{2,l}x_0-\phi_{2,l}\right), 
 \nonumber  \\
 M_2(x_0;\beta,\theta_0)&=& \sum_l w_{1,l} F_1(L_{1,l},\beta)\sin\left(L_{1,l}x_0-\xi_{1,l}\right) +\sum_l w_{2,l} F_2(L_{2,l},\beta)\sin\left(L_{2,l}x_0-\xi_{2,l}\right), 
 \label{M12}
 \end{eqnarray}
 where
 \begin{eqnarray}
  F_1(Y,\beta)&=&\frac{\pi Y}{2\sinh\left( \frac{\pi Y}{2\sqrt{\beta}}\right)}, \nonumber \\
  F_2(Y,\beta)&=&\frac{\pi Y(Y^2+4\beta)}{12\sinh\left( \frac{\pi Y}{2\sqrt{\beta}}\right)}. \label{F12}
 \end{eqnarray}
These functions depend strongly on the ratio of the two characteristic spatial length scales of the system: the width of the unperturbed soliton $(\sim \beta^{-1/2})$ and the period of the potential $(\sim Y^{-1})$, with $Y=K_{i,l}, L_{i,l}$. Moreover, it is clear that the components of the Melnikov vector have simple zeros since their partial derivatives with respect to $x_0$ (and $\beta$) are not vanishing at the zeros, so the latter correspond to homoclinic points where the stable and unstable manifolds intersect transversely. \

In the following, we focus our investigation to intermediate perturbation strength values ($\epsilon$) and to localized solutions with a transverse spatial extent comparable to the characteristic scales of the complex potentials corresponding to propagation constant $\beta=0.1$ for the case of $K_{i,l}, L_{i,l} = 1-2$; in such cases the effect of the potential on the solitary wave profile is more pronounced. Moreover, we consider cases where the amplitude of the imaginary part of the potential is smaller than the amplitude of the real part, since in the opposite case, not only the solitary waves but even the zero background solution is generally unstable.\

Prior to showing the properties of stationary soliton solutions of  Eq.~(\ref{p-Manakov}), we will briefly discuss the numerical methods used for attaining the spatial profile of such solitary waves. As it is well known (and also checked during this study), numerically exact (up to machine precision)  waveforms can be found in the case of PT-symmetric potentials by using fixed-point algorithms like the Newton-Raphson method. The most tricky issue arises when non-PT-symmetric potentials are considered, as fixed-point algorithms, at least in our experience, do not generally converge to machine precision. Thus, alternative methods must be used. First of all, we considered the Levenberg-Marquardt algorithm (LMA) which, for instance, has proved to be useful for finding capillary solitary waves \cite{Dutykh_2016}. The main drawback of such an algorithm (which is detailed in \cite{Cuevas_2018}) lies in the fact that it is an optimization method, and consequently, it looks for local minima of the residual which do not necessarily have to be zero. The other option is to use a boundary value problem solver which is implemented in Matlab by means of \texttt{bvp4c} command. From the description of Ref.~\cite{Shampine_2003} (in particular, from p.167 therein), in this case the solver residual is indicated to be found  by means of a five-point  Lobatto quadrature formula.\

We have  applied both the LMA and the boundary value problem solver \texttt{bvp4c} for getting stationary soliton solutions when the Newton-type algorithms fail because of the breaking of the PT-symmetry of NLS equation. In both methods, the initial guesses have been chosen with the utilization of the analytical results of the Melnikov's method, so that the unperturbed solutions (\ref{homoclinic}) with $x_0$ corresponding to the simple zeros of the Melnikov vector (\ref{M_12}) have been used. In particular, we have observed that generally the $L^2$-norm of the residual is larger --typically $\sim10^{-6}$-- for the former algorithm and smaller --typically can be made to be around $\sim10^{-10}$-- for the latter. The spatial profiles obtained by the two algorithms are almost identical. In order to study the propagation dynamics of the various stationary solutions we resort to direct numerical computations based on a standard beam propagation  (split-step Fourier) method. In all cases, a level of random noise has been superimposed to the initial profiles in order to trigger possible instabilities and check their robustness. 

\subsection{Linear potential}
Without loss of generality we consider a linear complex potential consisting of two sinusoidal modulations of different amplitude, period and phase, namely:
\begin{eqnarray}
V_1(x)&=&v_{1,1}\cos(K_{1,1}x)+v_{1,2}\cos(K_{1,2}x+\phi_{1,2}), \nonumber \\ W_1(x)&=&w_{1,1}\sin(L_{1,1}x)+w_{1,2}\sin(L_{1,2}x+\xi_{1,2}).
\end{eqnarray}
For a monochromatic real and imaginary part of the complex potential ($v_{1,2}=w_{1,2}=0$) we always have a PT-symmetric configuration. When $K_{1,1}, L_{1,1}$ are incommensurable $(K_{1,1}/L_{1,1}=\mbox{irrational})$ the equations (\ref{M_12}) have only one common solution at $x_0=0$, whereas when they are commensurable $(K_{1,1}/L_{1,1}=m/n)$,  they have common solutions at $x_0=l_1\pi/K_{1,1}=l_2\pi/L_{1,1}$ for integers $l_{1,2}$ such that $ml_2=nl_1$. The condition (\ref{condition}) is fulfilled when $K_{1,1}=L_{1,1}$; in such case we have the maximum number of common solutions $x_0=l\pi/K_{1,1}=l\pi/L_{1,1}$ with $l$ being any integer. For a non-monochromatic real and imaginary part of the potential ($v_{1,2},w_{1,2}\neq0$) with $\phi_{1,2}, \xi_{1,2}  \neq 0$ the configuration is not PT-symmetric, in general. The condition (\ref{condition}) is fulfilled when $K_{1,1}=L_{1,1}$, $K_{1,2}=L_{1,2}$, for any value of $\phi_{1,2}=\xi_{1,2}$ and all the solutions of the two equations (\ref{M_12}) are common, whereas in the general case where  $K_{1,1} \neq L_{1,1}$, $K_{1,2} \neq L_{1,2}$, common solutions may exist for some $x_0$, depending on the values of $K_{1,l}$ and $L_{1,l}$ and the propagation constant $\beta$.\

\begin{figure}[pt]
  \begin{center}
 \subfigure[]{\scalebox{\scl}{\includegraphics{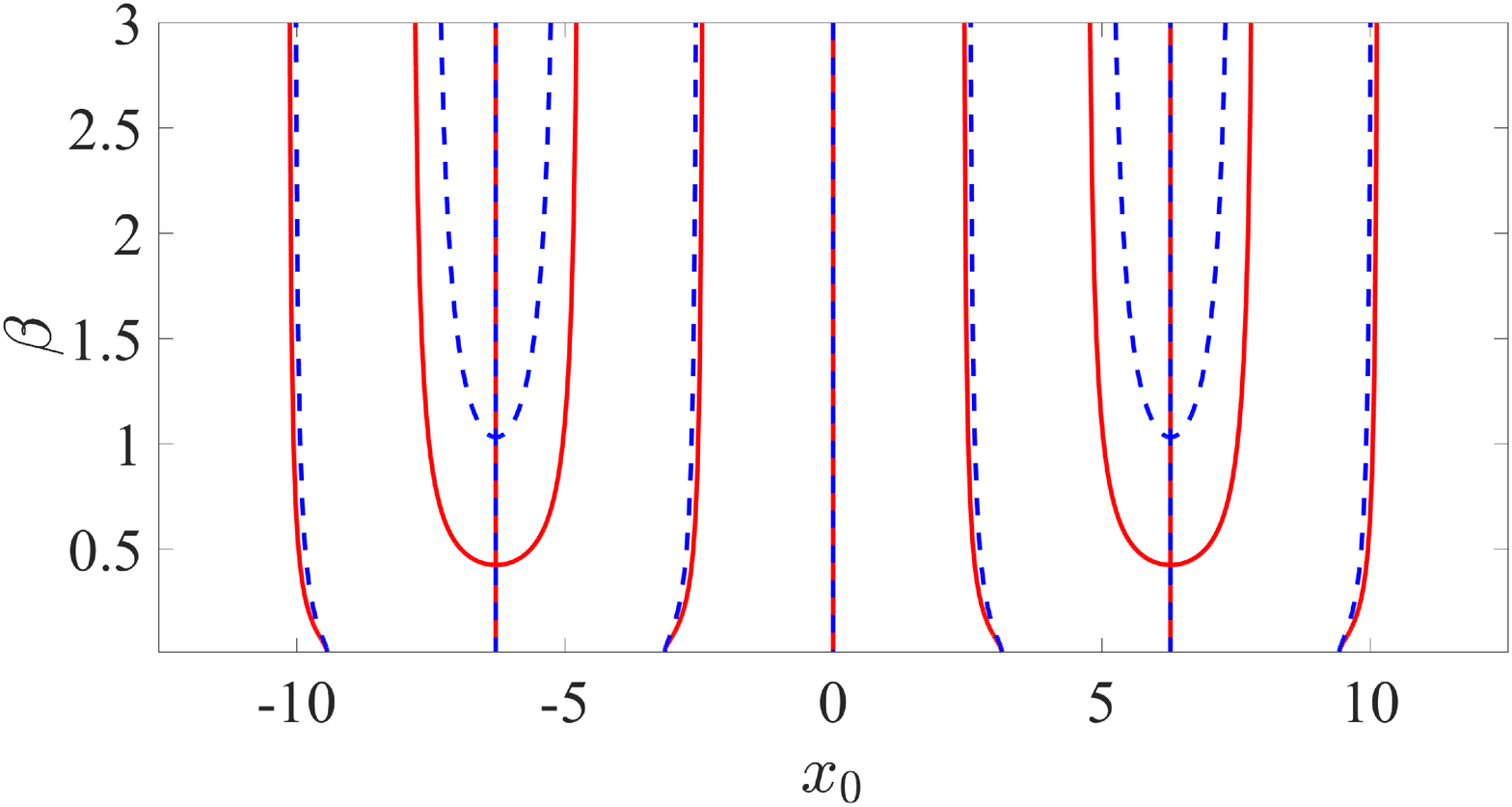}}}
 \subfigure[]{\scalebox{\scl}{\includegraphics{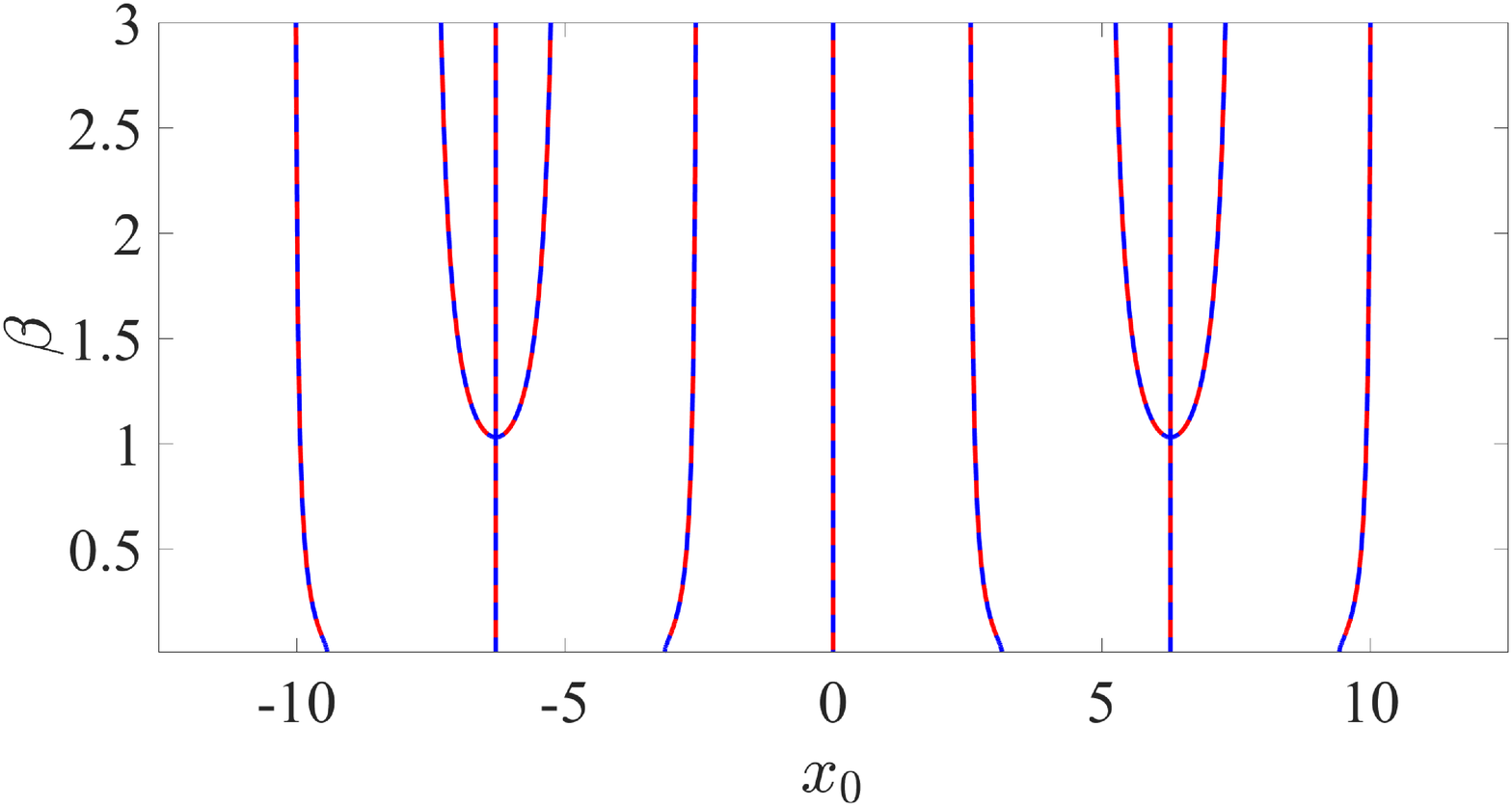}}}
 \subfigure[] {\scalebox{\scl}{\includegraphics{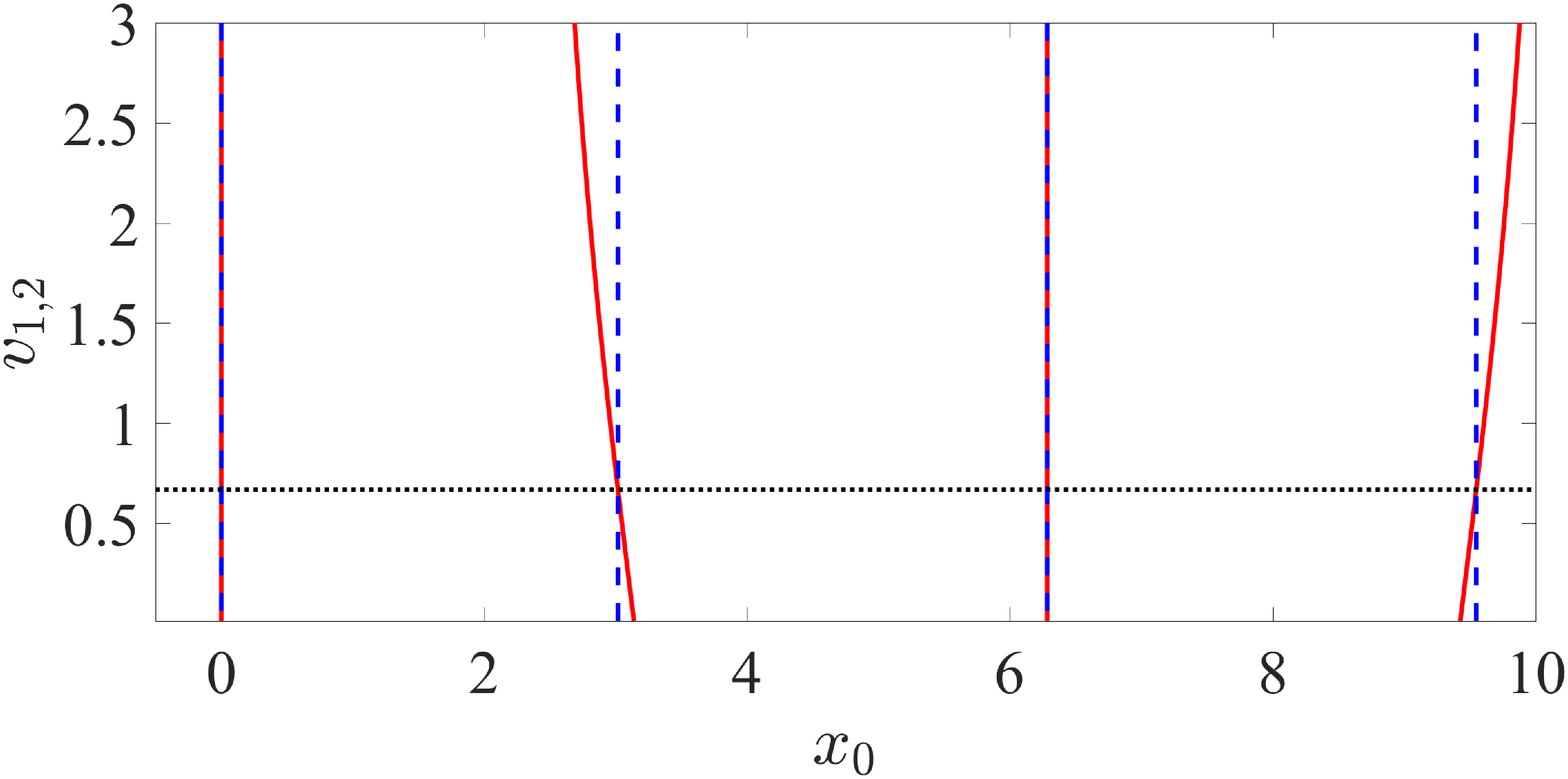}}}
  \caption{Zeros of the two components of the Melnikov vector $M_1$ (red, solid) and $M_2$ (blue, dashed), for a periodic PT-symmetric complex potential with $\epsilon=0.1$, $v_{1,1}=1$, $w_{1,1}=w_{1,2}=0.2$, $K_{1,1}=L_{1,1}=1$, $K_{1,2}=L_{1,2}=3/2$ and $\phi_{1,2}=\xi_{1,2}=0$. (a) $v_{1,2}=1$ (the condition (\ref{condition}) is not fulfilled), (b) $v_{1,2}=2/3$ (the condition (\ref{condition}) is fulfilled), (c) $\beta=0.1$ and varying $v_{1,2}$ (the black dotted line denotes the value of $v_{1,2}$ for which the condition (\ref{condition}) is fulfilled).} \label{fig_m1}
  \end{center}
\end{figure}

\begin{figure}[pt]
  \begin{center}
  \subfigure[]{\scalebox{\scla}{\includegraphics{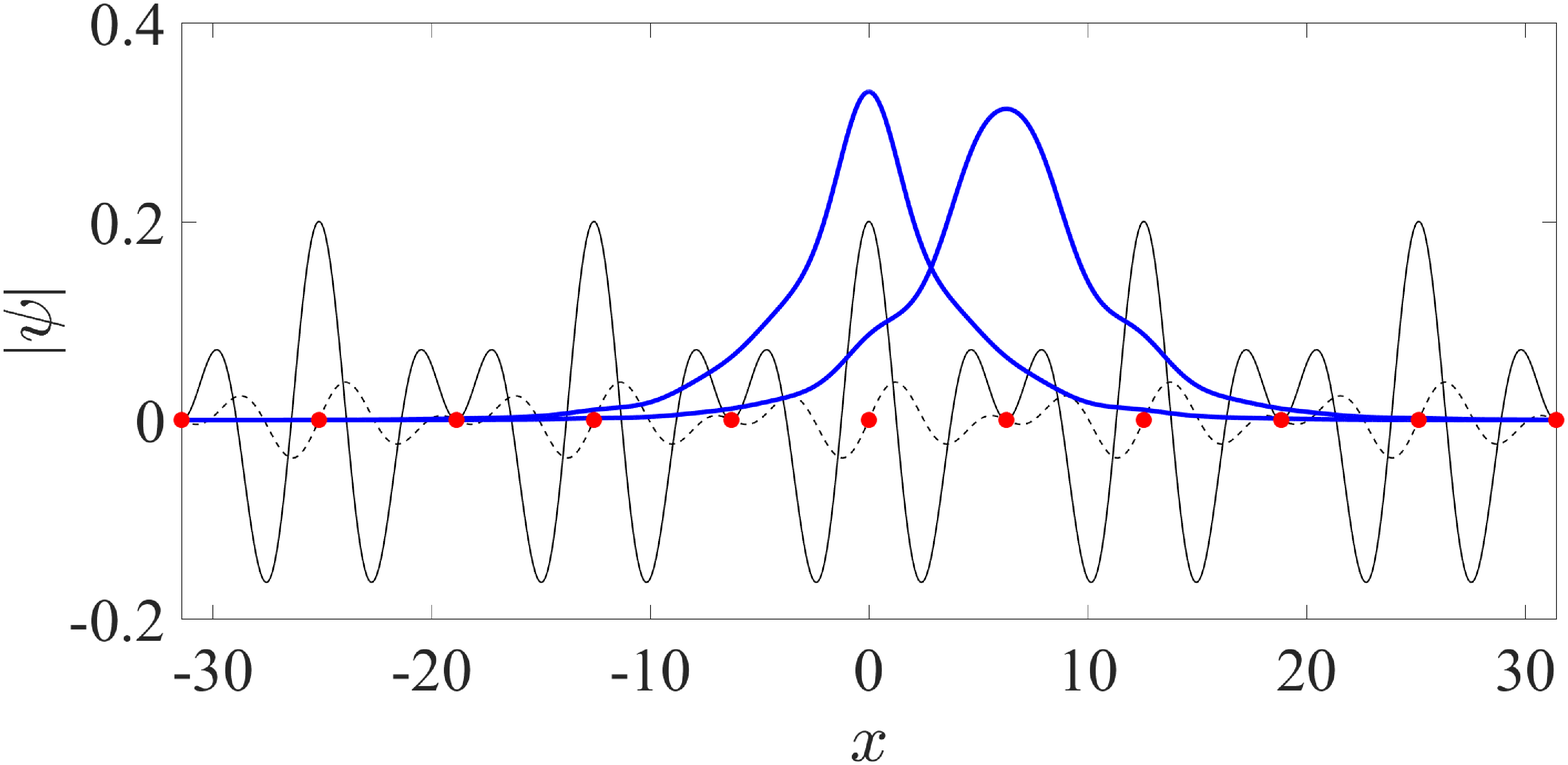}}}
  \subfigure[]{\scalebox{\scla}{\includegraphics{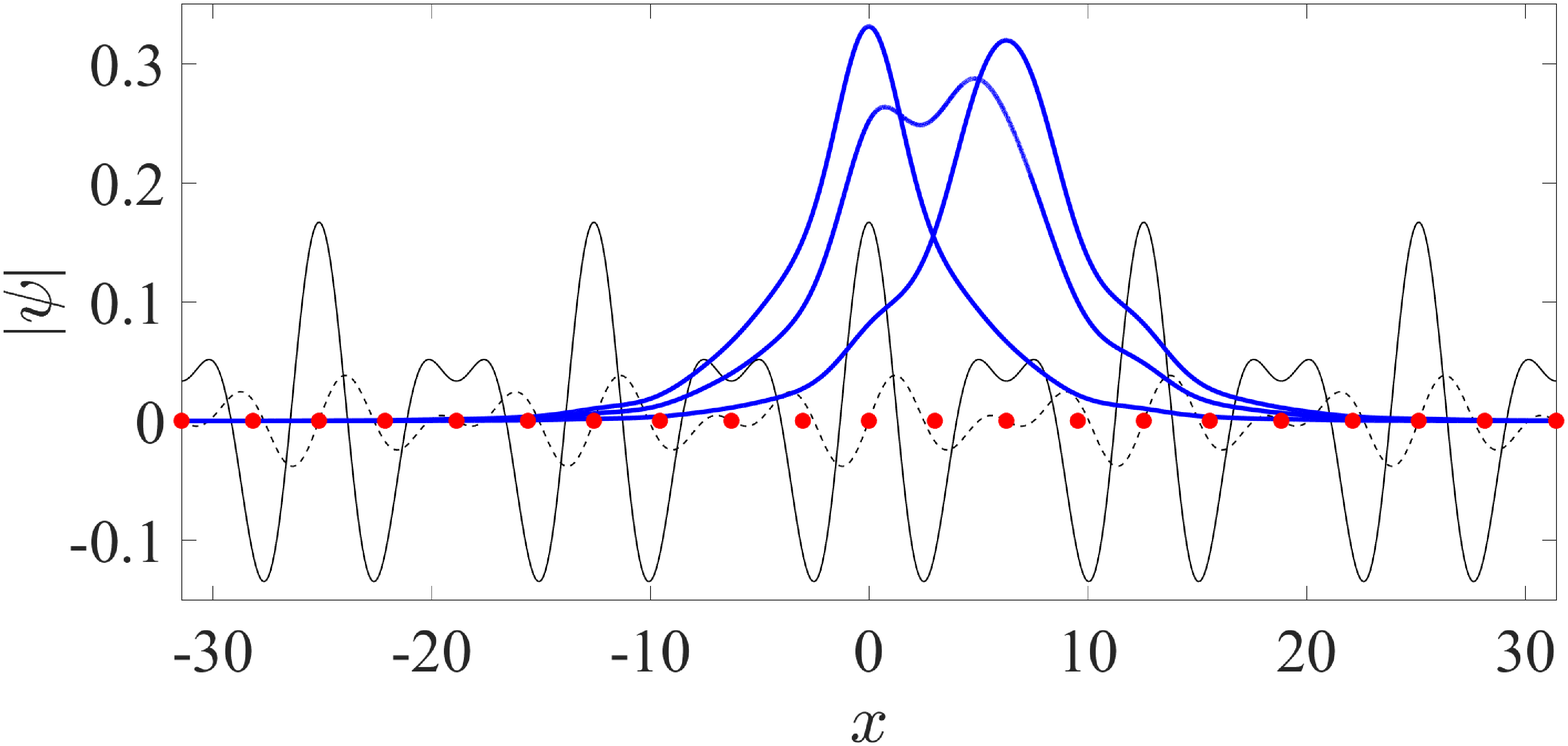}}}
   \caption{Transverse profiles of solitary waves with $\beta=0.1$ and centers corresponding to the zeros of the Melnikov vector  for the cases corresponding to Figs. \ref{fig_m1}(a) and (b), respectively. The black solid and dashed lines depict the real and the imaginary part of the potential and the red circles denote the location of the zeros of the Melnikov function.} \label{fig_12a}
  \end{center}
\end{figure}

\begin{figure}[pt]
  \begin{center}
  {\scalebox{\scl}{\includegraphics{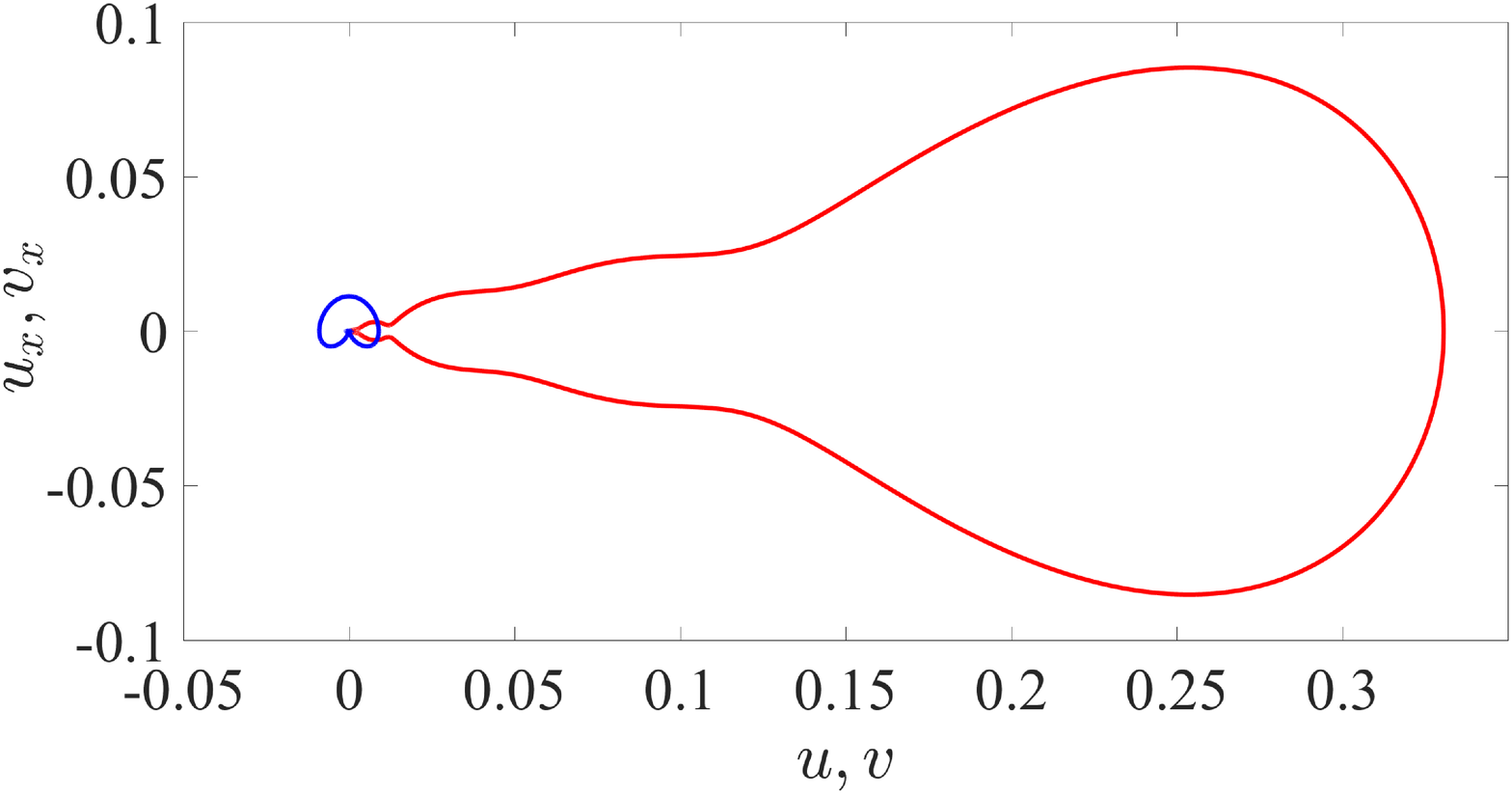}}}
  {\scalebox{\scl}{\includegraphics{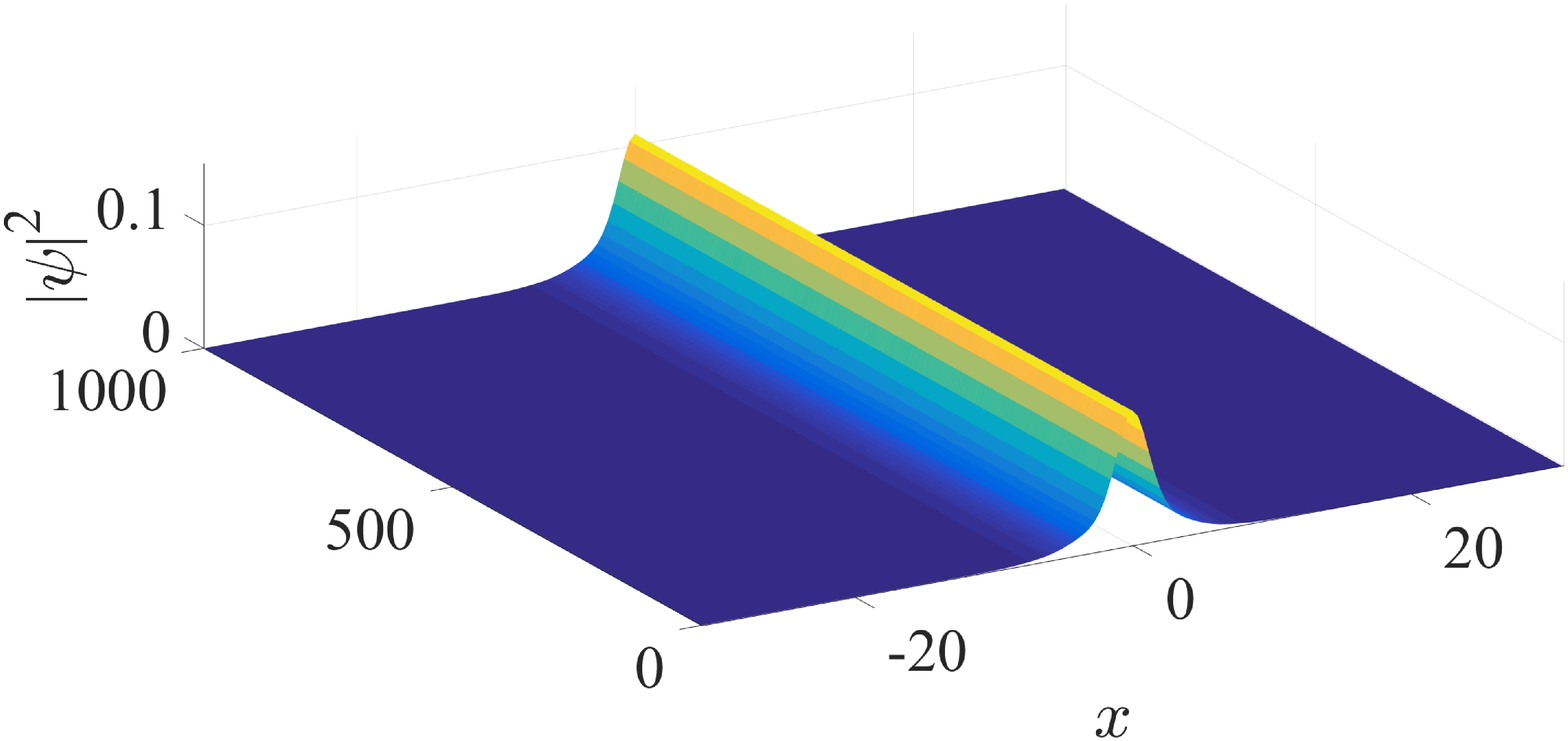}}}\\
  {\scalebox{\scl}{\includegraphics{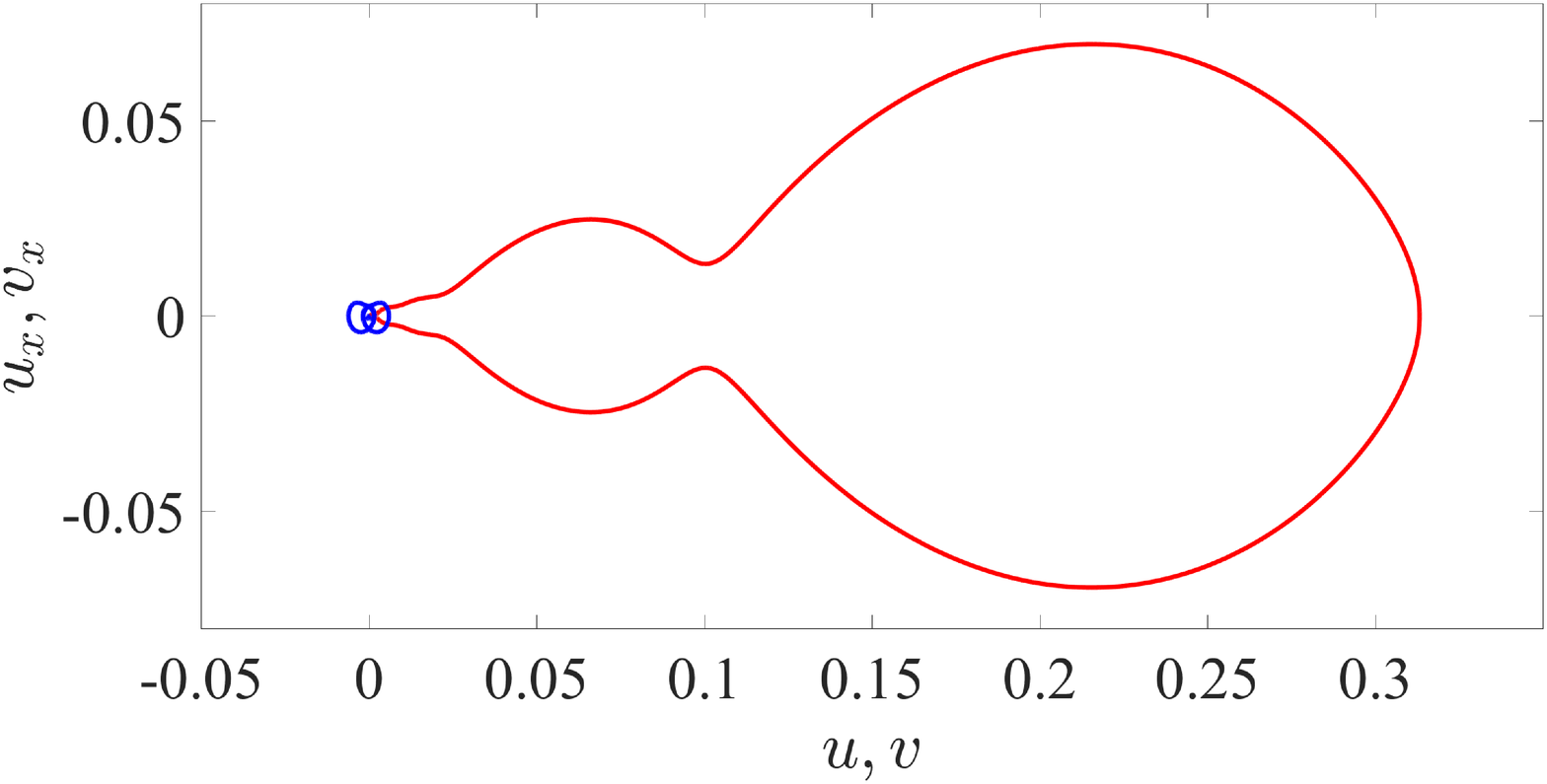}}}
  {\scalebox{\scl}{\includegraphics{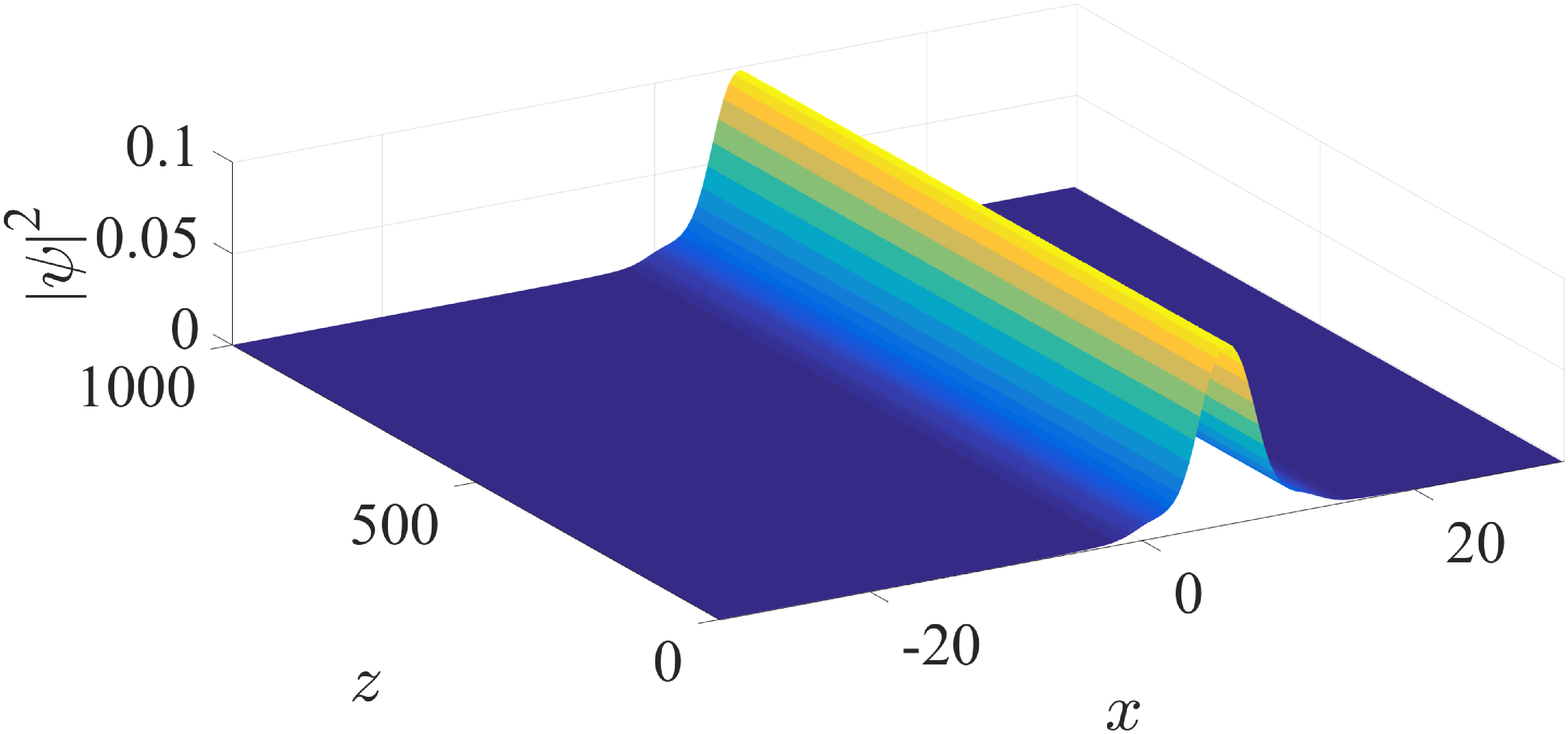}}}
   \caption{Projection of the real ($u$, red) and the imaginary ($v$, blue) parts the homoclinic solutions and propagation dynamics for the case corresponding to Fig. \ref{fig_m1}(a) and Fig. \ref{fig_12a}(a). } \label{fig_1bc}
  \end{center}
\end{figure}

\begin{figure}[pt]
  \begin{center}
  {\scalebox{\scl}{\includegraphics{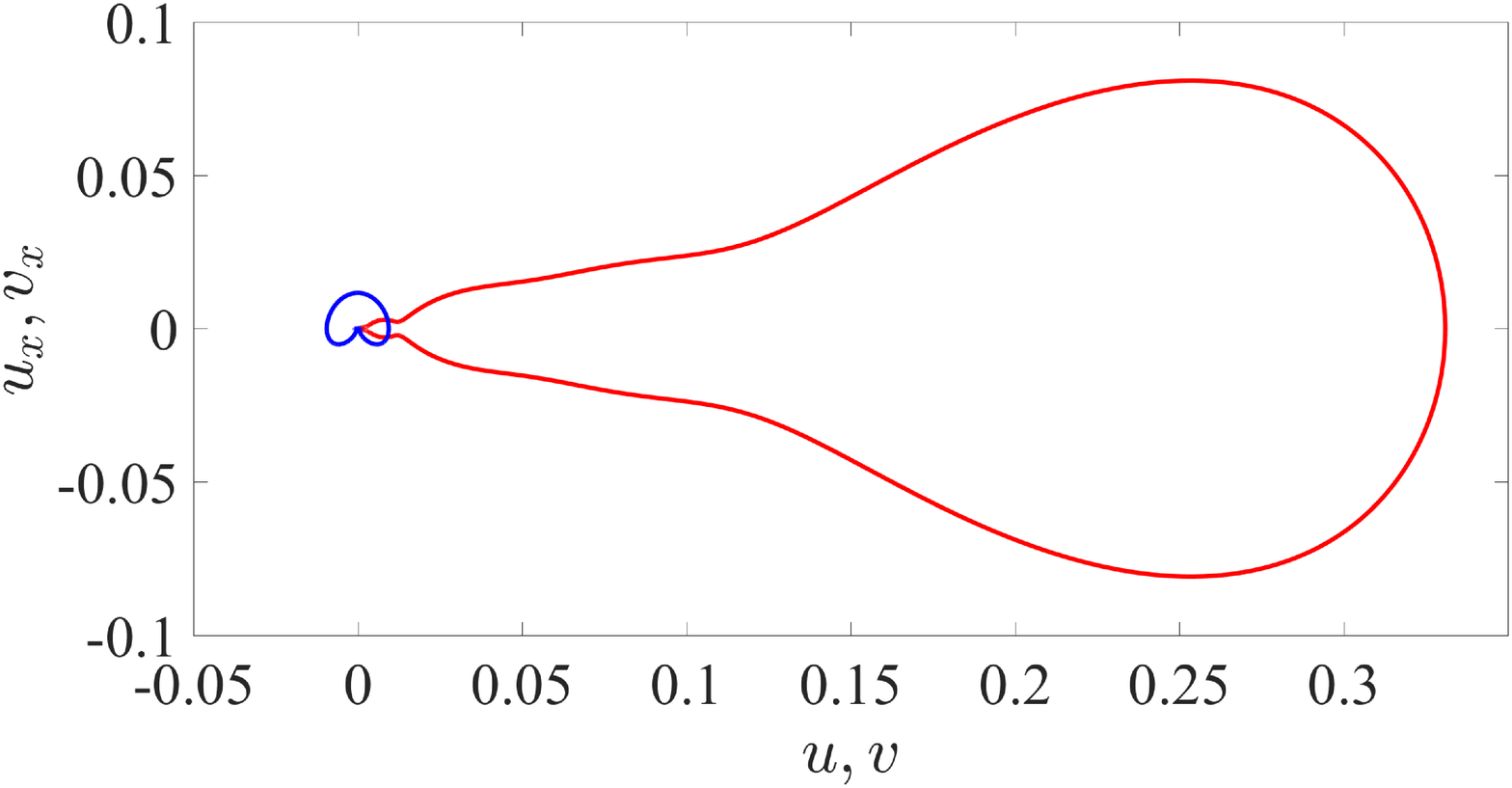}}}
  {\scalebox{\scl}{\includegraphics{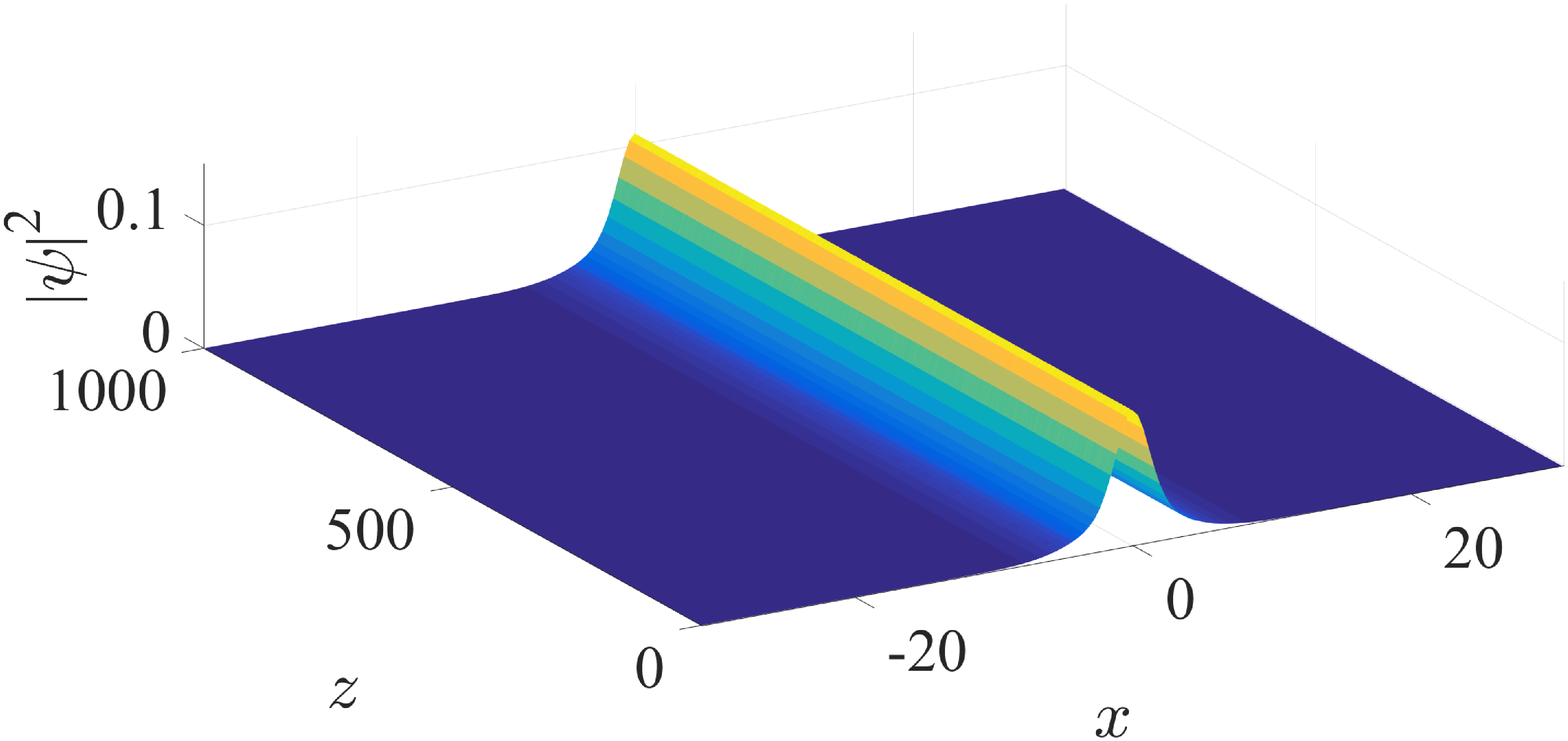}}}\\
  {\scalebox{\scl}{\includegraphics{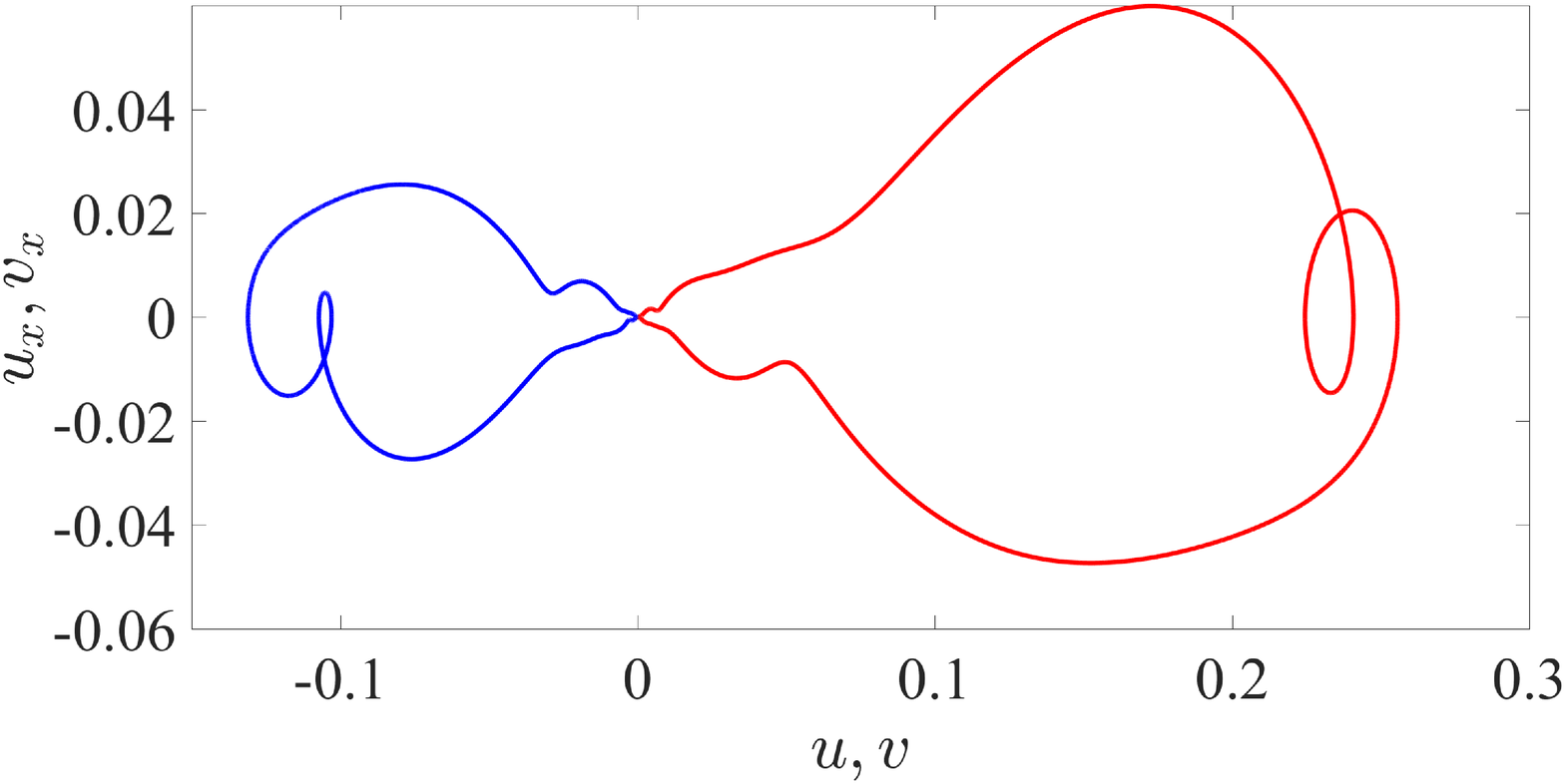}}}
  {\scalebox{\scl}{\includegraphics{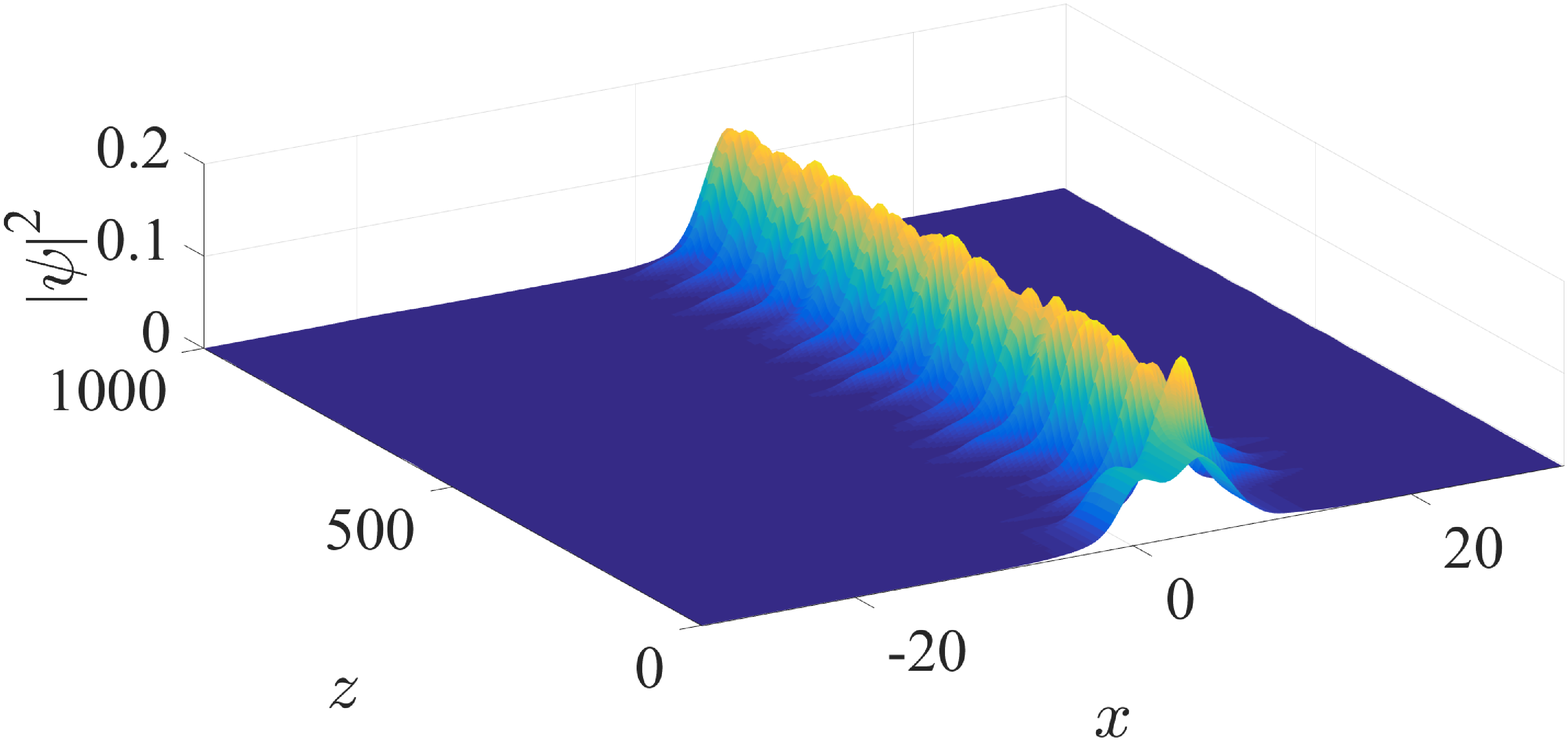}}}\\
  {\scalebox{\scl}{\includegraphics{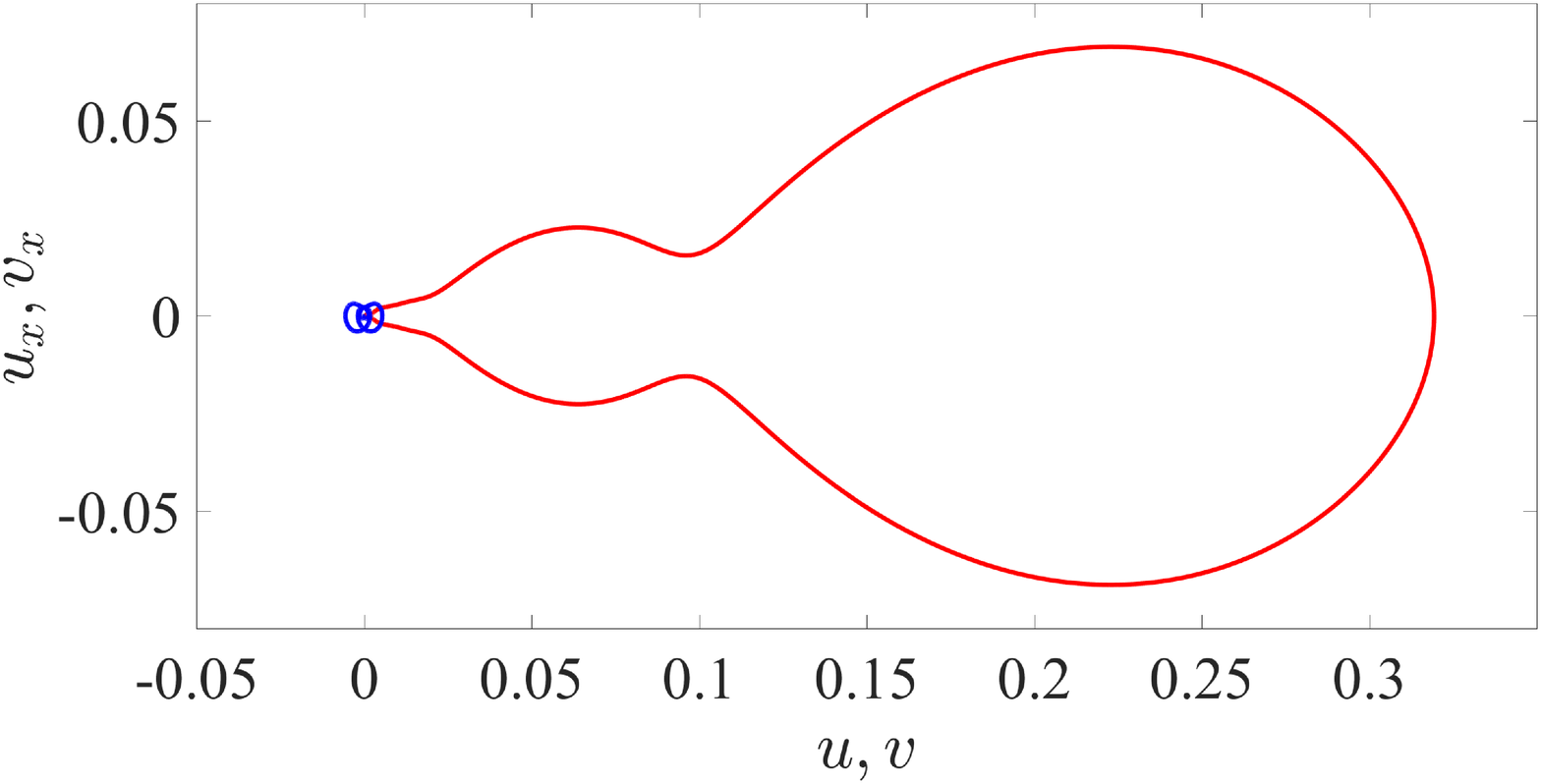}}}
  {\scalebox{\scl}{\includegraphics{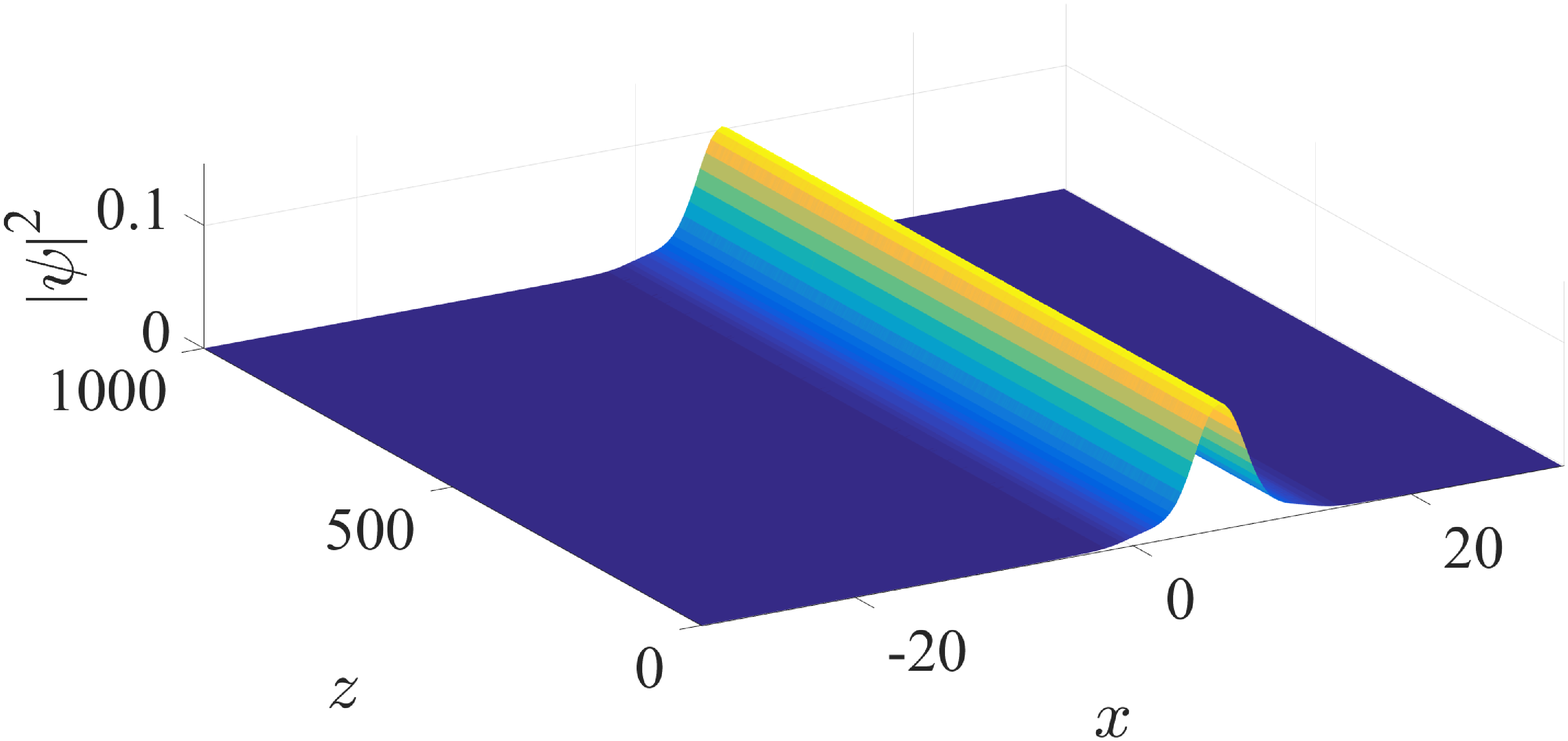}}}
  \caption{Projection of the real ($u$, red) and the imaginary ($v$, blue) parts the homoclinic solutions and propagation dynamics for the case corresponding to Fig. \ref{fig_m1}(b) and Fig. \ref{fig_12a}(b). } \label{fig_2bcd}
  \end{center}
\end{figure}

In Fig. \ref{fig_m1} we show the zeros of the components of the Melnikov vector for a PT-symmetric periodic complex potential with parameters, $\epsilon=0.1$, $v_{1,1}=1$, $w_{1,1}=w_{1,2}=0.2$, $K_{1,1}=L_{1,1}=1$, $K_{1,2}=L_{1,2}=3/2$ and $\phi_{1,2}=\xi_{1,2}=0$. For the case where $v_{1,2}=1$ the condition (\ref{condition}) is not fulfilled and the common zeros of $M_1,M_2$ are located at $x_0=2\pi l, l=0,\pm1,\pm2,...$ for all propagation constants $\beta$, as shown in Fig. \ref{fig_m1}(a). When $v_{1,2}=2/3$ the condition (\ref{condition}) is fulfilled and the zeros of $M_1$ and $M_2$ coincide for all $\beta$ as shown in Fig.  \ref{fig_m1}(b). In comparison to the previous case, additional common zeros exist, with their position depending on $\beta$ and there exist zeros bifurcating for higher values of $\beta$. The fullfilment of the condition (\ref{condition}) in terms of $v_{1,2}$ is depicted in Fig. \ref{fig_m1}(c) where the respective value of $v_{1,2}$ is denoted with a horizontal line. The transverse profiles of solitary waves with $\beta=0.1$ and centers corresponding to the zeros of the Melnikov vector are shown in Fig. \ref{fig_12a}(a,b) for the cases corresponding to Figs. \ref{fig_m1}(a) and (b), respectively. The projection of the real ($u$) and the imaginary ($v$) parts the homoclinic solutions as well as the propagation dynamics for these cases are shown in Fig. \ref{fig_1bc} and Fig. \ref{fig_2bcd}. It is worth emphasizing the existence of an asymmetric solitary wave of two-humped profile in this PT-symmetric case \cite{Yang_2014b} shown in Fig. \ref{fig_2bcd}; this solitary wave is unstable and it evolves to a solution oscillating around one of the stable symmetric solutions.\

\begin{figure}[pt]
  \begin{center}
  \subfigure[]{\scalebox{\scl}{\includegraphics{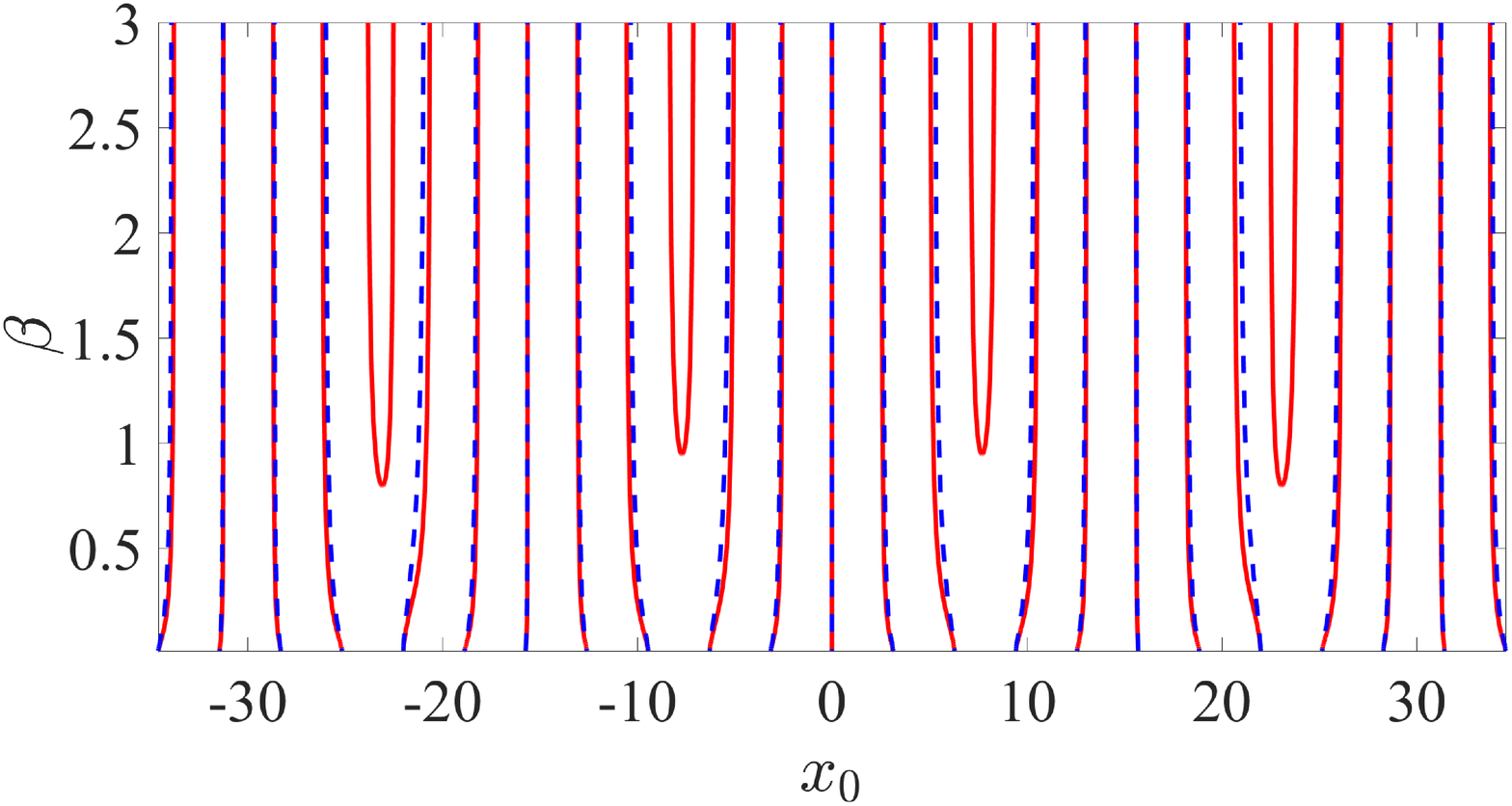}}}
  \subfigure[]{\scalebox{\scl}{\includegraphics{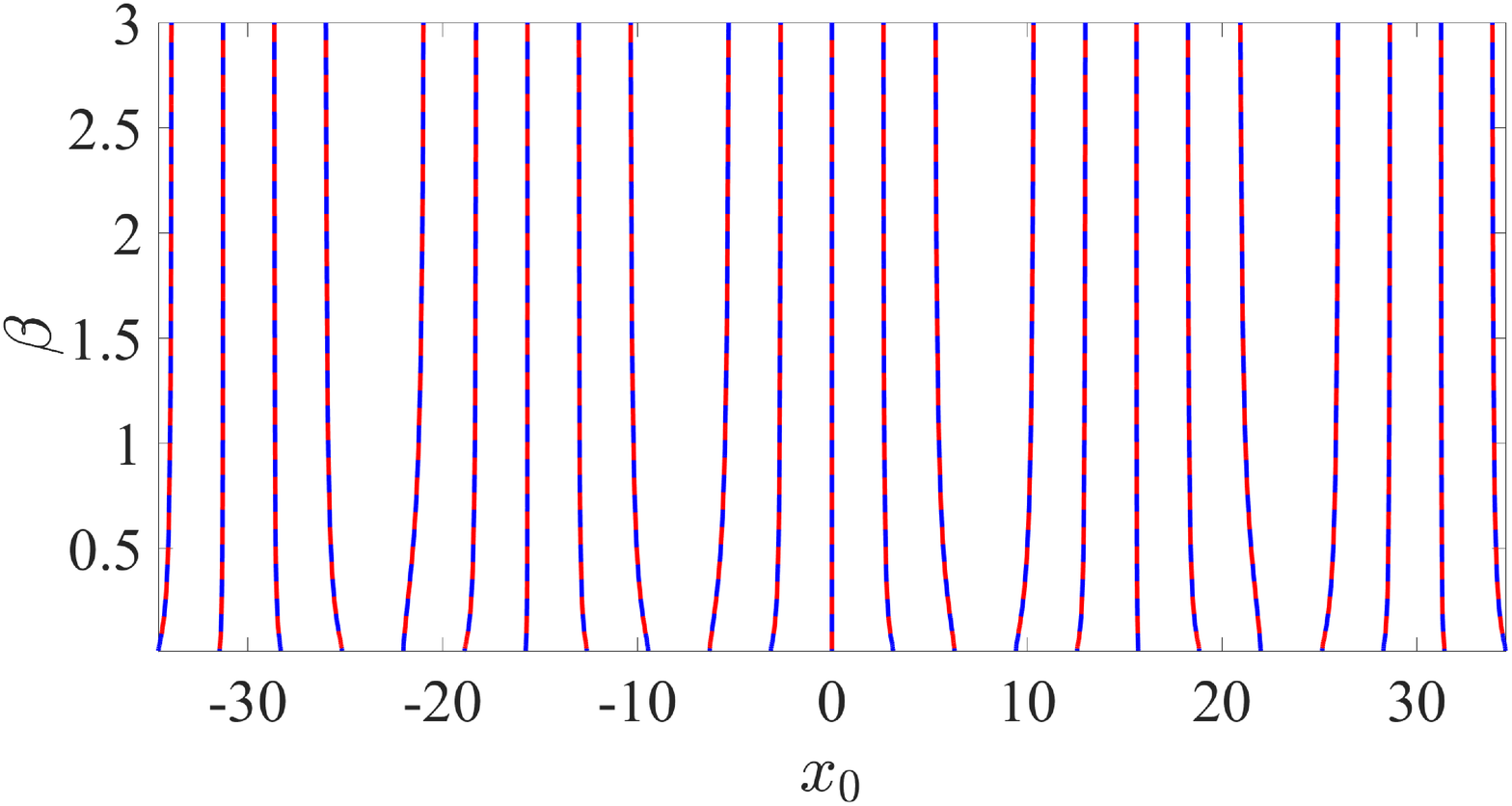}}}
  \subfigure[]{\scalebox{\scl}{\includegraphics{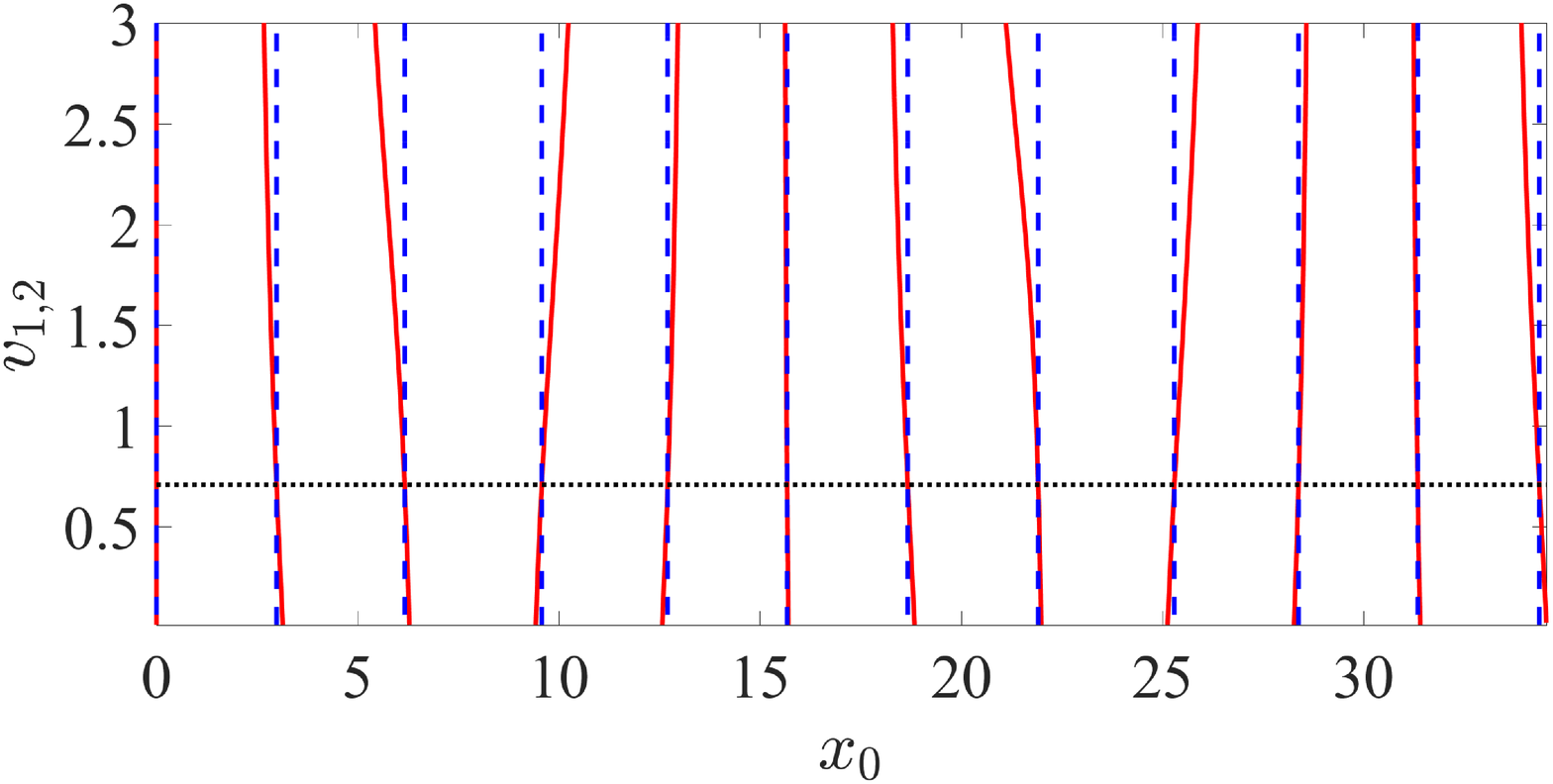}}}
   \caption{Zeros of the two components of the Melnikov vector $M_1$ (red, solid) and $M_2$ (blue, dashed), for a quasiperiodic PT-symmetric complex potential with $v_{1,1}=1$, $w_{1,1}=w_{1,2}=0.2$, $K_{1,1}=L_{1,1}=1$, $K_{1,2}=L_{1,2}=\sqrt{2}$ and $\phi_{1,2}=\xi_{1,2}=0$. (a) $v_{1,2}=1$ (the condition (\ref{condition}) is not fulfilled), (b) $v_{1,2}=1/ \sqrt{2}$ (the condition (\ref{condition}) is fulfilled), (c) $\beta=0.1$ and varying $v_{1,2}$ (the black dotted line denotes the value of $v_{1,2}$ for which the condition (\ref{condition}) is fulfilled).} \label{fig_m2}
  \end{center}
\end{figure}

\begin{figure}[pt]
  \begin{center}
  \subfigure[]{\scalebox{\scla}{\includegraphics{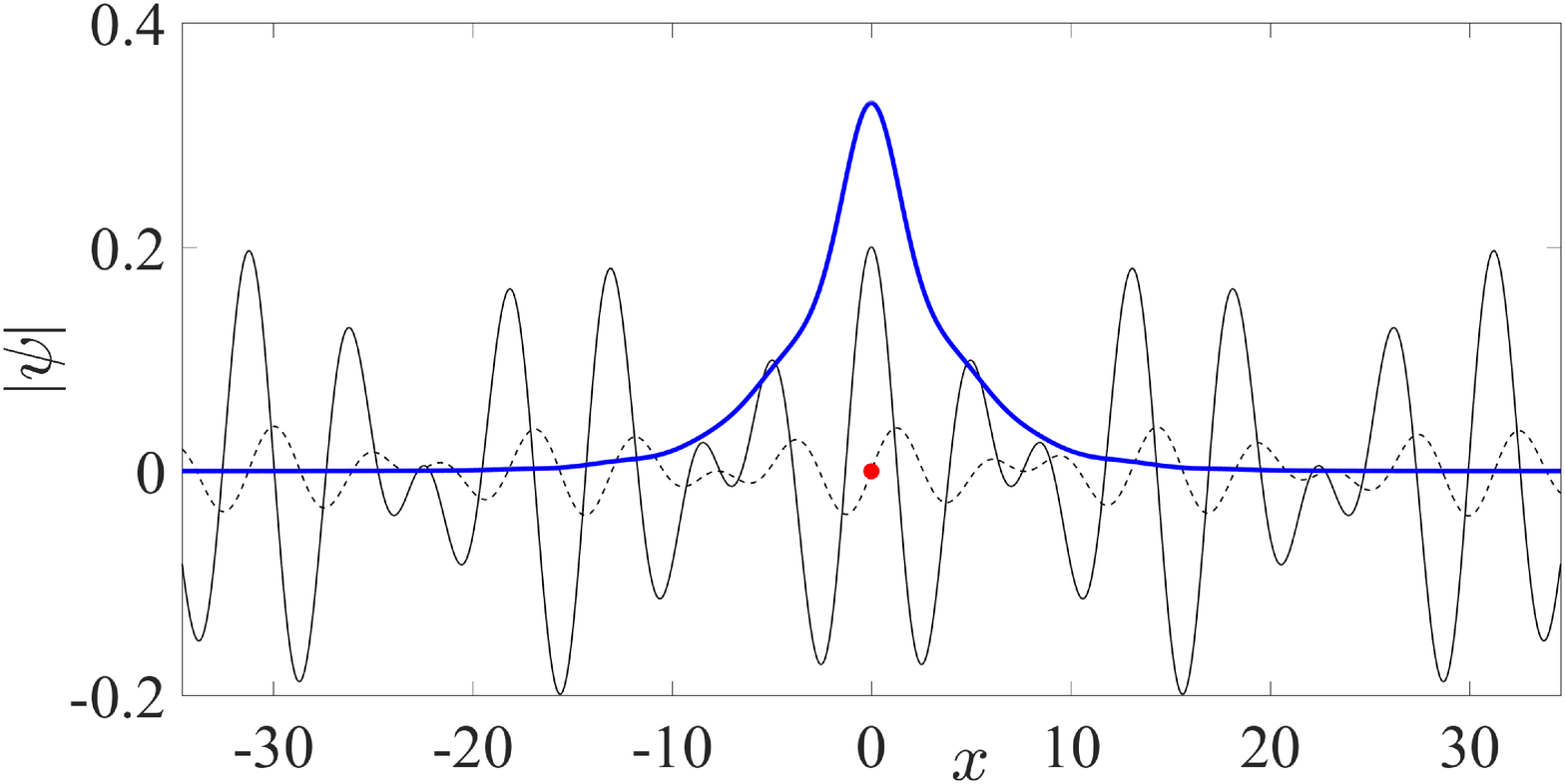}}}
  \subfigure[]{\scalebox{\scla}{\includegraphics{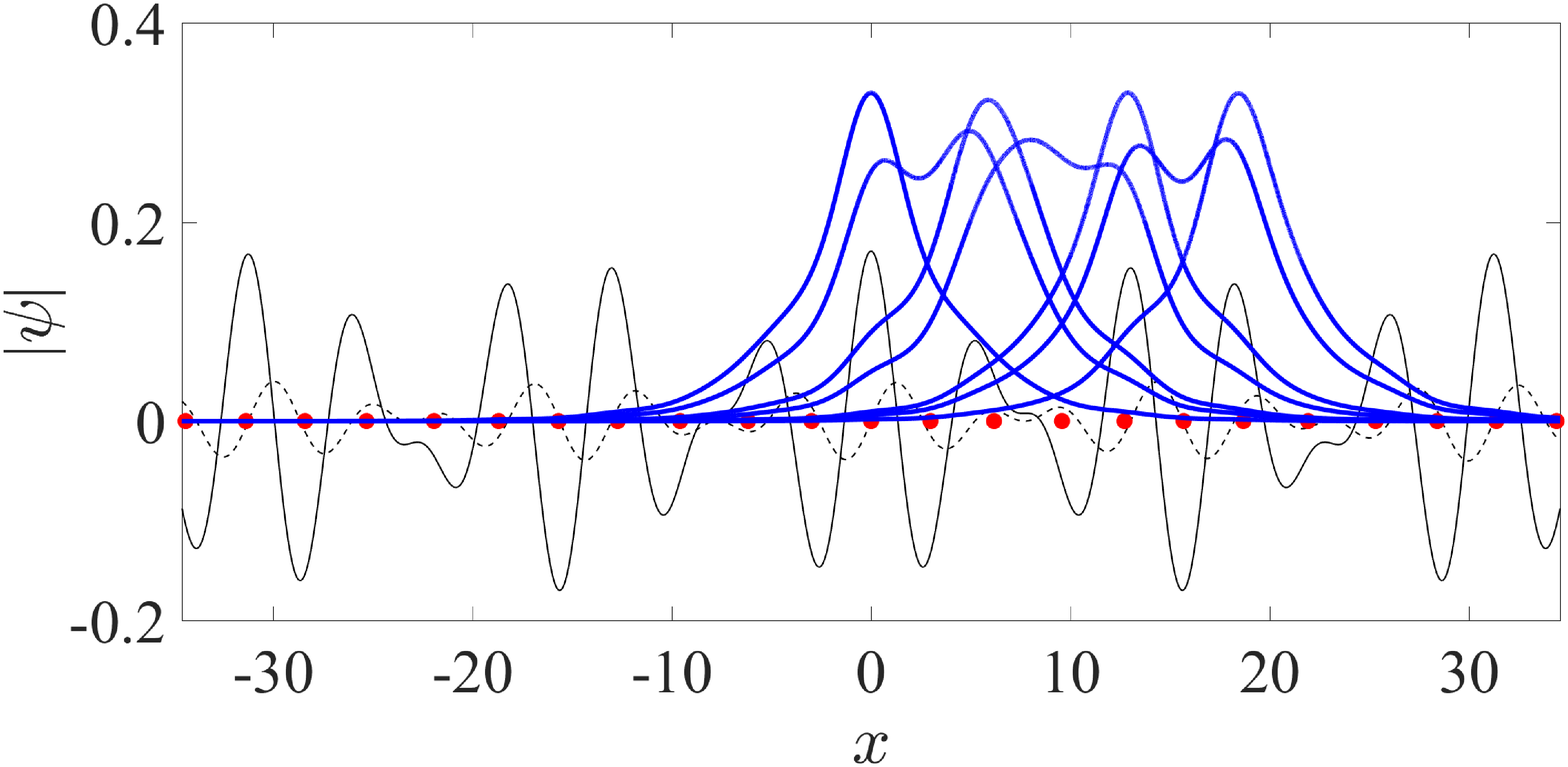}}}
 \caption{Transverse profiles of solitary waves with $\beta=0.1$ and centers corresponding to the zeros of the Melnikov vector  for the cases corresponding to Figs. \ref{fig_m2}(a) and (b), respectively. The black solid and dashed lines depict the real and the imaginary part of the potential and the red circles denote the location of the zeros of the Melnikov function.} \label{fig_34a}
 \end{center}
\end{figure}

\begin{figure}[pt]
  \begin{center}
  {\scalebox{\scl}{\includegraphics{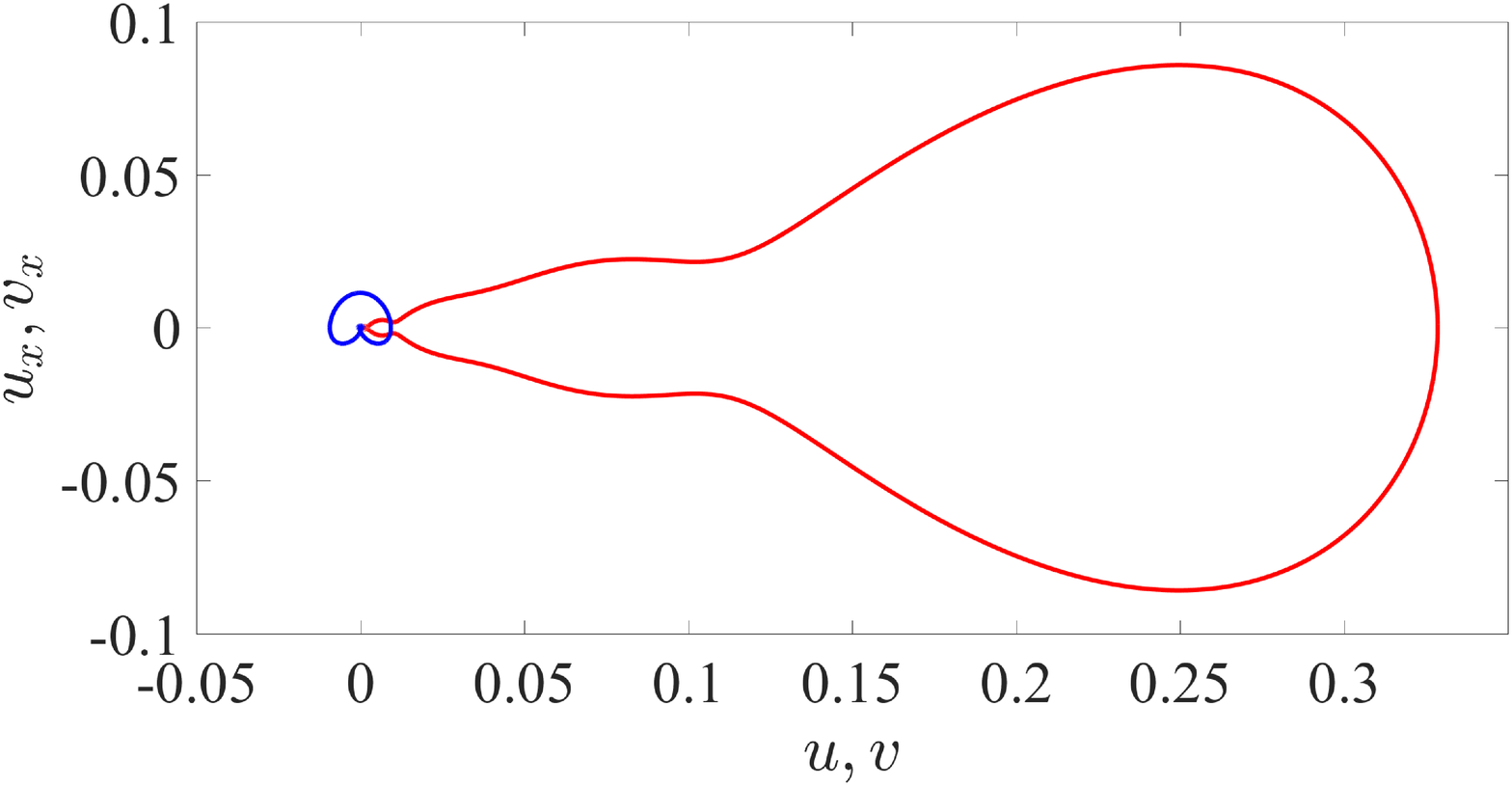}}}
  {\scalebox{\scl}{\includegraphics{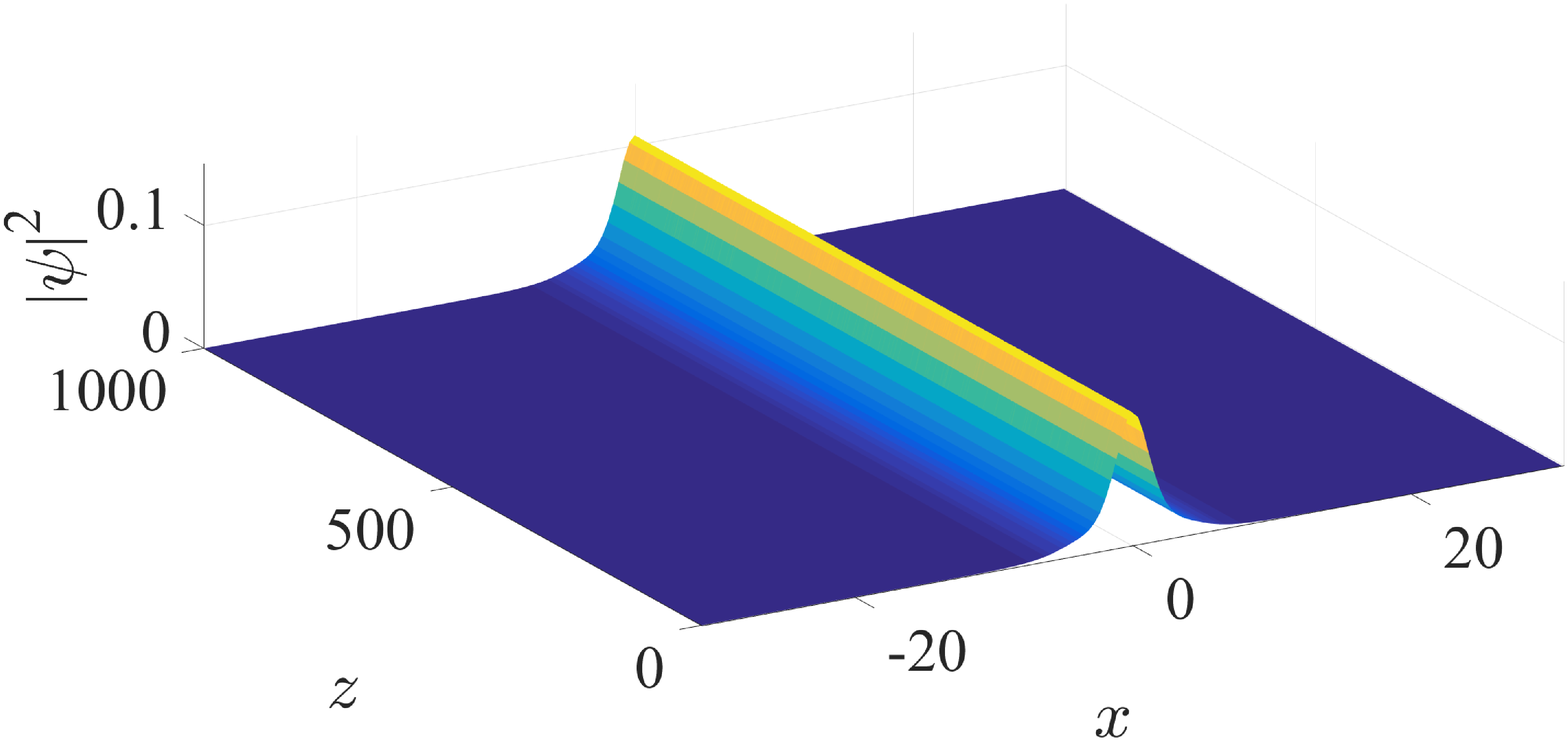}}}
    \caption{Projection of the real ($u$, red) and the imaginary ($v$, blue) parts the homoclinic solutions and propagation dynamics for the case corresponding to Fig. \ref{fig_m2}(a) and Fig. \ref{fig_34a}(a). } \label{fig_3b}
  \end{center}
\end{figure}

\begin{figure}[pt]
  \begin{center}
  {\scalebox{\scl}{\includegraphics{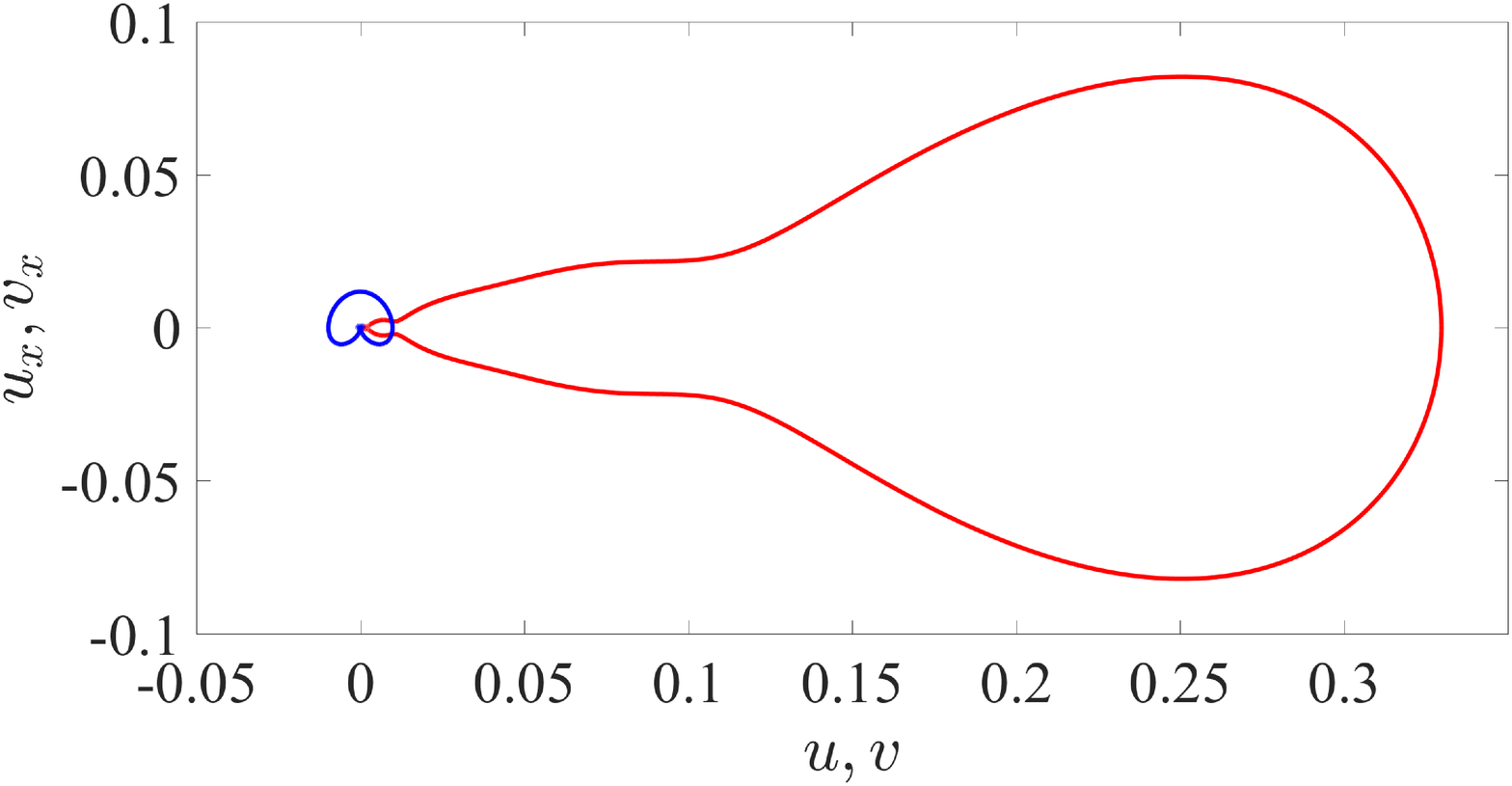}}}
  {\scalebox{\scl}{\includegraphics{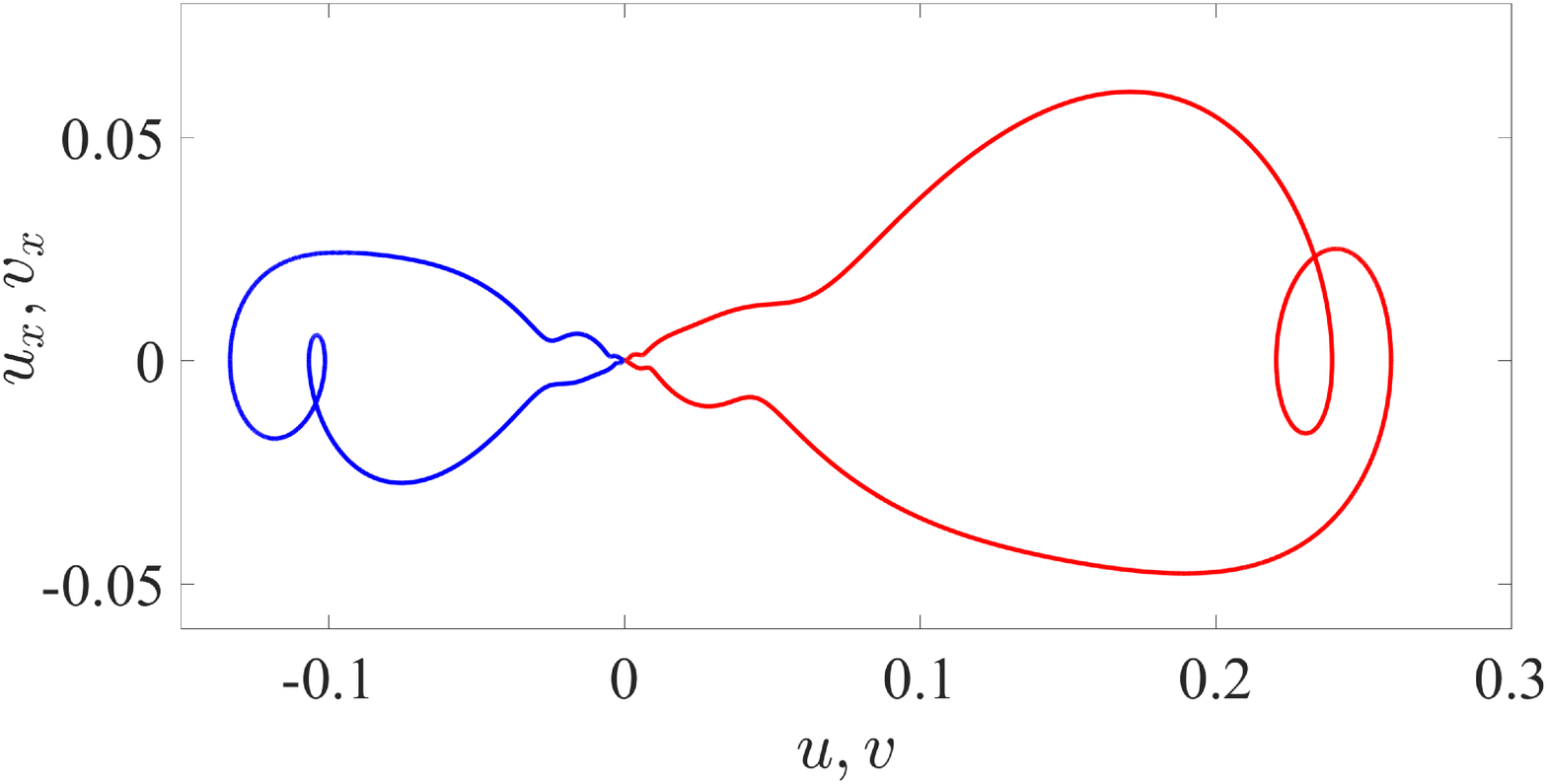}}}
  {\scalebox{\scl}{\includegraphics{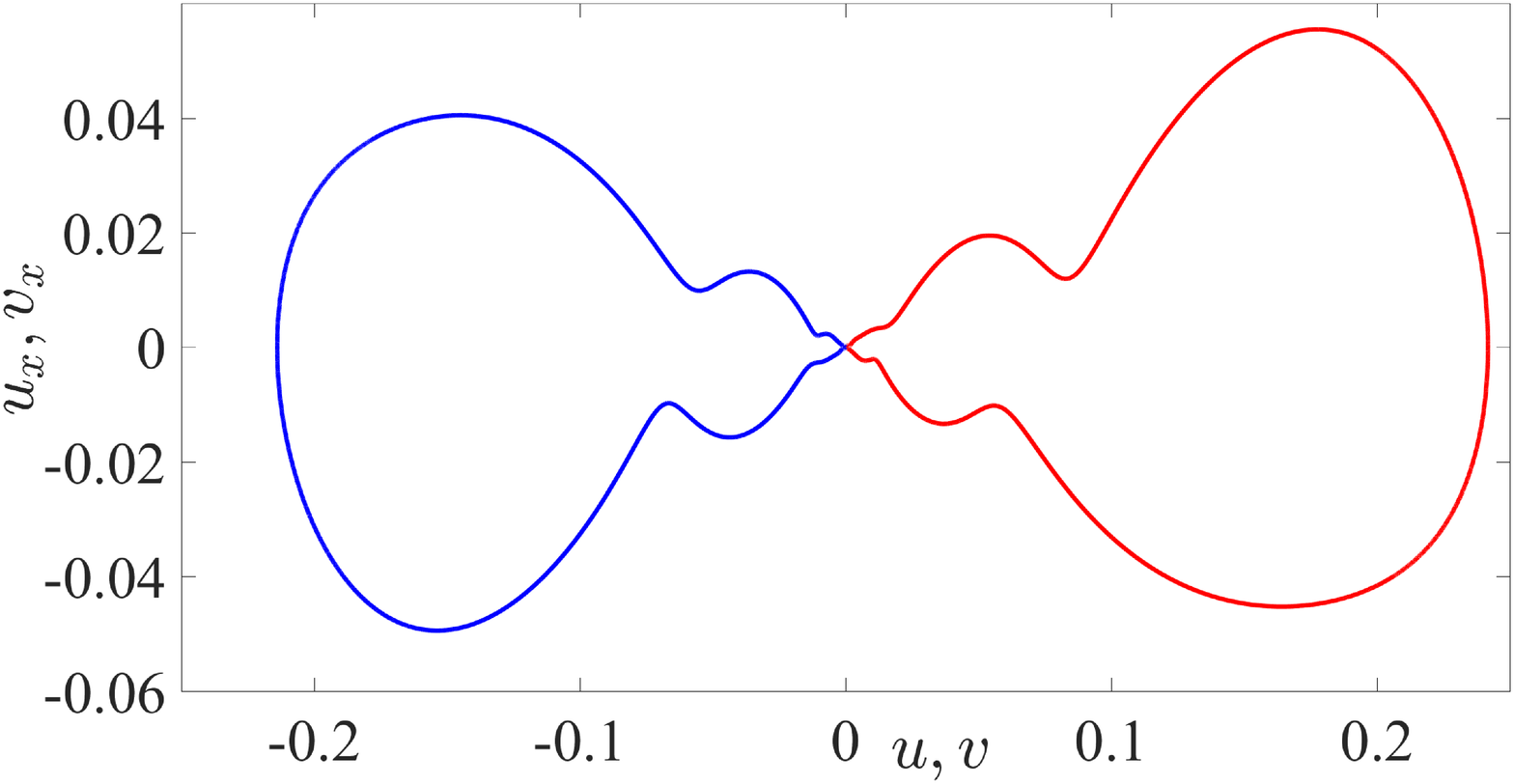}}}
  {\scalebox{\scl}{\includegraphics{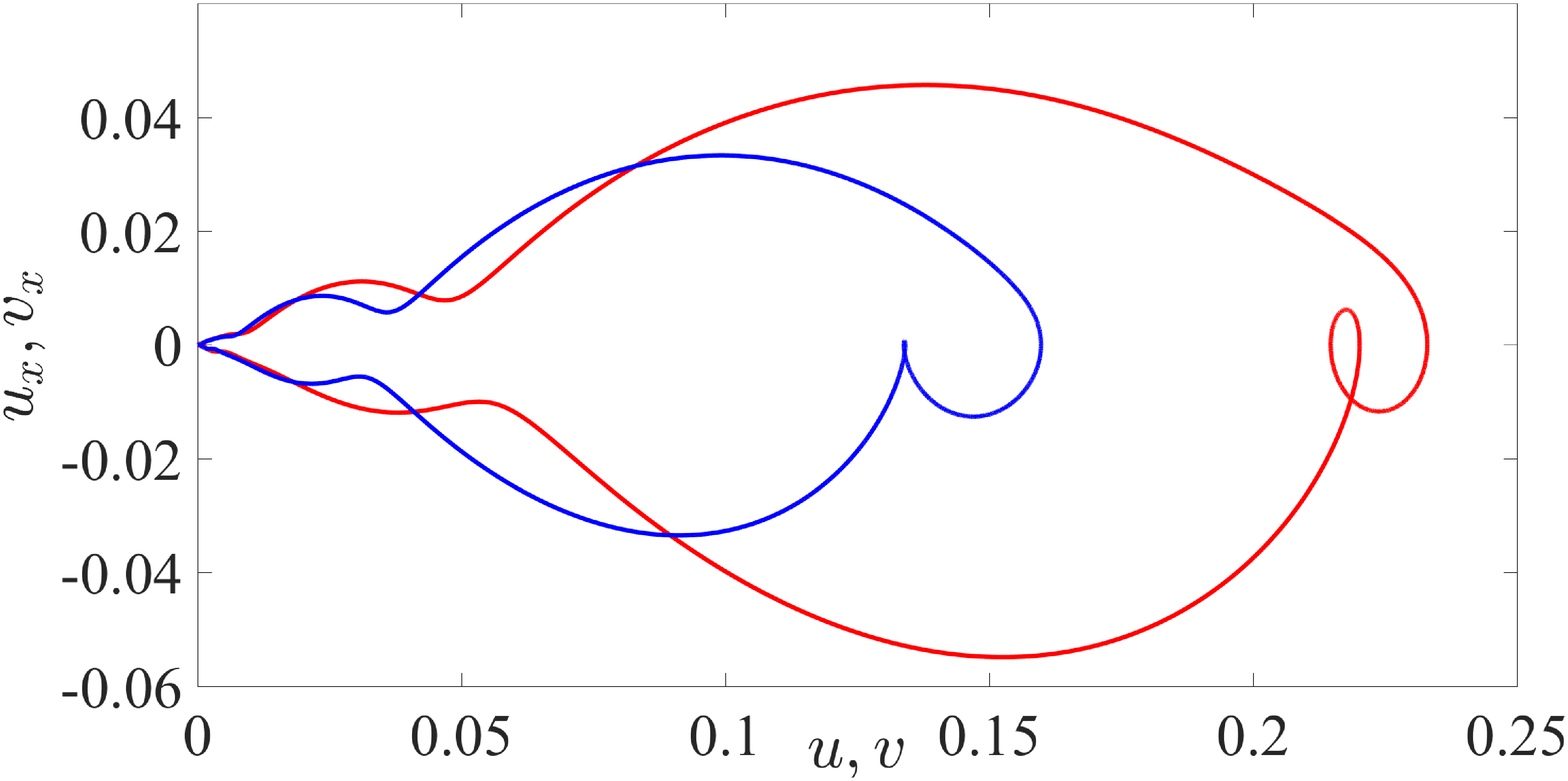}}}
  {\scalebox{\scl}{\includegraphics{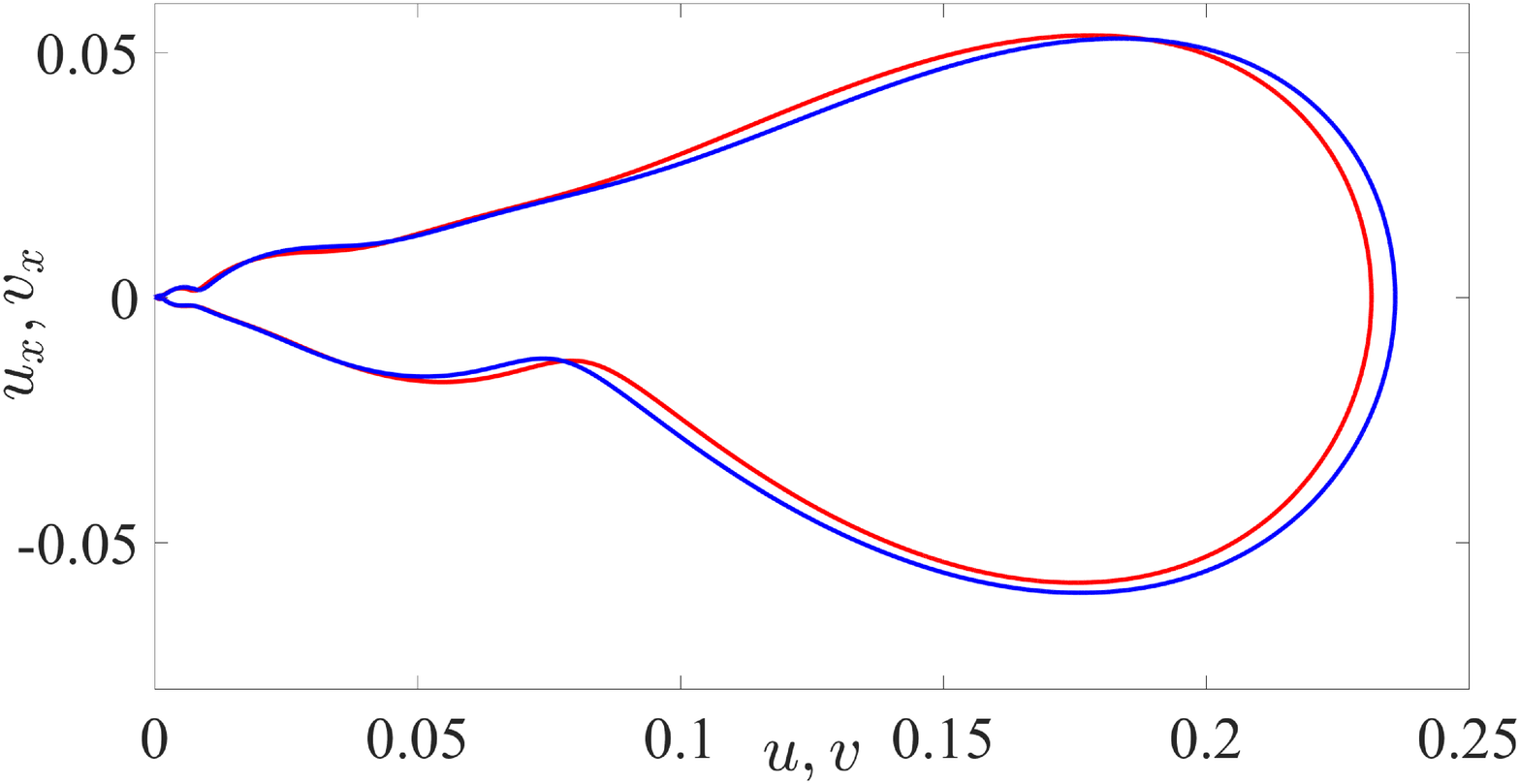}}}
  {\scalebox{\scl}{\includegraphics{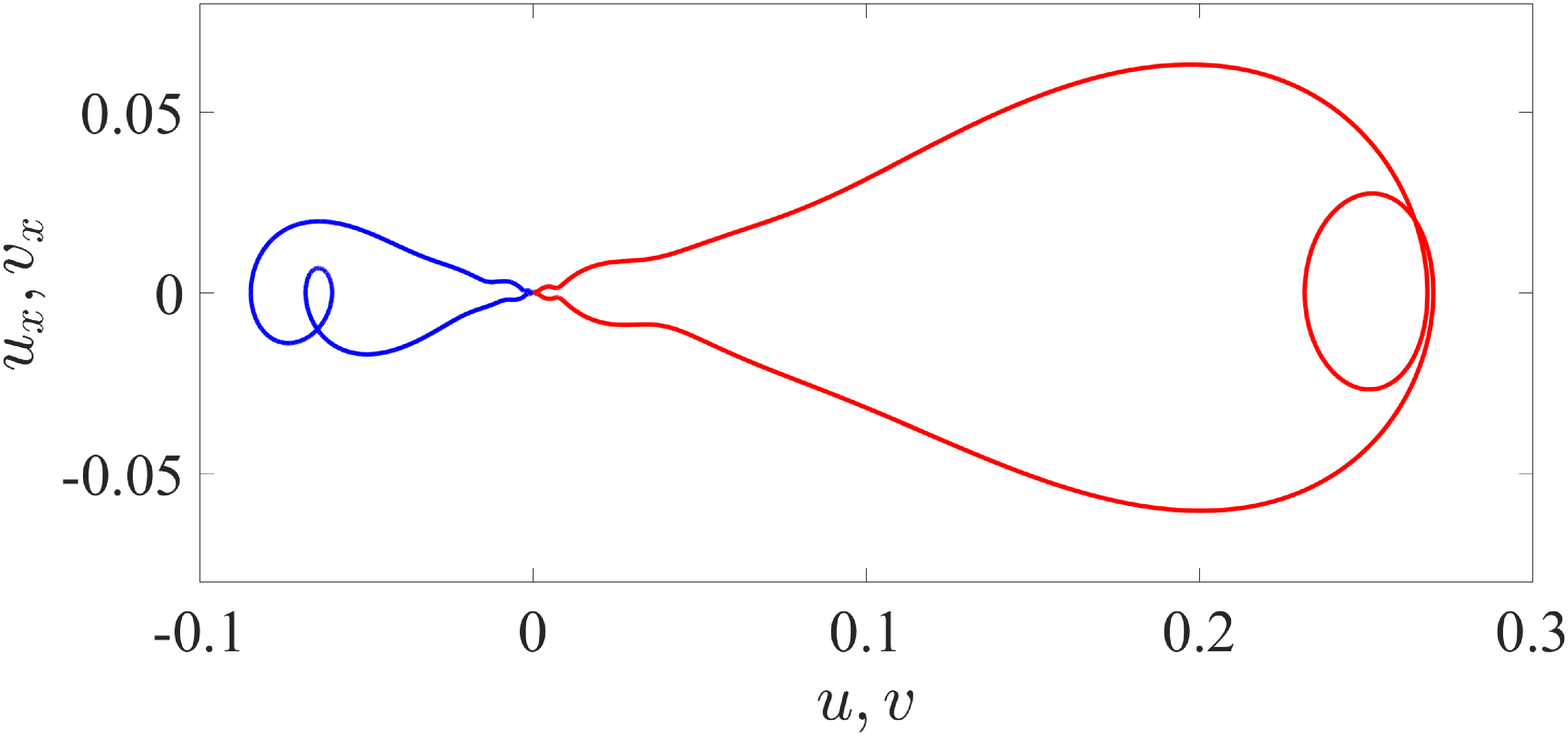}}}
  {\scalebox{\scl}{\includegraphics{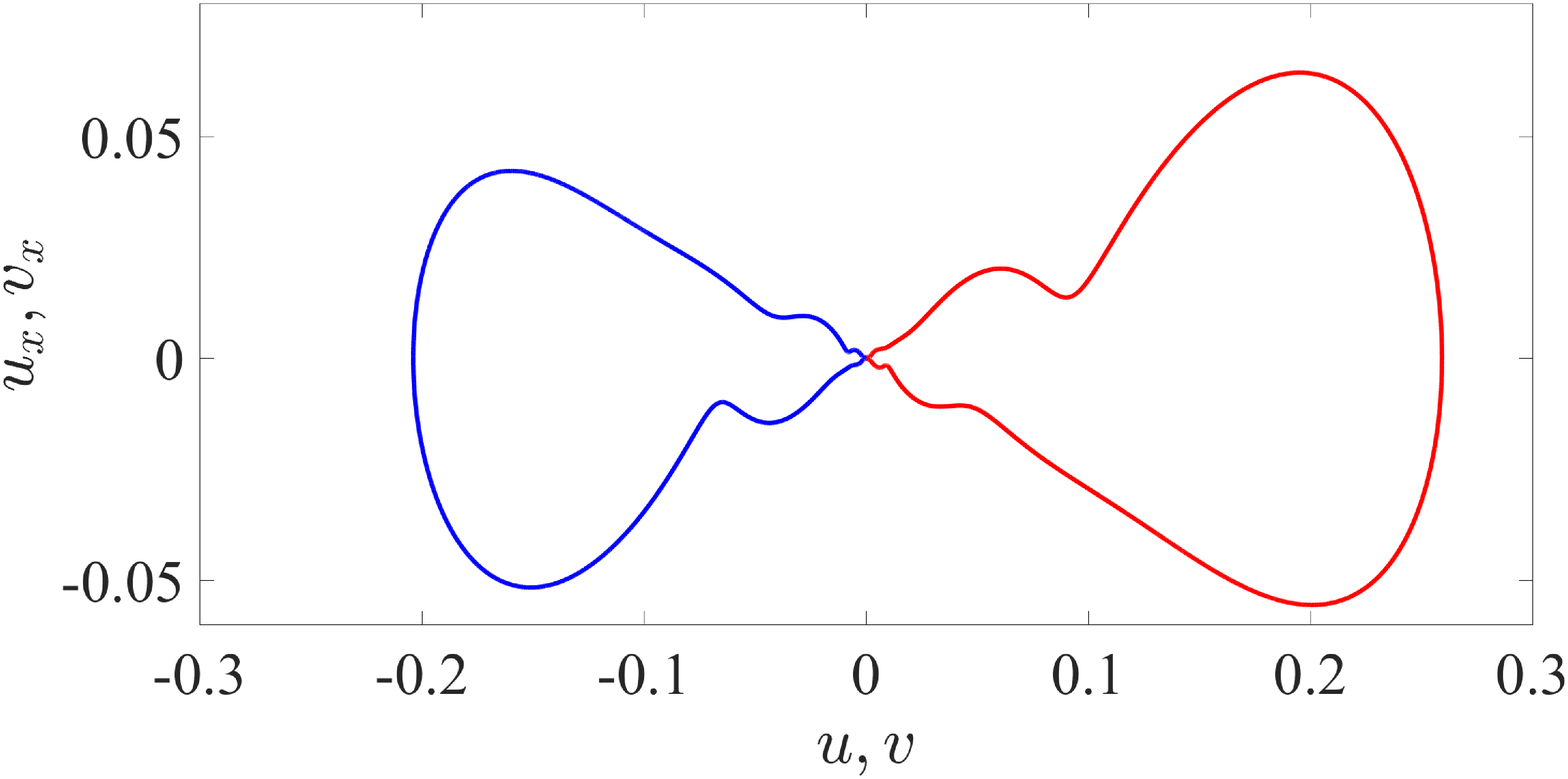}}}
  \caption{Projection of the real ($u$, red) and the imaginary ($v$, blue) parts the homoclinic solutions for the case corresponding to Fig. \ref{fig_m2}(b) and Fig. \ref{fig_34a}(b). } \label{fig_4bcdefgh}
  \end{center}
\end{figure}

For $K_{1,2}=L_{1,2}=\sqrt{2}$ (and all the other parameters having their aforementioned values) the potential is still PT-symmetric but also quasiperiodic. When $v_{1,2}=1$ the condition (\ref{condition}) is not fullfiled and the only common zero of $M_1$ and $M_2$ is at $x_0=0$ for all values of $\beta$ as shown in Fig. \ref{fig_m2}(a); the corresponding solitary wave has a symmetric profile [Fig. \ref{fig_34a}(a)] and it appears to remain robust under the dynamical propagation shown in Fig. \ref{fig_3b}. On the contrary, when $v_{1,2}=1/\sqrt{2}$ the condition (\ref{condition}) is fulfilled and all the zeros of $M_1$ and $M_2$ are common [Fig. \ref{fig_m2}(b,c)]. In that case there
exists a multitude of solitary waves with asymmetric profiles centered at the zeros of the Melnikov vector, as shown in Fig. \ref{fig_34a}(b) and Figs. \ref{fig_4bcdefgh}. \

\begin{figure}[pt]
  \begin{center}
  \subfigure[]{\scalebox{\scl}{\includegraphics{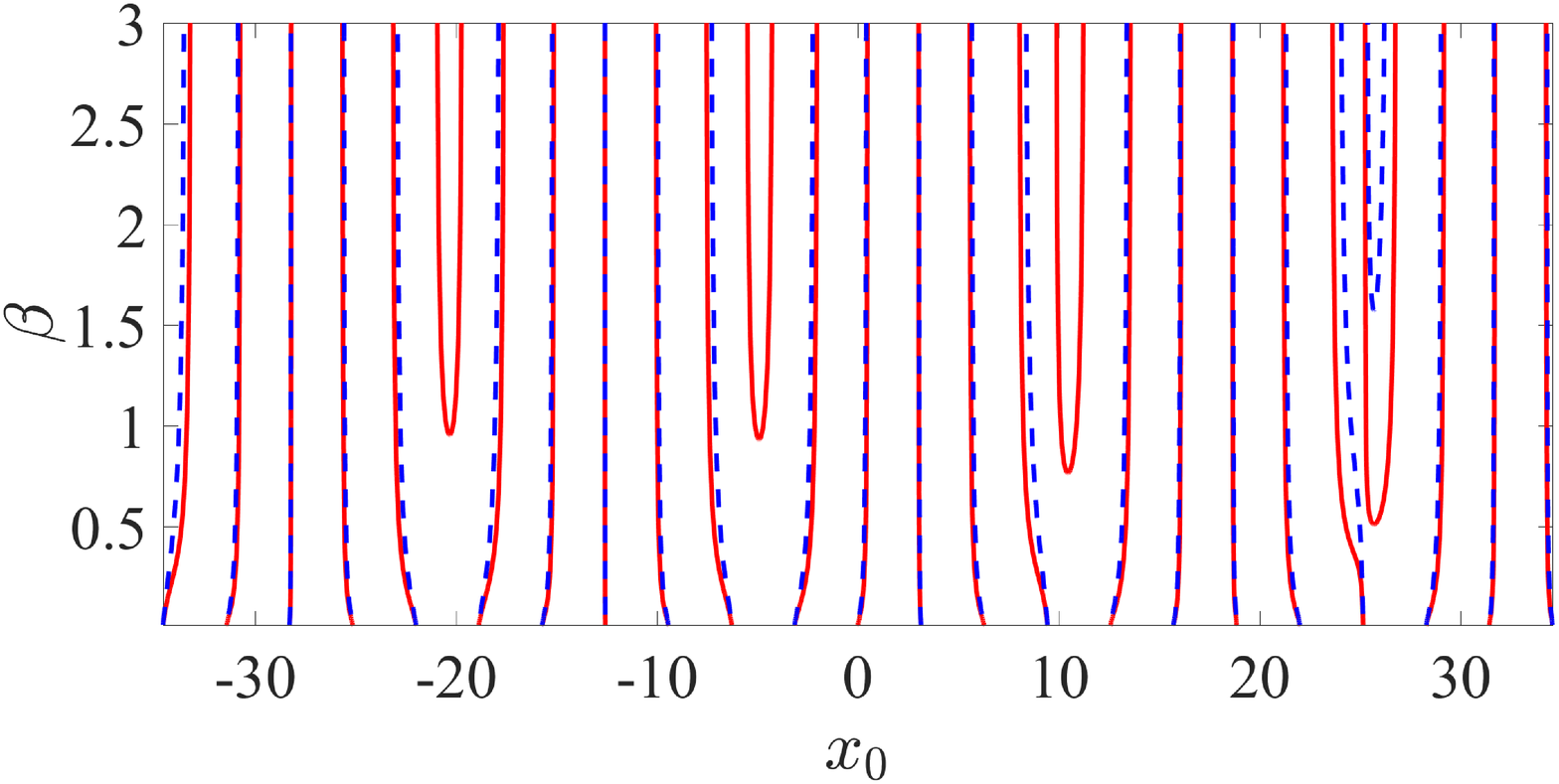}}}
  \subfigure[]{\scalebox{\scl}{\includegraphics{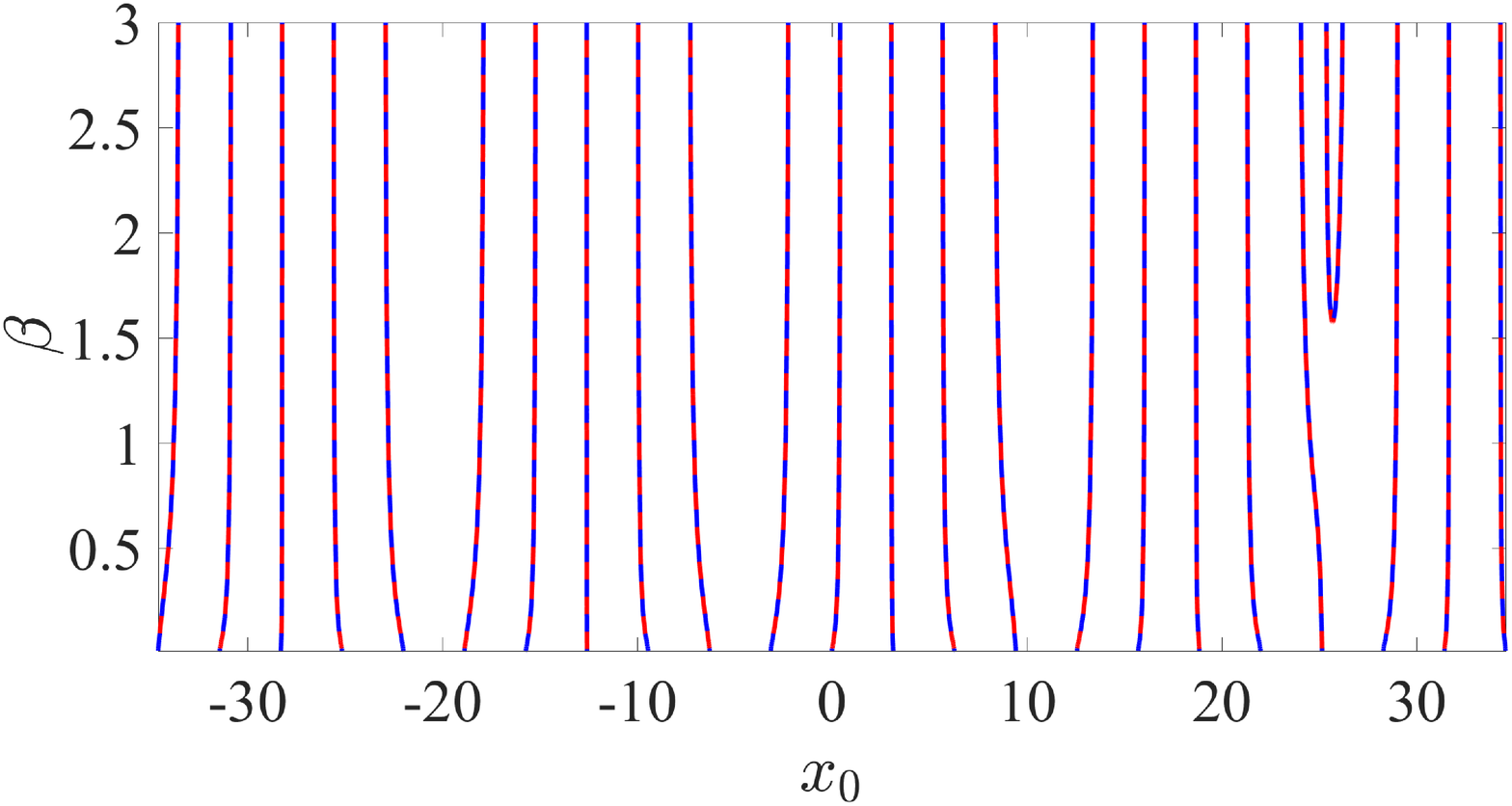}}}
   \subfigure[]{\scalebox{\scl}{\includegraphics{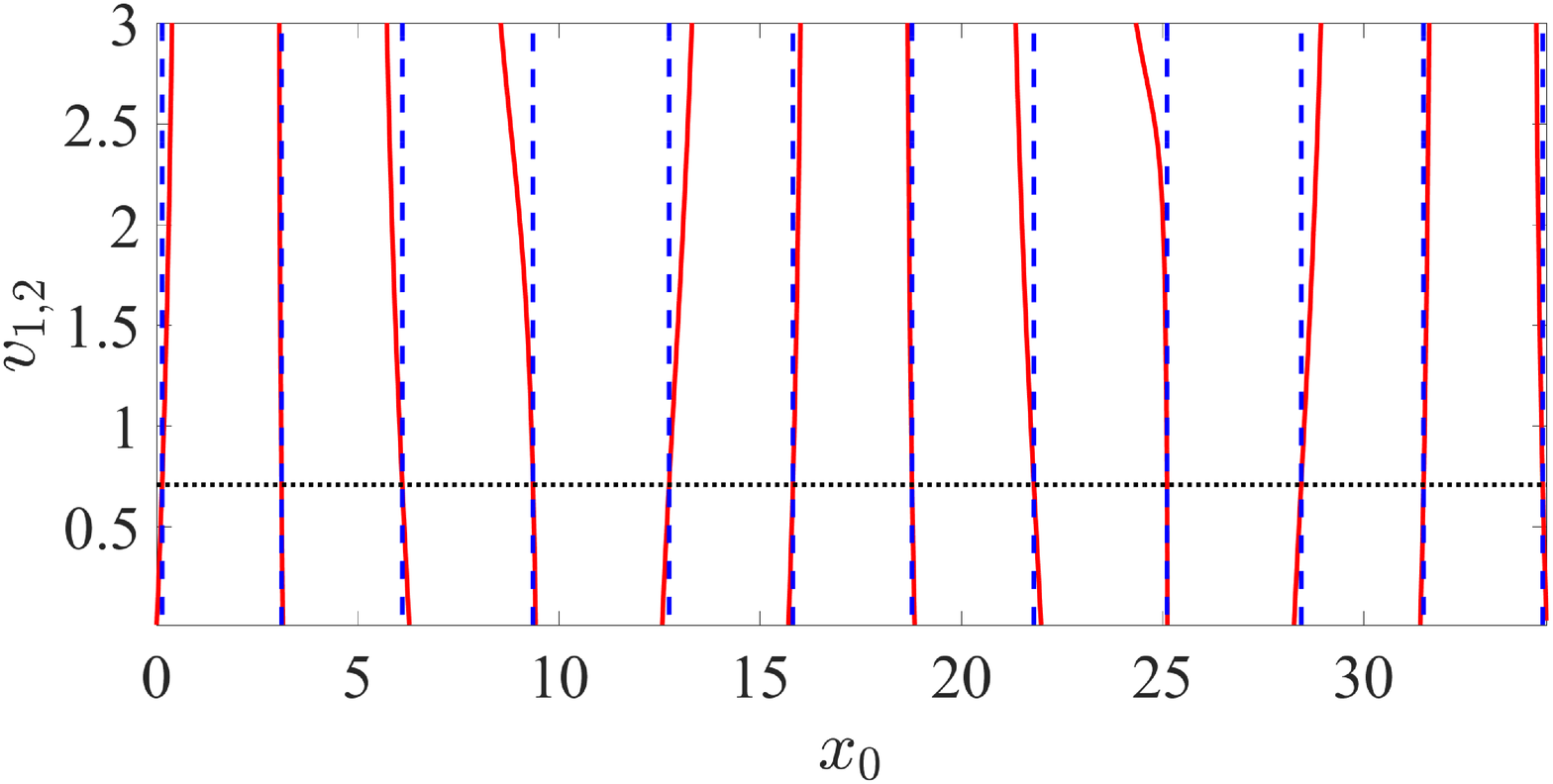}}}
   \caption{Zeros of the two components of the Melnikov vector $M_1$ (red, solid) and $M_2$ (blue, dashed), for a quasiperiodic non PT-symmetric complex potential with $v_{1,1}=1$, $w_{1,1}=w_{1,2}=0.2$, $K_{1,1}=L_{1,1}=1$, $K_{1,2}=L_{1,2}=\sqrt{2}$ and $\phi_{1,2}=\xi_{1,2}=\pi/3$. (a) $v_{1,2}=1$ (the condition (\ref{condition}) is not fulfilled), (b) $v_{1,2}=1/ \sqrt{2}$ (the condition (\ref{condition}) is fulfilled), (c) $\beta=0.1$ and varying $v_{1,2}$ (the black dotted line denotes the value of $v_{1,2}$ for which the condition (\ref{condition}) is fulfilled).} \label{fig_m3}
  \end{center}
\end{figure}

\begin{figure}[pt]
  \begin{center}
  {\scalebox{\scla}{\includegraphics{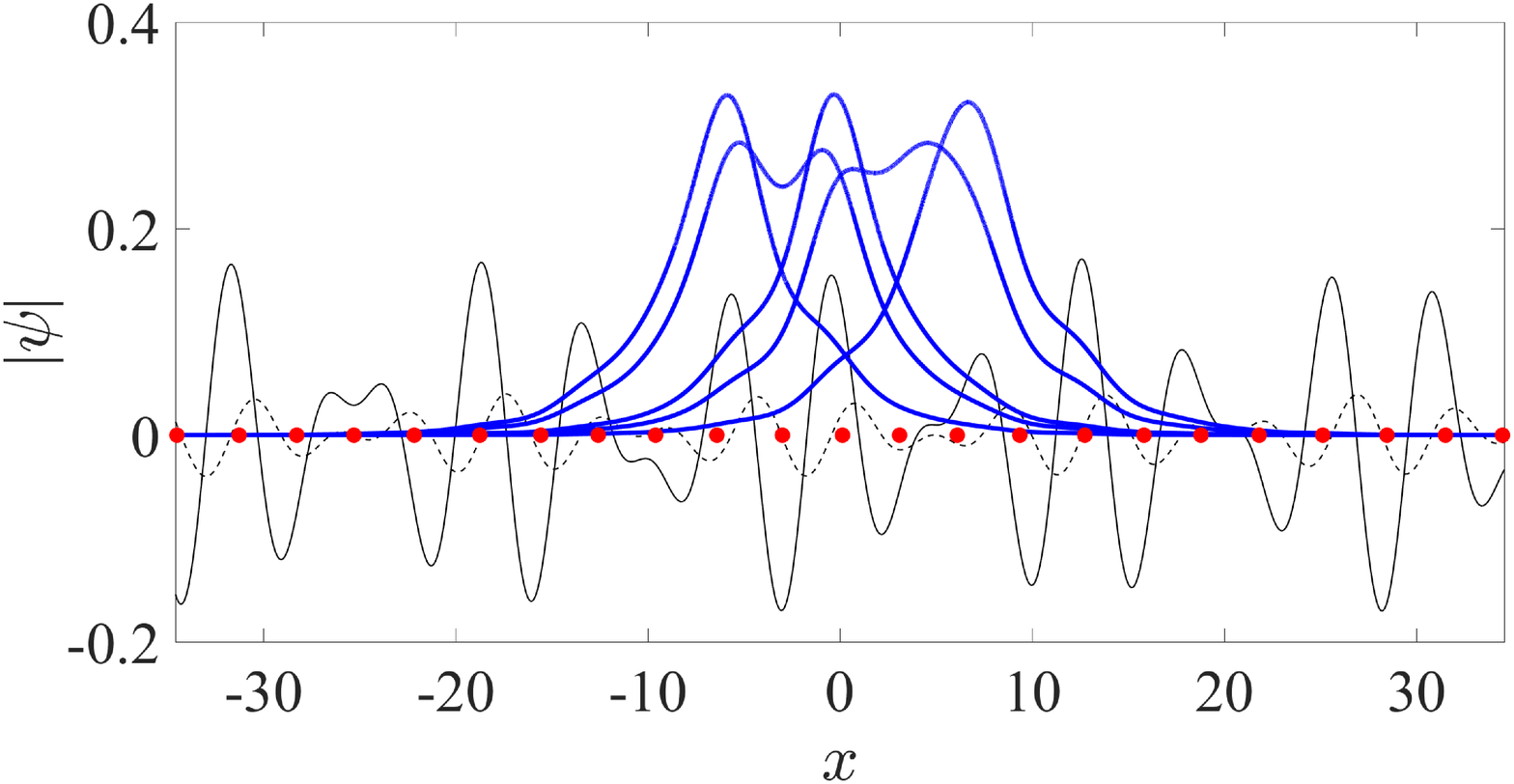}}}
  \caption{Transverse profiles of solitary waves with $\beta=0.1$ and centers corresponding to the zeros of the Melnikov vector  for the case corresponding to Figs. \ref{fig_m3}(b). The black solid and dashed lines depict the real and the imaginary part of the potential and the red circles denote the location of the zeros of the Melnikov function.} \label{fig_5a}
  \end{center}
\end{figure}

\begin{figure}[pt]
  \begin{center}
  {\scalebox{\scl}{\includegraphics{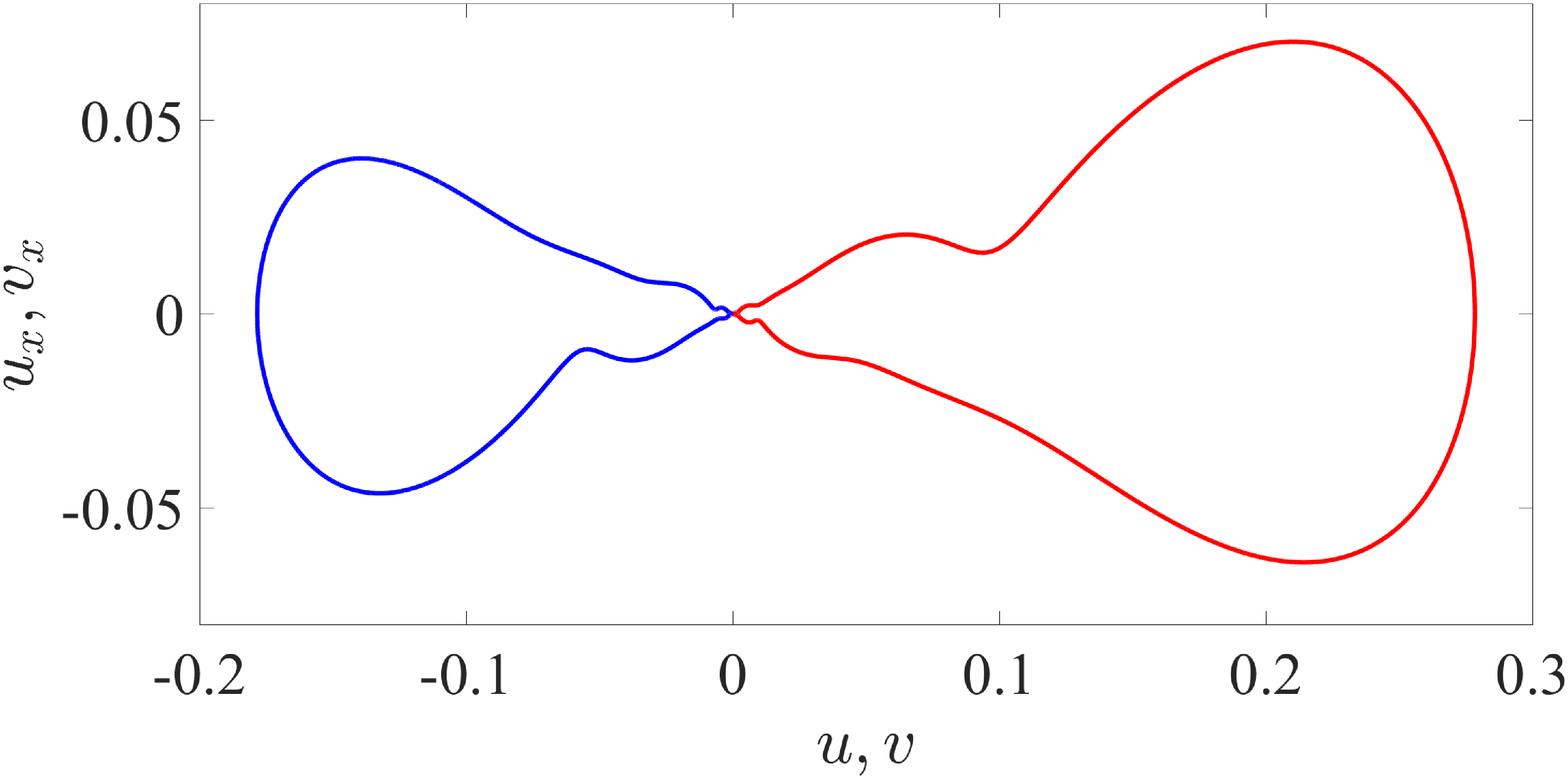}}}
  {\scalebox{\scl}{\includegraphics{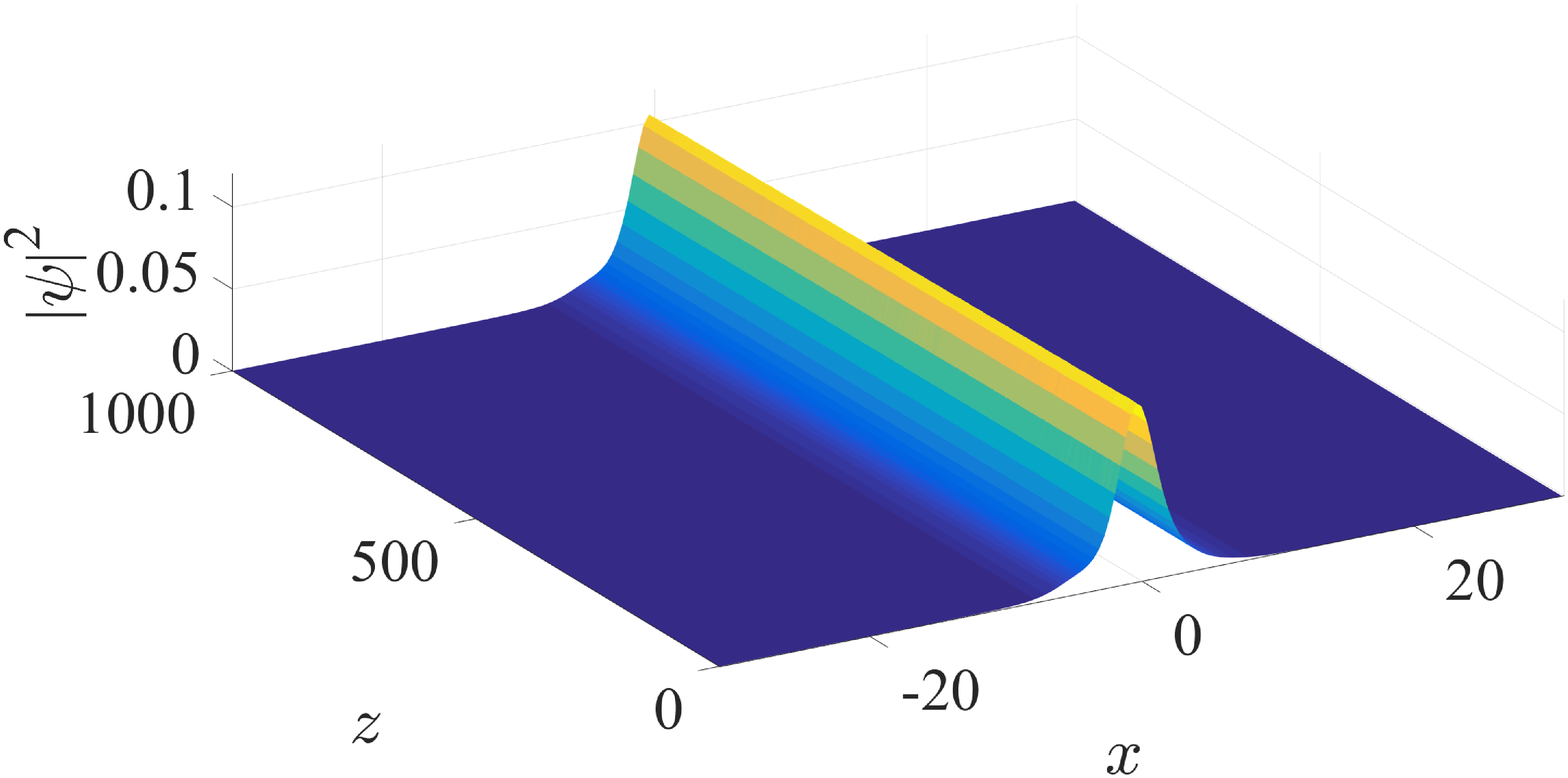}}}\\
  {\scalebox{\scl}{\includegraphics{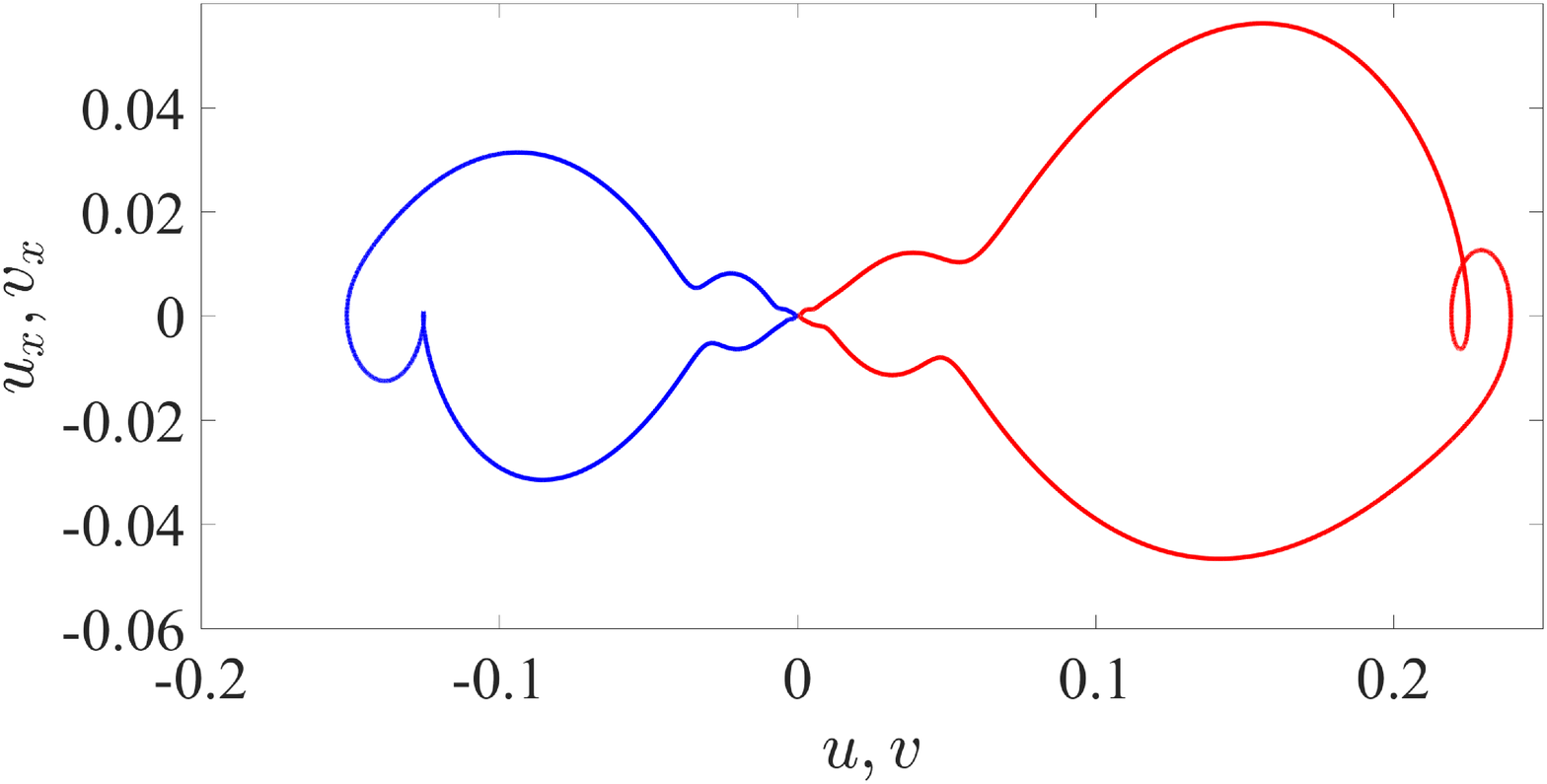}}}
  {\scalebox{\scl}{\includegraphics{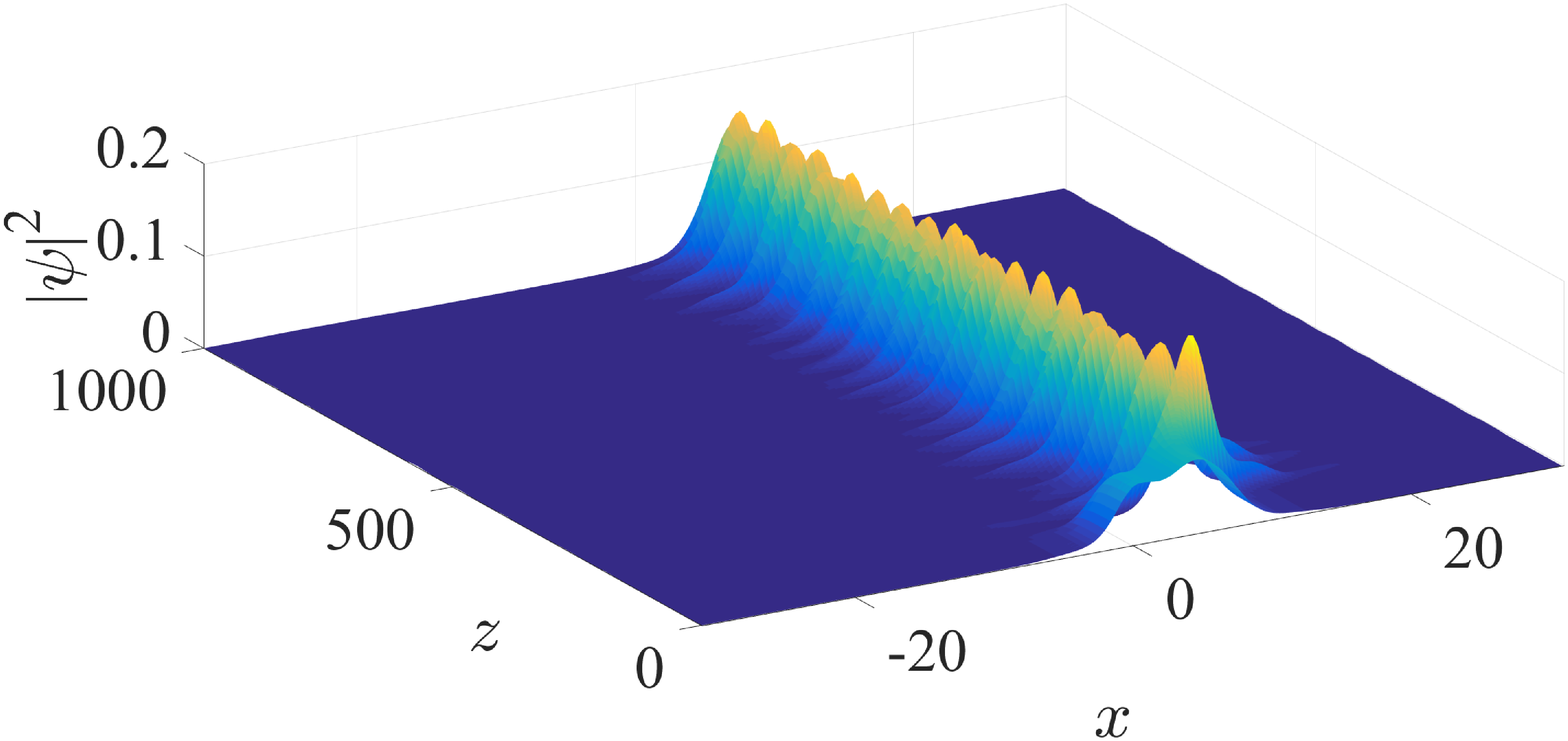}}}\\
  {\scalebox{\scl}{\includegraphics{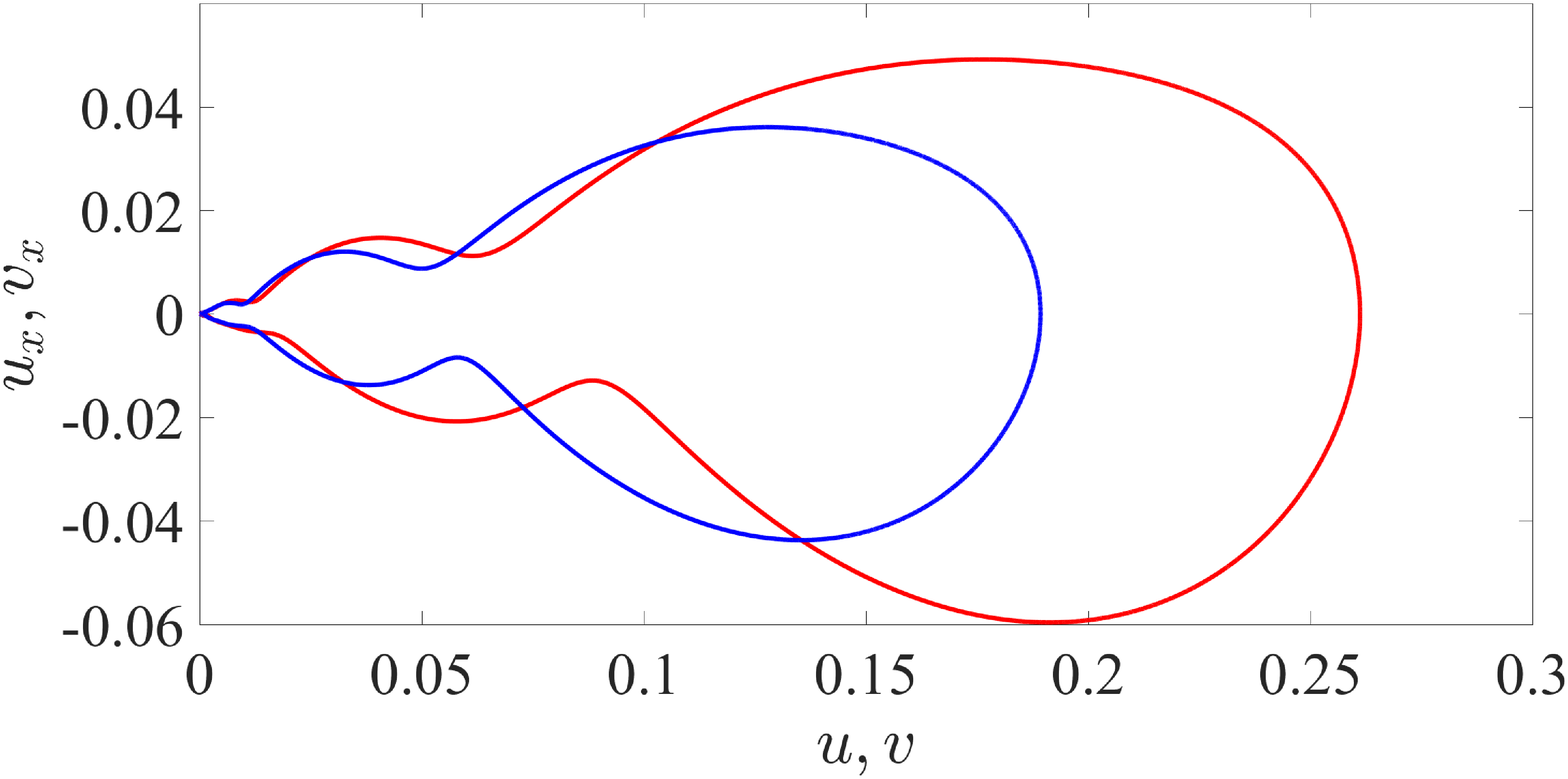}}}
  {\scalebox{\scl}{\includegraphics{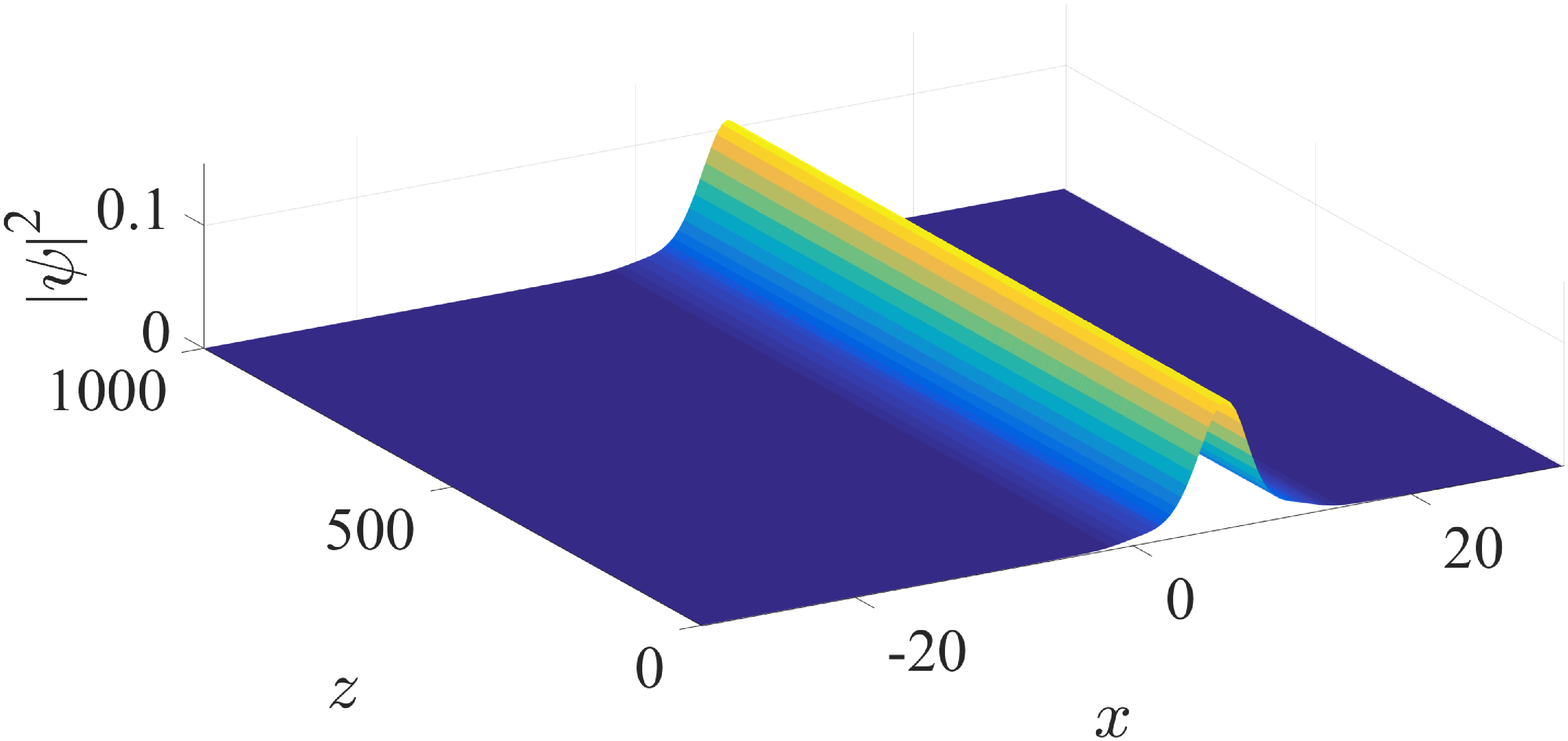}}}\\
  {\scalebox{\scl}{\includegraphics{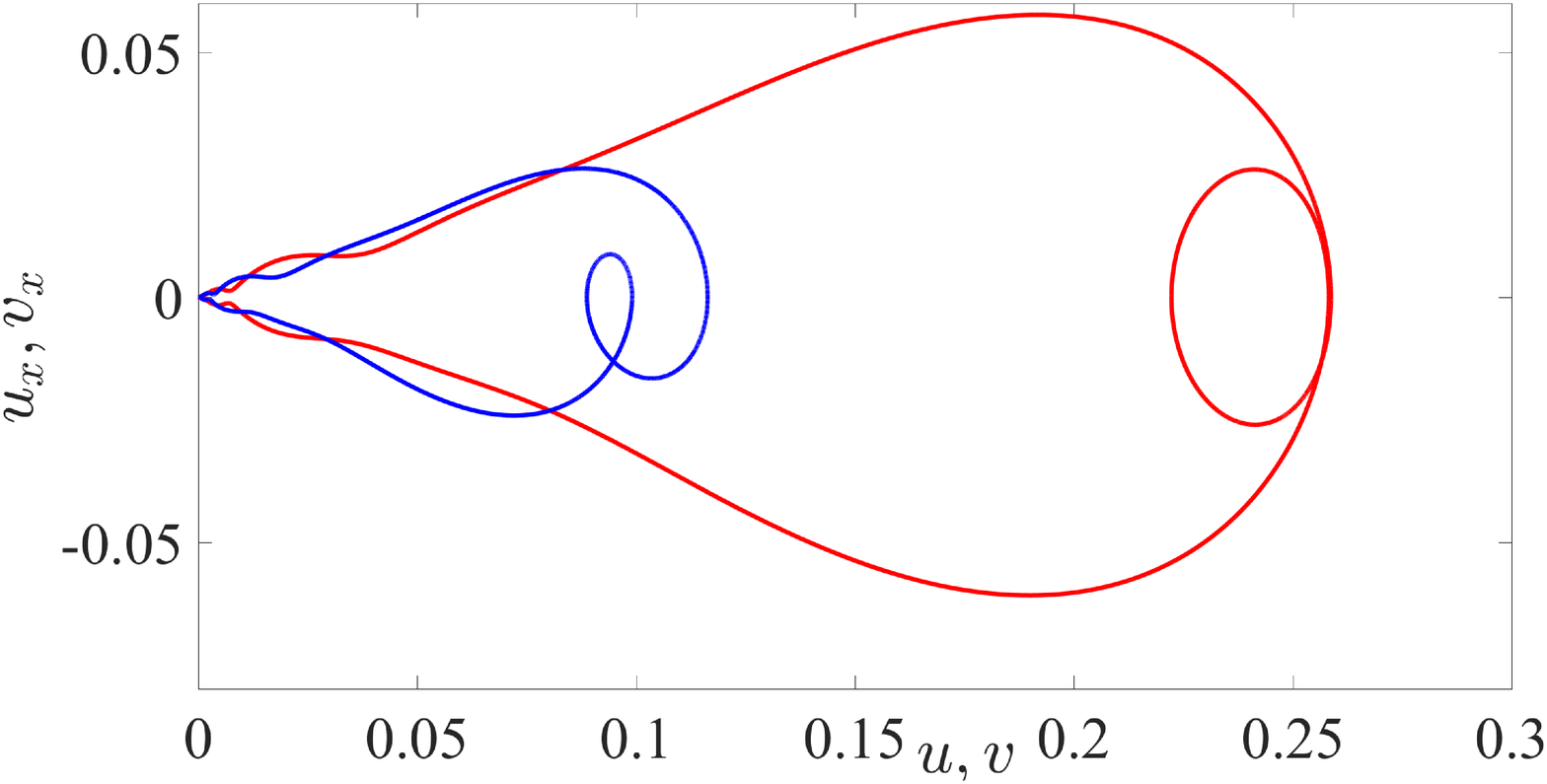}}}
  {\scalebox{\scl}{\includegraphics{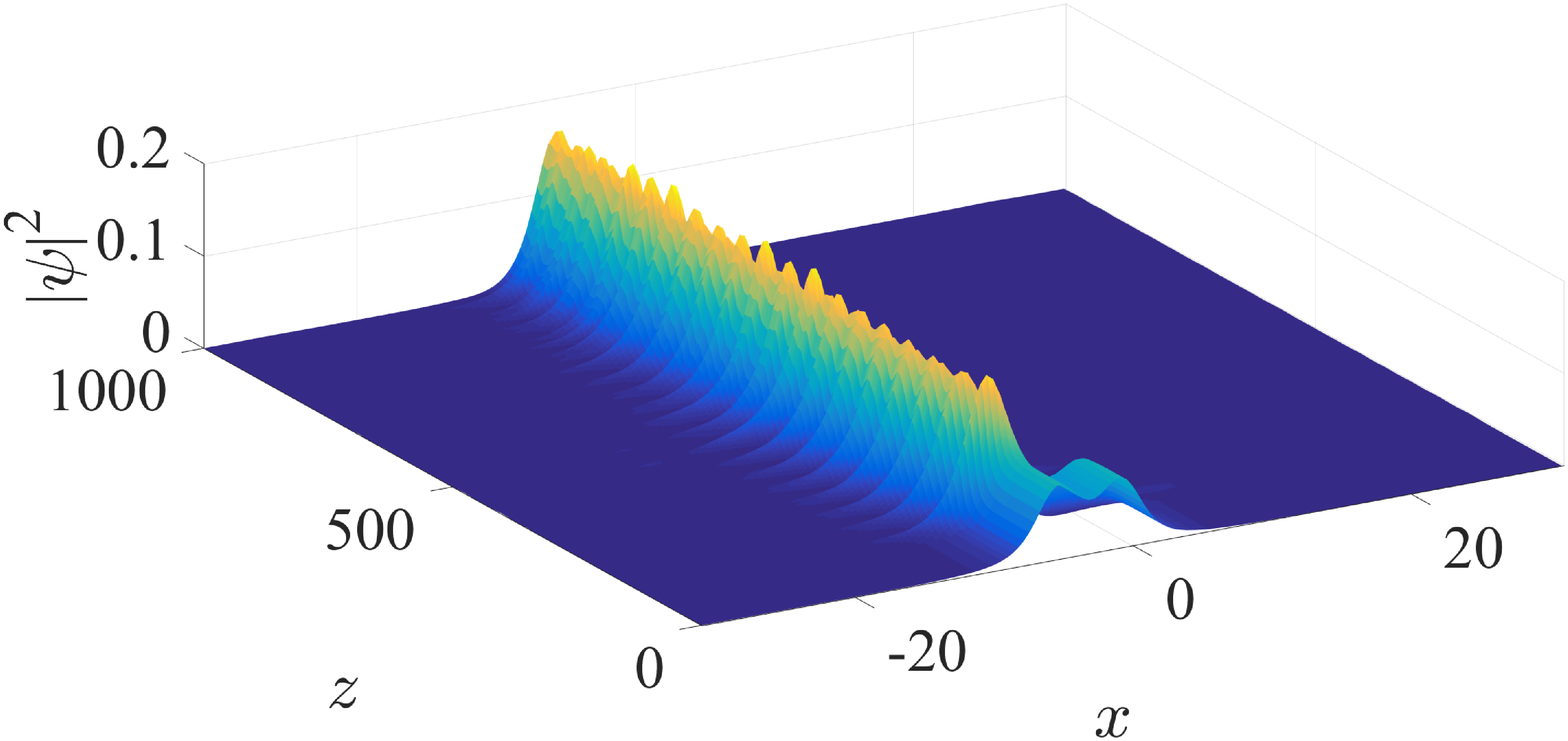}}}\\
  {\scalebox{\scl}{\includegraphics{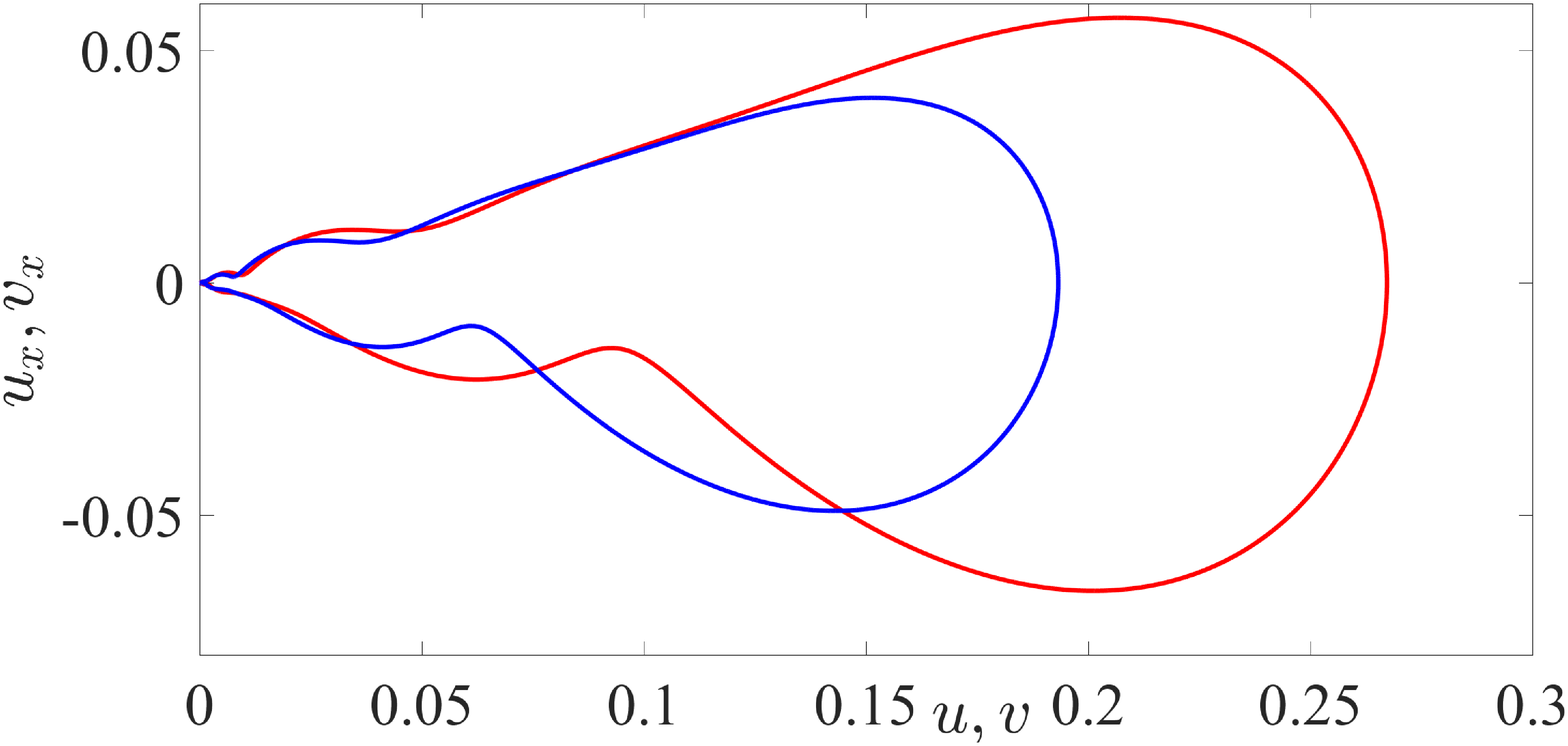}}}
  {\scalebox{\scl}{\includegraphics{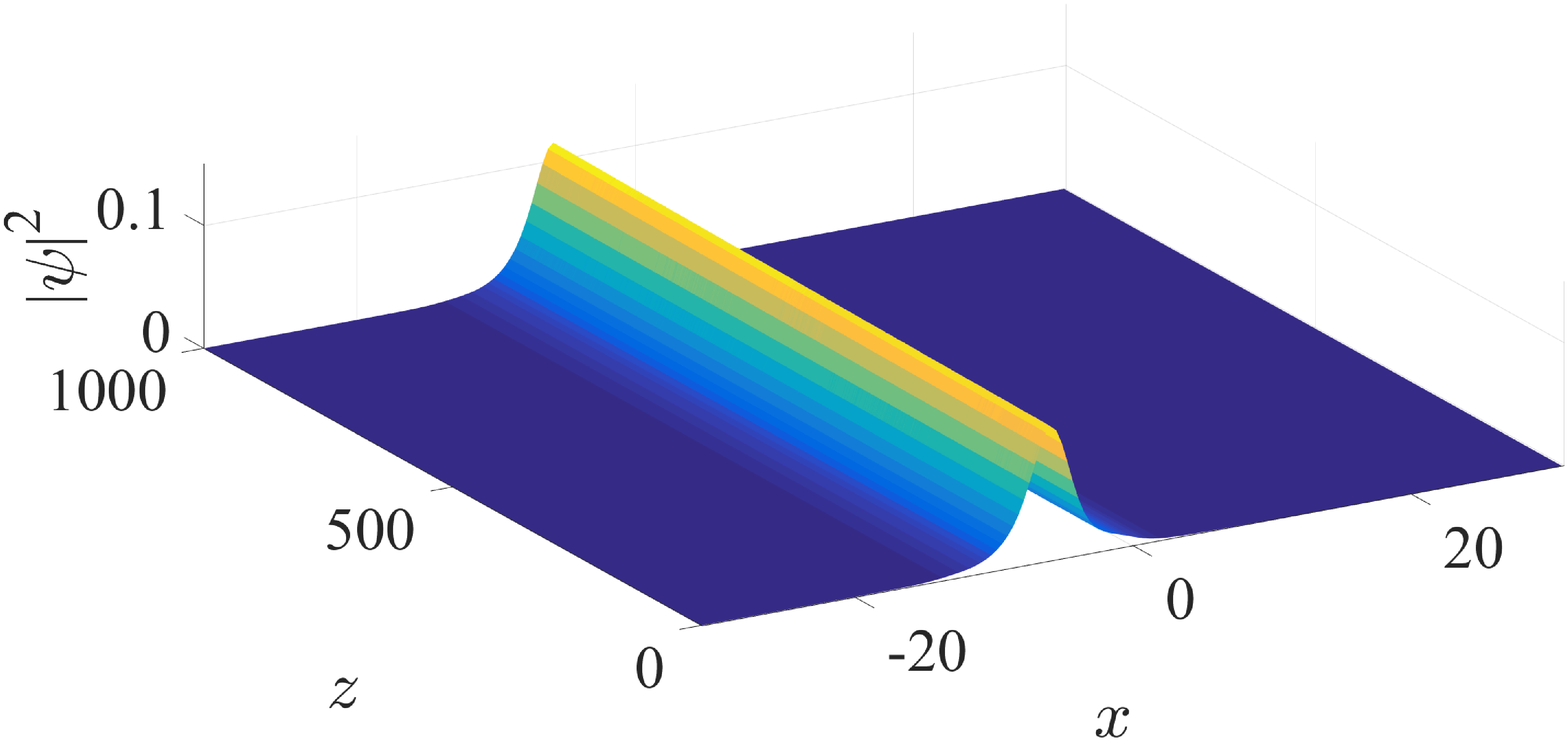}}}
  \caption{Projection of the real ($u$, red) and the imaginary ($v$, blue) parts the homoclinic solutions and propagation dynamics for the case corresponding to Fig. \ref{fig_m3}(b) and Fig. \ref{fig_5a}. } \label{fig_5bcdef}
  \end{center}
\end{figure}

The most general case occurs when a non-zero phase difference ($\phi_{1,2}=\xi_{1,2}=\pi/3$) is considered (all the other parameters having their aforementioned values). In such case the complex potential is not PT-symmetric, and when the condition (\ref{condition}) is not fullifiled there are no zeros of the Melnikov vector [Fig. \ref{fig_m3}(a)] and therefore no solitary waves. When the condition is fullfiled, although the potential is still non-PT-symmetric, the zeros of $M_1$ and $M_2$ coincide [Figs. \ref{fig_m3}(b,c)] for all $\beta$ and this yields a multitude of continuous families of solitary waves with asymmetric profiles centered at the zeros of the Melnikov vector as shown in Fig. \ref{fig_5a} and Fig. \ref{fig_5bcdef}, with the two-humped solution profiles being typically unstable evolving to a single-humped profile undergoing position and amplitude oscillations and the single-humped solution profiles being asymmetric and typically pinned at a fixed transverse position and ungergoing very small amplitude growth or decay.

\subsection{Nonlinear potential}
We consider now the case of a complex potential having both linear and nonlinear parts \cite{Kominis_2008, Kartashov_2016}, and more specifically
\begin{align}
V_1(x)&=v_{1,1}\cos(K_{1,1}x+\phi_{1,1}),  & W_1(x)&=w_{1,1}\sin(L_{1,1}x+\xi_{1,1}), \nonumber \\
V_2(x)&=v_{2,1}\cos(K_{2,1}x+\phi_{2,1}),  & W_2(x)&=w_{2,1}\sin(L_{2,1}x+\xi_{2,1}). \label{nonlinear}
\end{align}
For $\phi_{1,1} \neq \xi_{1,1}$ and $\phi_{2,1} \neq \xi_{2,1}$ neither the linear nor the nonlinear part of the potential are PT-symmetric. An appropriate parameter selection for the fullfilment of the condition (\ref{condition}) is the following:
\begin{equation}
 K_{1,1}=L_{2,1},~~ K_{2,1}=L_{1,1},~~ \phi_{1,1}=\xi_{2,1},~~ \phi_{2,1}=\xi_{1,1},
\end{equation}
and
\begin{equation}
 \frac{v_{2,1}}{v_{1,1}}=\frac{w_{1,1}}{w_{2,1}}\frac{K_{1,1}}{K_{2,1}}\frac{36}{(K_{1,1}^2+4\beta)(K_{2,1}^2+4\beta)}, \label{v21}
\end{equation}
as it can be verified by Eqs. (\ref{M12}) and  (\ref{F12}). This parameter selection ensures that the zeros of the first and the second term of $M_1$ coincide with the zeros of the second and the first term of $M_2$, respectively. It is worth mentioning that, for this particular case of nonlinear potential, the parameter values that enable the fulfillment of the condition (\ref{condition}) depend on the specific solitary wave through the propagation constant $\beta$, in contrast to the linear case where the condition is fulfilled simultaneously for all values of $\beta$ under appropriate conditions. Thus, in this case there do {\it not} exist continuous families of solitary waves for any set of fixed values of the potential parameters.  It is worth noting that this non-existence of continuous families refers to the specific form of Eq. (\ref{nonlinear}) and does not characterize nonlinear complex potentials in general; e.g. for a purely nonlinear complex potential (with $U_1=0$) continuous families can be found similarly to the previous case of a linear potential.

\begin{figure}[pt]
  \begin{center}
  \subfigure[]{\scalebox{\scl}{\includegraphics{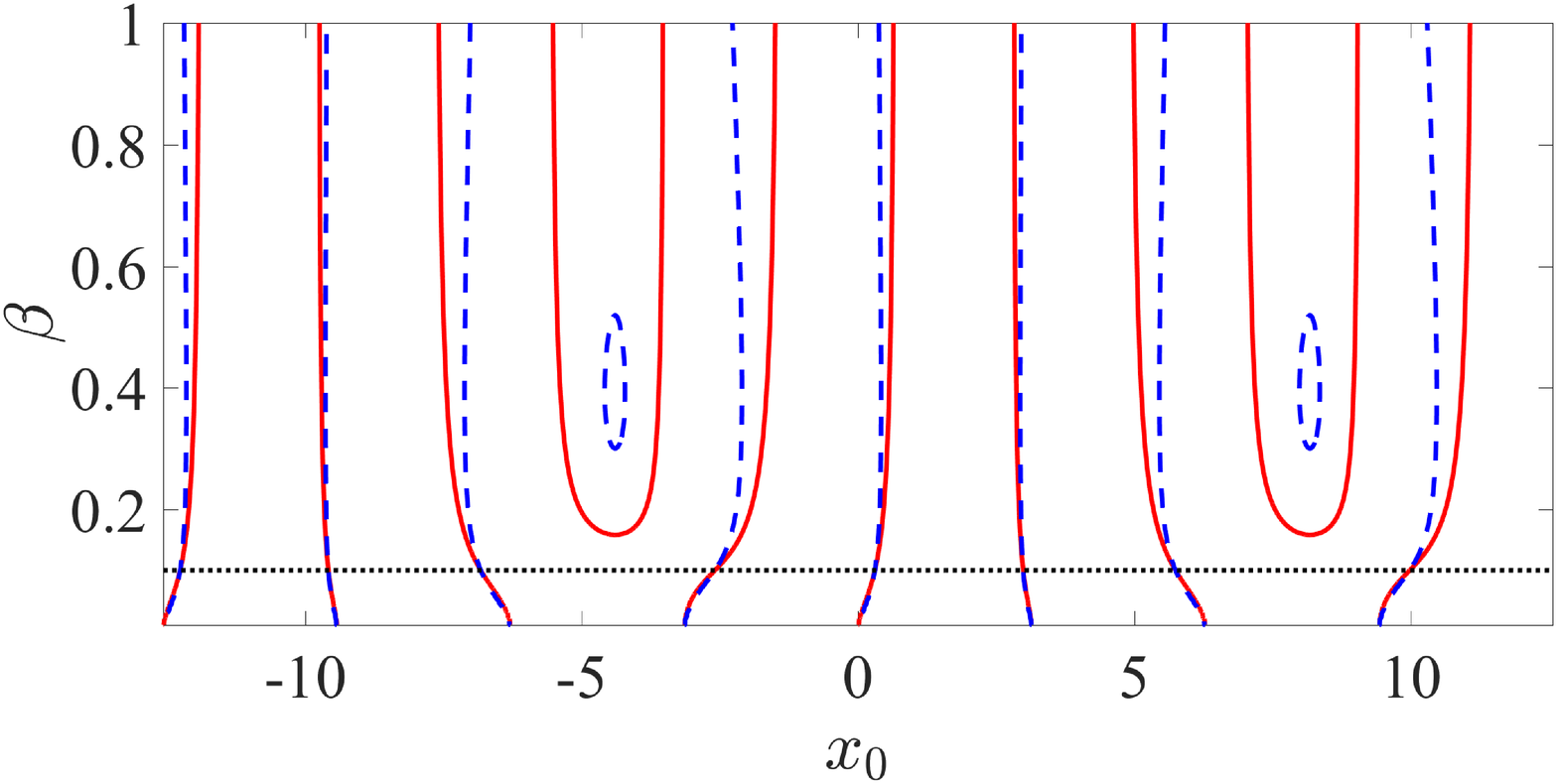}}}
  \subfigure[]{\scalebox{\scl}{\includegraphics{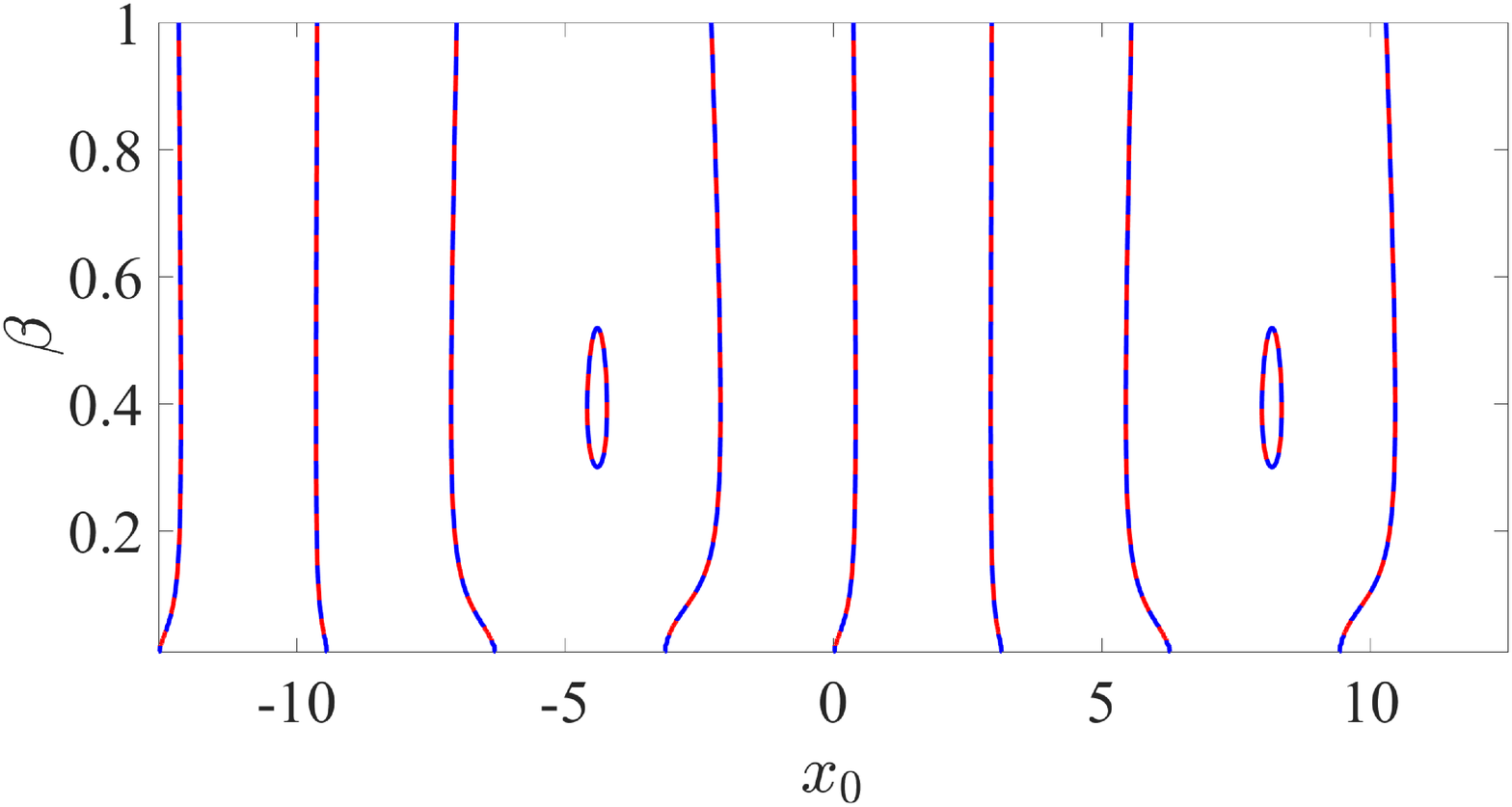}}}
   \caption{Zeros of the two components of the Melnikov vector $M_1$ (red, solid) and $M_2$ (blue, dashed), for a non-symmetric nonlinear complex potential of the form (\ref{nonlinear}) with $\epsilon=0.02$, $v_{1,1}=1$, $w_{1,1}=w_{2,1}=0.2$, $K_{1,1}=L_{2,1}=1$, $K_{2,1}=L_{1,1}=3/2$,  $\phi_{1,1}=\xi_{2,1}=0$, $\phi_{2,1}=\xi_{1,1}=\pi/3$ and $v_{2,1}$ given by Eq. (\ref{v21}) for $\beta=0.1$ (a), and every $\beta$ (b). The condition (\ref{condition}) is fulfilled only for $\beta=0.1$ (black dotted line) in (a) and all values of $\beta$ in (b), where $v_{21}$ is different for every $\beta$ according to Eq. (\ref{v21}).} \label{fig_m4}
  \end{center}
\end{figure}

\begin{figure}[pt]
  \begin{center}
  {\scalebox{\scla}{\includegraphics{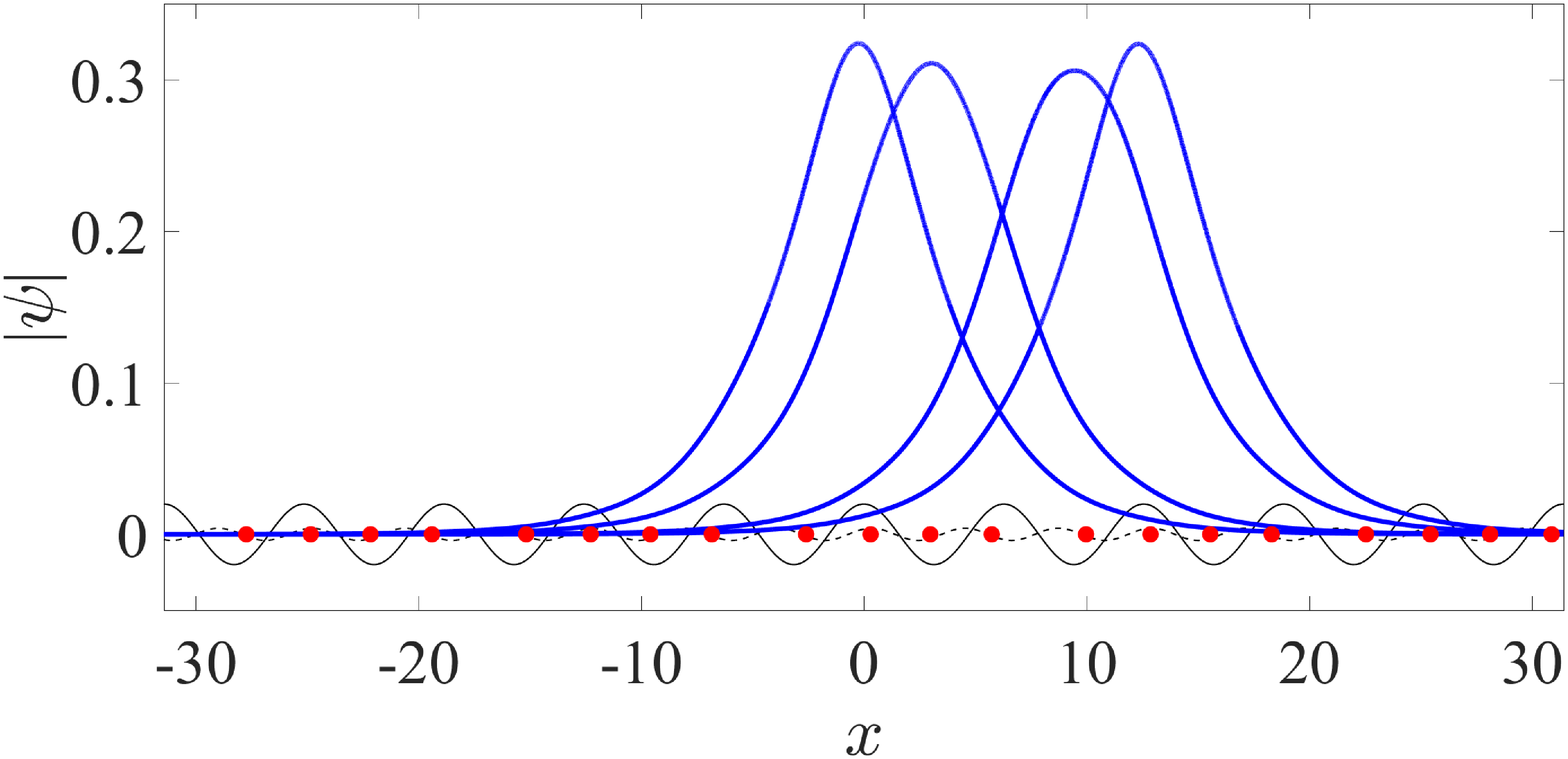}}}
  \caption{Transverse profiles of solitary waves with $\beta=0.1$ and centers corresponding to the zeros of the Melnikov vector  for the case corresponding to Figs. \ref{fig_m4}. The black solid and dashed lines depict the real and the imaginary part of the linear part of the potential and the red circles denote the location of the zeros of the Melnikov function. Two succesive zeros of the Melnikov vector correspond to stationary solutions [Fig. \ref{fig_6bcdef}(c,d)] with identical amplitude profile ($|\psi|$).} \label{fig_6a}
  \end{center}
\end{figure}

\begin{figure}[pt]
  \begin{center}
  {\scalebox{\scl}{\includegraphics{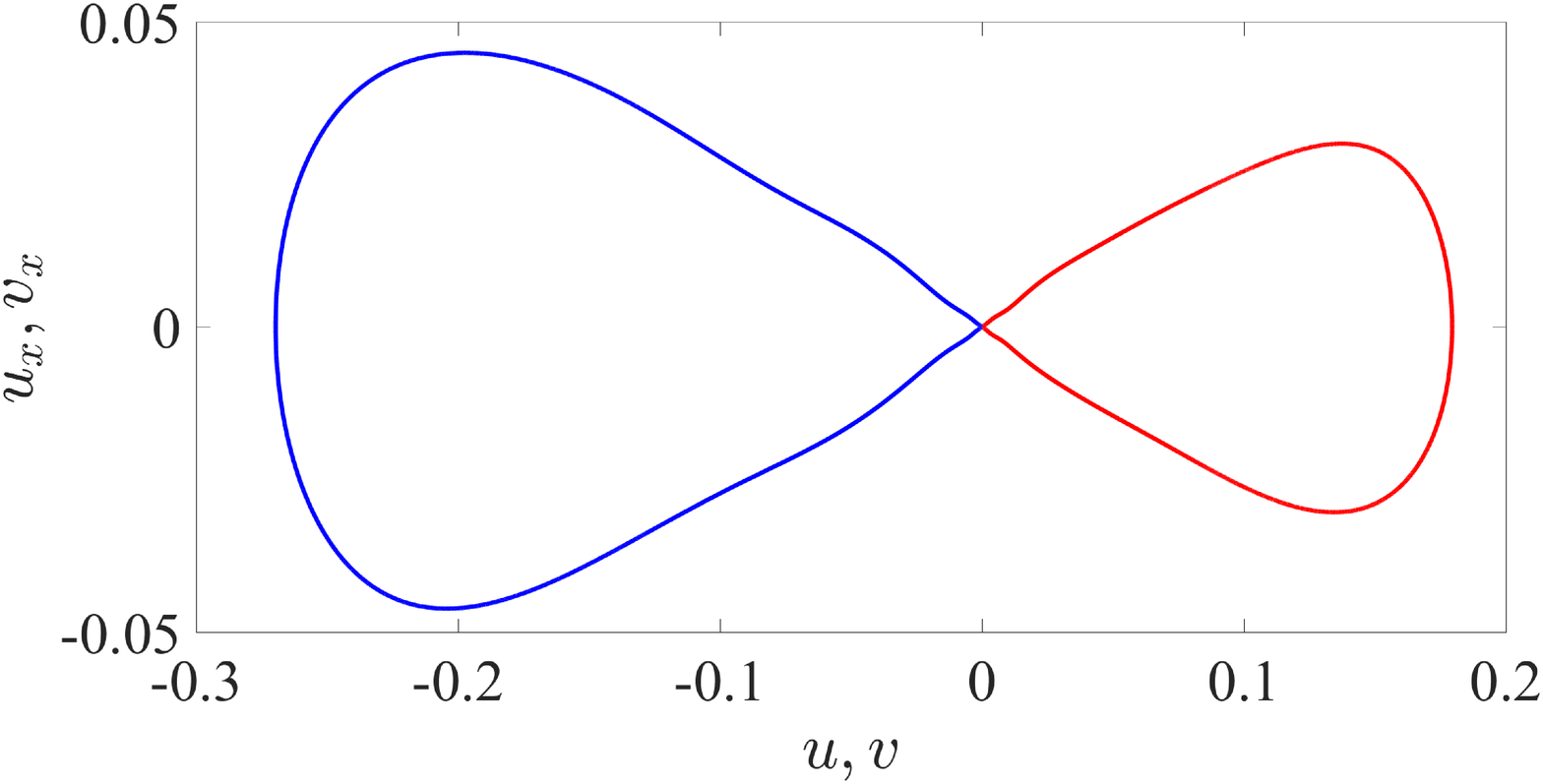}}}
  {\scalebox{\scl}{\includegraphics{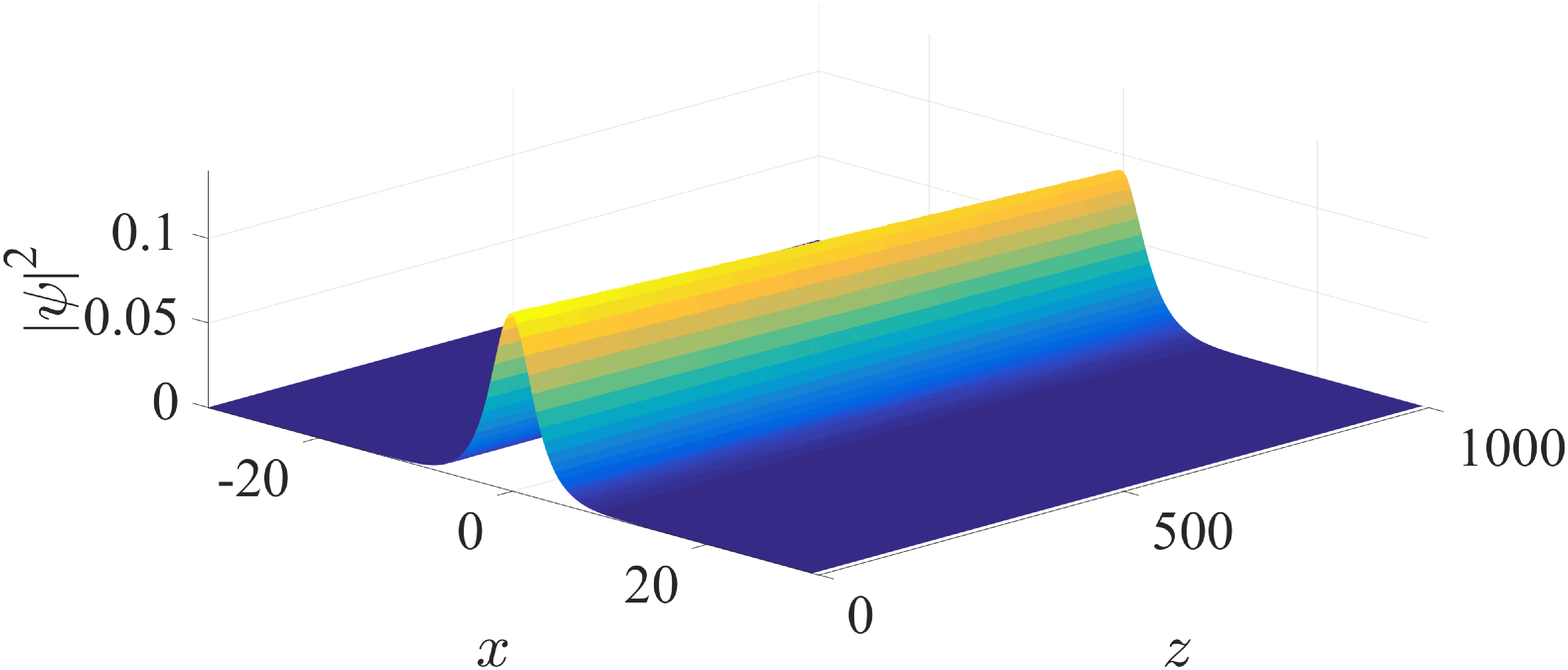}}}  \\
  {\scalebox{\scl}{\includegraphics{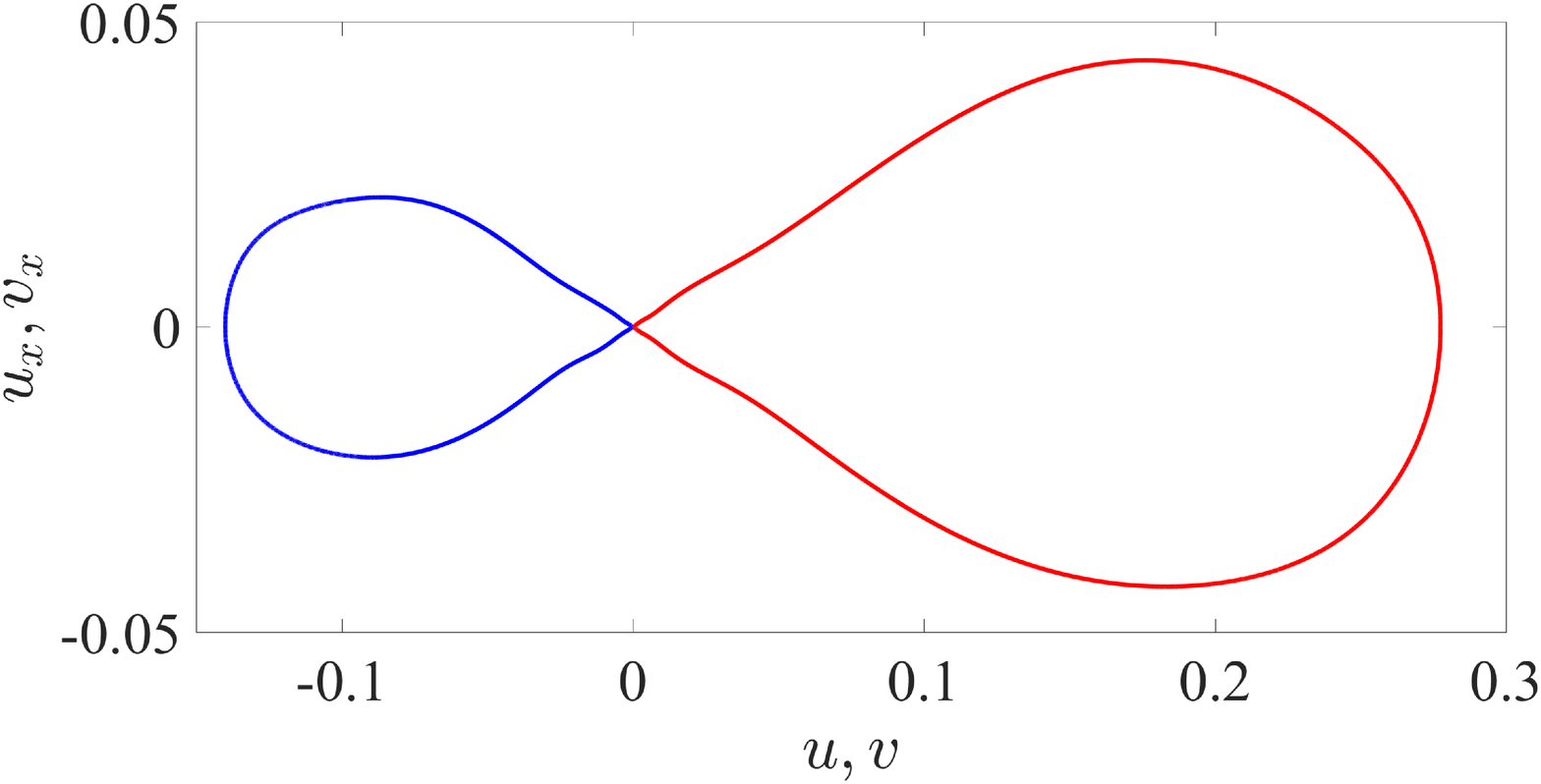}}}
  {\scalebox{\scl}{\includegraphics{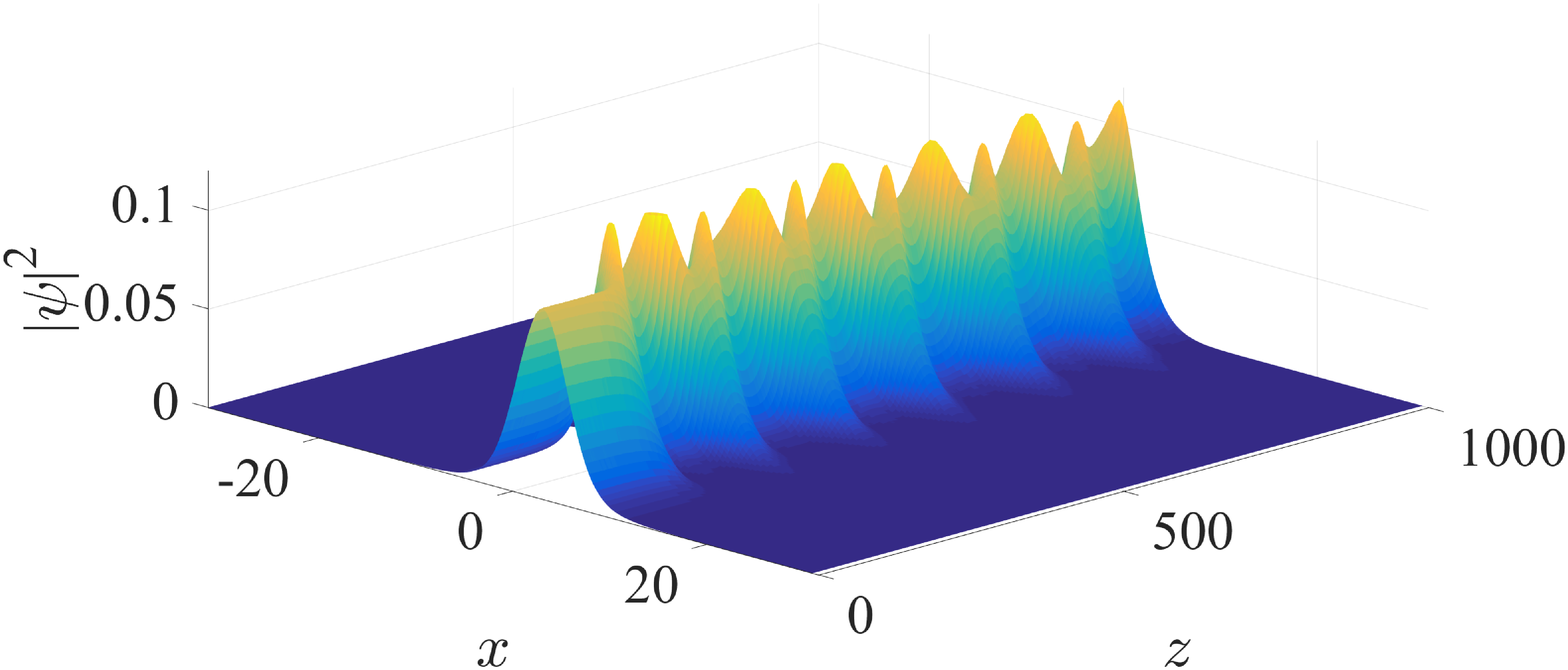}}} \\
  {\scalebox{\scl}{\includegraphics{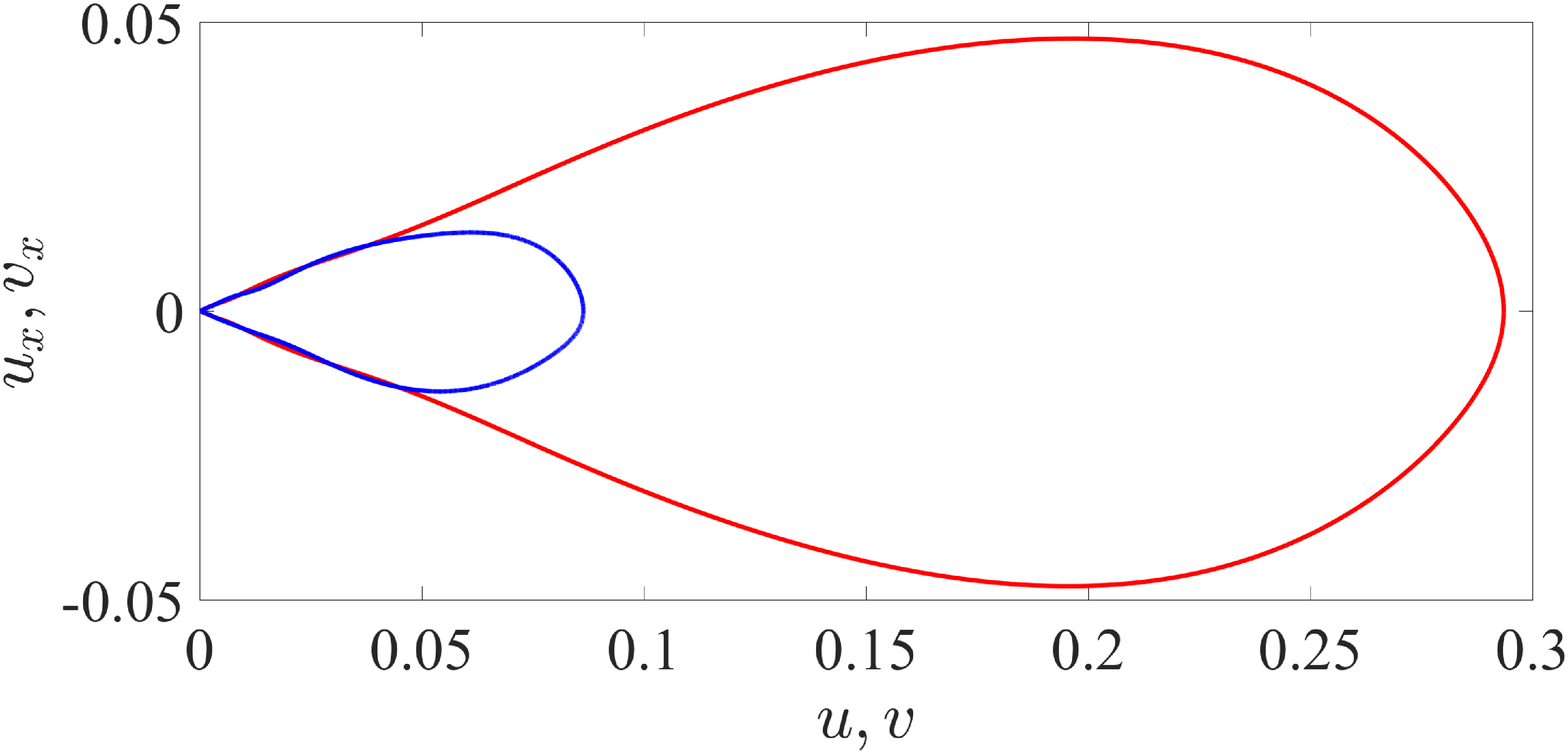}}}
  {\scalebox{\scl}{\includegraphics{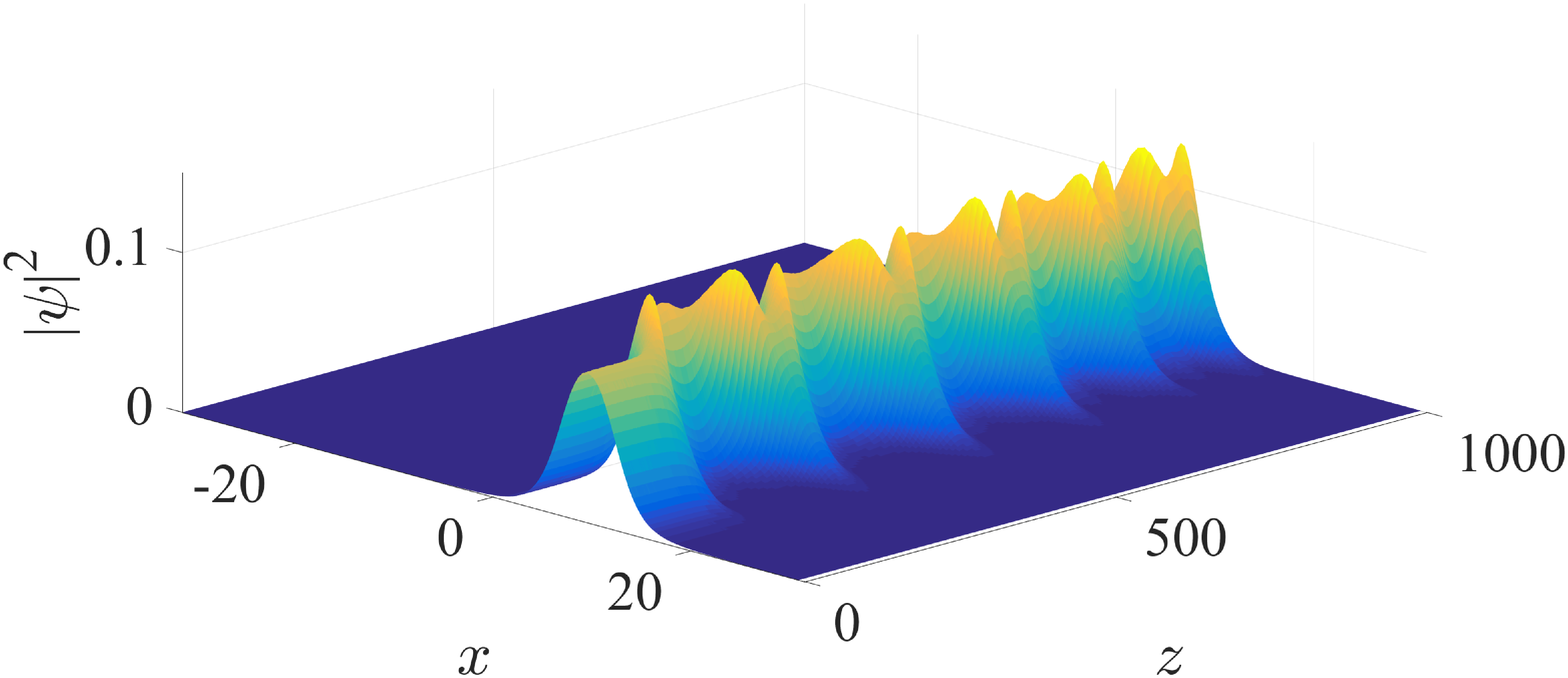}}} \\
  {\scalebox{\scl}{\includegraphics{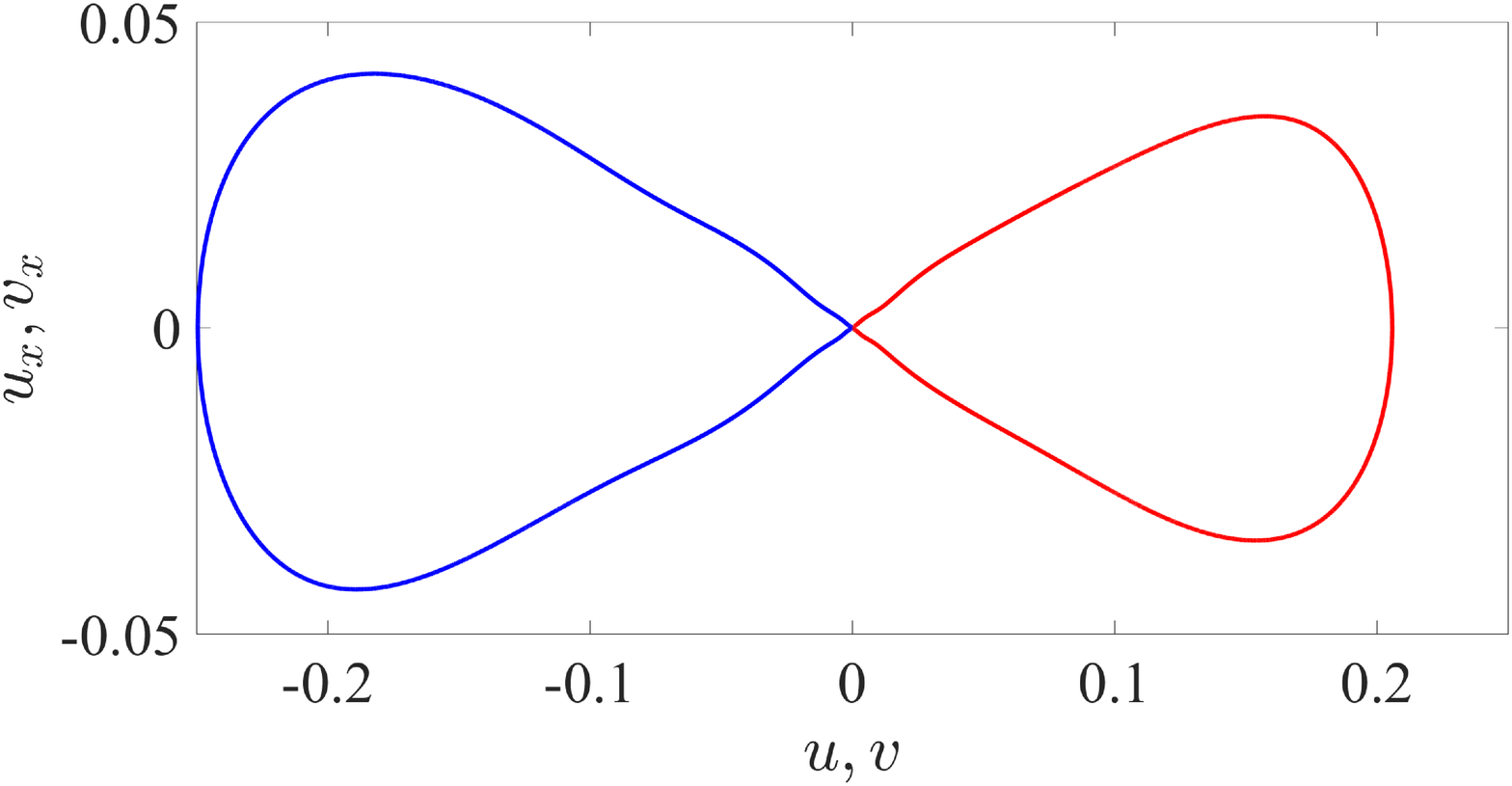}}}
  {\scalebox{\scl}{\includegraphics{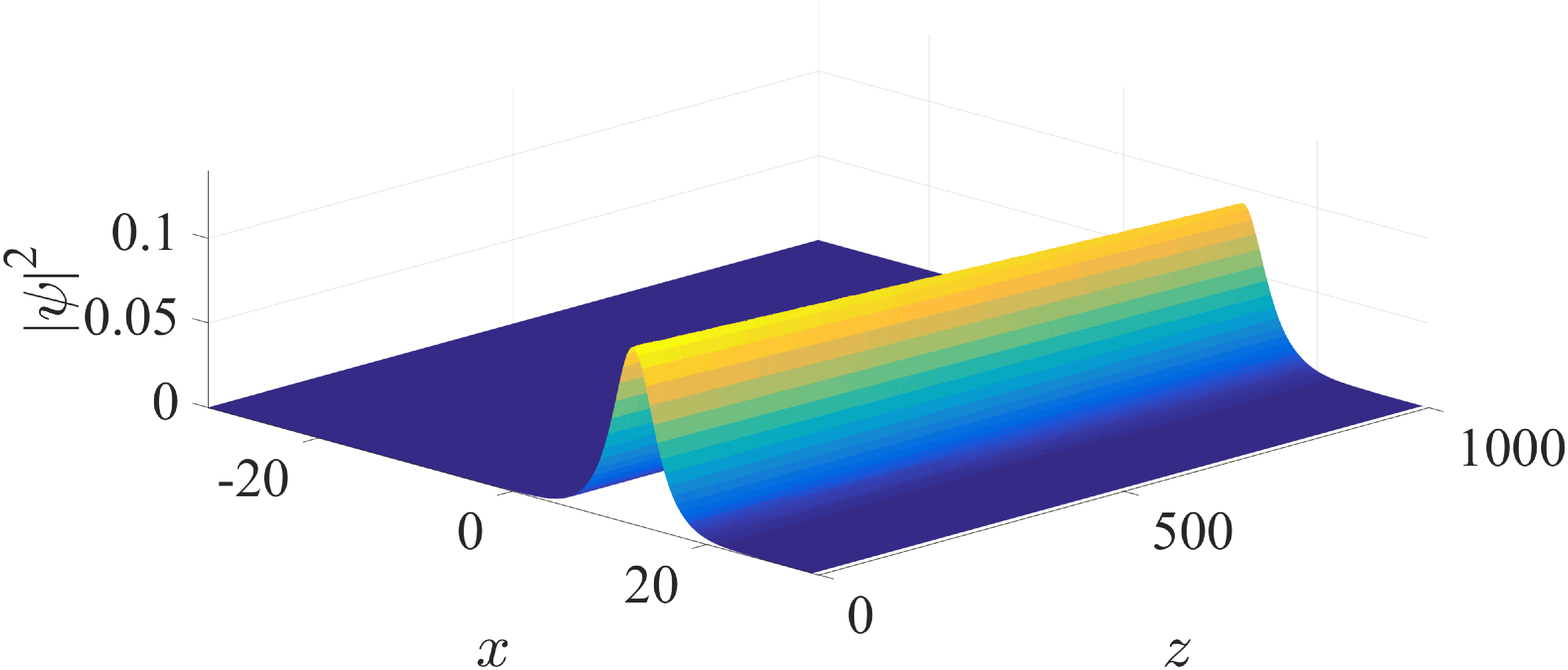}}}
  \caption{Projection of the real ($u$, red) and the imaginary ($v$, blue) parts the homoclinic solutions and propagation dynamics for the case corresponding to Fig. \ref{fig_m4} and Fig. \ref{fig_6a}. } \label{fig_6bcdef}
  \end{center}
\end{figure}

We consider a nonlinear complex potential of the form ($\ref{nonlinear}$) with parameters $\epsilon=0.02$, $v_{1,1}=1$, $w_{1,1}=w_{2,1}=0.2$, $K_{1,1}=L_{2,1}=1$, $K_{2,1}=L_{1,1}=3/2$,  $\phi_{1,1}=\xi_{2,1}=0$, $\phi_{2,1}=\xi_{1,1}=\pi/3$ and $v_{2,1}$ provided by Eq. (\ref{v21}). The value of $\epsilon$ has been chosen smaller than in the case of the linear potential due to the fact that larger values of $v_{2,1}$ result from the use of Eq. (\ref{v21}) and, hence, the effective magnitude of the perturbation ($\sim \epsilon max(v_{i,j}, w_{i,j})$) is larger in this case, in comparison to the ones discussed previously. It is worth emphasizing that this nonlinear potential has no spatial symmetry. The zeros of the two components of the Melnikov vector are depicted in Fig. \ref{fig_m4} for the case where $v_{2,1}$ is given by Eq. (\ref{v21}) for $\beta=0.1$ [Fig. \ref{fig_m4}(a)] and for the case where it is different for every $\beta$ according to Eq. (\ref{v21}) [Fig. \ref{fig_m4}(b)]. In the first case the condition (\ref{condition}) is fulfilled and the zeros of $M_1$ and $M_2$ coincide only for $\beta=0.1$, whereas in the second case this happens for all values of $\beta$. Solitary wave profiles along with their position with respect to the linear part of the potential and the zeros of the Melnikov vector are depicted in Fig. \ref{fig_6a}. The projection of the real ($u$) and the imaginary ($v$) parts the homoclinic solutions as well as the propagation dynamics for cases shown in Fig. \ref{fig_6a} are presented in Fig. \ref{fig_6bcdef}, where it is shown that the solitary waves may have instabilities related to amplitude decay as well as position oscillations.


\section{Conclusions and Future Challenges}
Solitary wave formation in symmetric and non-symmetric, linear and nonlinear complex potentials has been studied by means of Melnikov's perturbation method. Conditions for the existence of stationary solitary waves, bifurcating from the nonlinear modes of the homogenous system, have been obtained analytically. The conditions are expressed as relations involving the real and the imaginary part of the potential, as well as the unperturbed nonlinear solution. This Melnikov analysis suggests that continuous families of stationary solitary waves may exist for complex potentials not restricted by being PT-symmetric or of Wadati-type, for linear as well as  nonlinear complex potentials. In a concrete (yet, generic in its nature due to its Fourier mode nature) example, specific positions around where these solitary waves are located, with respect to the prescribed underlying potential.\

The analytical predictions were numerically tested and the propagation dynamics of the solitary waves has also been demonstrated. Returning to the point of steady-state numerical computations, the use of different numerical methods yielded results of different accuracy, as detailed in the main text. In that light, it would be desirable to obtain a deeper numerical analysis of standing wave problems in such non-Hermitian potentials, as both the LMA and other Newton--Krylov-type methods (such as \texttt{nsoli}) \cite{nsoli} that we used only allowed convergence up to $10^{-6}$ rather than machine precision, contrary to what was the case with \texttt{bvp4c} (which typically indicated convergence to $10^{-10}$). This suggests that the identification of efficient numerical methods for this type of computation may be highly desirable. Also, systematic investigations on the spectral stability of the solutions via eigenvalue computations would be of interest.

Aside, however, from some of these intriguing numerical challenges, the analytical results based on the Melnikov method provide a fundamental understanding of the essentially necessary features of a complex potential that can support the existence of continuous families of stationary solitary waves and are applicable to generic classes of potentials of both theoretical and practical interest. It would naturally be of relevance from such a practical perspective to generalize the present considerations to other settings, including self-defocusing nonlinearities (see relevant work in the context of PT-symmetric systems, e.g., in Ref.~\cite{vas}), that typically bear ``dark'' structures, such as dark solitons, vortices, vortex rings, and so on~\cite{siambook}, as well as multi-dimensional systems (where vortical patterns are also of wide interest~\cite{pismen}). Such studies will be deferred to future publications.

\section*{Acknowledgements}
J.C.-M. thanks financial support from MAT2016-79866-R project (AEI/FEDER, UE). P.G.K. gratefully acknowledges support from NSF-PHY-1602994. A.B. acknowledges partial support from an ORAU research grant from Nazarbayev University 2017-2020.

\clearpage

\end{document}